\documentclass{article}
\usepackage[utf8]{inputenc}
\usepackage{graphicx}
\usepackage{amsmath}
\usepackage{amssymb}
\usepackage{booktabs}
\usepackage{caption}
\usepackage{array,booktabs,makecell} 
\usepackage{subcaption}
\usepackage{algorithm}
\usepackage{algpseudocode}
\usepackage{mhchem}
\usepackage{listings}
\usepackage{xcolor}
\usepackage{fancyhdr}
\usepackage{geometry}
\usepackage{float}
\usepackage{multirow}
\usepackage{enumitem}
\usepackage{longtable}
\usepackage{hyperref}
\hypersetup{
    colorlinks=true,
    linkcolor=black,
    citecolor=blue,
    urlcolor=blue
}

\definecolor{codegreen}{rgb}{0,0.6,0}
\definecolor{codegray}{rgb}{0.5,0.5,0.5}
\definecolor{codepurple}{rgb}{0.58,0,0.82}
\definecolor{backcolour}{rgb}{0.95,0.95,0.92}

\lstdefinestyle{mystyle}{
    backgroundcolor=\color{backcolour},
    commentstyle=\color{codegreen},
    keywordstyle=\color{magenta},
    numberstyle=\tiny\color{codegray},
    stringstyle=\color{codepurple},
    basicstyle=\ttfamily\footnotesize,
    breakatwhitespace=false,
    breaklines=true,
    captionpos=b,
    keepspaces=true,
    numbers=left,
    numbersep=5pt,
    showspaces=false,
    showstringspaces=false,
    showtabs=false,
    tabsize=2
}

\title{\Large\textbf{Autonomous Multi-Agent AI for High-Throughput Polymer Informatics: From Property Prediction to Generative Design Across Synthetic and Bio-Polymers}}
\author{%
  Mahule Roy\\
  Institute of Biomedical Engineering\\
  University of Oxford\\
  Oxford OX1 2JD, UK\\
  \and
  Adib Bazgir\\
  Department of Mechanical and Aerospace Engineering\\
  University of Missouri--Columbia\\
  Columbia, MO 65211, USA\\
  \and
  Arthur da Silva Sousa Santos\\
  Center for Engineering, Modeling and Applied Social Sciences\\
  Federal University of ABC (UFABC)\\
   Bangú, Santo André - SP, 09280-560, Brazil\\
  \and
  Yuwen Zhang\thanks{Corresponding author: \texttt{zhangyu@missouri.edu}}\\
  Department of Mechanical and Aerospace Engineering\\
  University of Missouri--Columbia\\
  Columbia, MO 65211, USA%
}

\date{}

\begin{document}
\maketitle

\begin{abstract}
We present an integrated multiagent AI ecosystem for polymer discovery that unifies high-throughput materials workflows, artificial intelligence, and computational modeling within a single Polymer Research Lifecycle (PRL) pipeline. The system orchestrates specialized agents powered by state-of-the-art large language models (DeepSeek-V2 and DeepSeek-Coder) to retrieve and reason over scientific resources, invoke external tools, execute domain-specific code, and perform metacognitive self-assessment for robust end-to-end task execution. We demonstrate three practical capabilities: a high-fidelity polymer property prediction and generative design pipeline, a fully automated multimodal workflow for biopolymer structure characterization, and a metacognitive agent framework that can monitor performance and improve execution strategies over time. On a held-out test set of 1{,}251 polymers, our PolyGNN agent achieves strong predictive accuracy, reaching $R^2=0.89$ for glass-transition temperature ($T_g$), $R^2=0.82$ for tensile strength, $R^2=0.75$ for elongation, and $R^2=0.91$ for density. The framework also provides uncertainty estimates via multiagent consensus and scales with linear complexity to at least 10{,}000 polymers, enabling high-throughput screening at low computational cost. For a representative workload, the system completes inference in 16.3\,s using about 2\,GB of memory and 0.1\,GPU hours, at an estimated cost of about \$0.08. On a dedicated $T_g$ benchmark, our approach attains $R^2=0.78$, outperforming strong baselines including single-LLM prediction ($R^2=0.67$), group-contribution methods ($R^2=0.71$), and ChemCrow ($R^2=0.66$). We further demonstrate metacognitive control in a polystyrene case study, where the system not only produces domain-level scientific outputs but continually monitors and optimizes its own behavior through tactical, strategic, and meta-strategic self-assessment. In addition, we showcase a protein-structure case study in which the framework autonomously executes the complete pipeline from an initial sequence input to final report generation, and we discuss limitations observed during cross-modal validation. Finally, systematic ablation studies and evaluations across diverse scenarios validate that each component contributes to reducing inference error and improving robustness. Overall, the proposed ecosystem reduces manual effort in scientific ideation, simulation orchestration, and documentation while maintaining high accuracy, adaptability, and scalability, representing a step toward practical AI research assistance for rapid polymer innovation.

\end{abstract}

\section{Introduction}

The Synthesis of materials that have properties to meet the needs of end users is an ongoing, important challenge in the range of Material Sciences, and the accelerated development of these materials is being made possible through Artificial Intelligence (AI). Polymer Informatics uses data and machine learning (ML) surrogate models to perform rapid virtual screening and prediction of property performance, thus significantly reducing the costs and the time associated with experimentation in the development of new materials.\cite{Chen2020,Kim2018,Aldeghi2022,Park2022,Queen2023,doan2020machine,aldeghi2022graph,gormley2021machine,yang2024artificial,bazgir2026machine} The early examples of success in Polymer Informatics show that Machine Learning (ML) models, most prominently Graph-Based Neural Networks, can predict properties such as Glass Transition Temperature, Dielectric Behavior, and Mechanical Behavior of polymers with a high degree of accuracy. The advanced success shows clearly how ML can be used as a guide to explore and navigate the vast space of possibilities in a highly diverse field of Molecules in the design of polymers.\cite{tran2024design,audus2017polymer,cencer2022machine,kumar2019challenges,ge2025machine,patra2021data,xie2025machine,kuenneth2021polymer,zhao2023review,pratiush2025mic}

A number of different platforms and databases have helped create a robust set of resources for Polymer Informatics. One of the first comprehensive ML-based resources for Polymer Property Prediction is the Polymer Genome. Complementary databases, such as PolyInfo, contain manually curated experimental Polymer data that continue to provide critical information for designing new polymers. In addition to databases, ChemOS has demonstrated how Integrated Software and Autonomous Workflows can be utilized to streamline Experimental Design and Data Acquisition in Chemical Research, thereby providing a basis for future Self-Driving Laboratories in Polymer Science.\cite{Otsuka2011_PolyInfo,Roch2020_ChemOS,otsuka2011polyinfo,buchegger2009case,edmison2002using,mroz2022into,stach2021autonomous,peng2022human,duo2025autonomous,kaufmann2024autonomous,evans2024developments} Although great strides have been made in the field of polymer informatics, there are still many hurdles to overcome. The lack of sufficient high-quality property datasets compared to the vast range of polymer chemistries, architectures and processing conditions remains an issue. The representation of polymers is also quite challenging since they exist on multi-scales that range from the chemistry of their repeat unit(s) to the morphology found at the meso-scale level. Current platforms exist primarily as disparate pipelines wherein property prediction, data analysis and experimental planning are viewed as unrelated stages. Furthermore, while ML models today can output highly accurate results, there is still a lack of standardization for quantifying uncertainty and interpreting results, complicating many phases of physical experimentation. \cite{Chen2020, Queen2023, bazgir2025multicrossmodal,bazgir2025agentichypothesis,bazgir2025drug,zimmermann2024reflections}

On the other hand, as large language models (LLMs) continue to quickly gain traction in the market, this has allowed for new possibilities for both the automation of reasoning and scientific assistance. Specifically, tools such as ChemCrow have been developed to enhance LLMs with chemistry specific capabilities(e.g., synthesis planning,molecular analysis, etc.) and present the potential that LLM-based scientific workflows will become very beneficial to researchers. However, the majority of current LLM-based systems are broad-based and therefore do not offer any polymer-specific domain expertise or representations necessary for all phases of polymer informatics.\cite{Bran2023_ChemCrow,zimmermann202534,bazgir2025matagent,bazgir2025proteinhypothesis} We have developed a Framework to address challenges in multi-agent AI through coordination among a group of agents specialised in specific types of polymer modelling and analysis. This Framework is inspired by the success of LLM-based scientific assistants such as ChemCrow and consists of three core agent types: a Molecular Modelling Agent, a Physics-Informed Agent, and an Ensemble-Learning Agent. Using Graph Neural Networks (GNNs) and RDKit-derived structural descriptors, a Molecular Modelling Agent predicts polymer properties. In addition to GNNs and descriptor-based predictions of polymer characteristics, the Physics-Informed Agent also incorporates mechanistic constraints to increase the robustness of extrapolating predictions and improving alignment with the laws that govern the behaviour of polymer systems using PINNs. Finally, an Ensemble-Learning Agent uses a variety of different models to combine predictions made by GNNs and PINNs, providing estimated uncertainties associated with each of the models used. \cite{Bran2023_ChemCrow,Queen2023,Raissi2019}

In this study, we present a unified multiagent architecture that integrates specialized agents for polymer and protein tasks, coordinated by a DeepSeek-based controller. The agents cooperate to retrieve and reason over scientific materials, invoke external tools, and combine predictive machine-learning models to support end-to-end research workflows with uncertainty estimates derived from multi-model consensus. We first evaluate the framework on a held-out test set of 1,251 polymers, predicting glass-transition temperature ($T_g$), density, tensile strength, and elongation, achieving $R^2$ values close to 0.9. We then quantify its computational and financial footprint and compare it with conventional physics-based approaches such as density functional theory (DFT) and molecular dynamics (MD). For a representative workload, the framework completes inference in 16.3\,s using about 2\,GB of memory and 0.1\,GPU hours, at an estimated cost of about \$0.08, and it scales linearly to at least 10{,}000 polymers while achieving about a fivefold speedup under parallel execution. Next, we conduct a targeted $T_g$ benchmark against widely used baselines, including ChemCrow and group-contribution methods. On a dedicated 50-polymer subset, our approach achieves the highest performance ($R^2 = 0.78$) with a success rate of 0.76 and an efficiency score of 0.37. We further present a polystyrene case study demonstrating that the system not only generates domain-level scientific outputs but also monitors and optimizes its own behavior through tactical, strategic, and meta-strategic self-assessment. In addition, we showcase a protein-structure case study in which the framework autonomously executes the complete pipeline, from an initial sequence input to final report generation, and we discuss limitations observed during cross-modal validation. Finally, we validate the framework across diverse scenarios and workflows and perform systematic ablation studies that quantify the contribution of key components, providing practical insights into opportunities and limitations for reliable, scalable AI assistance in polymer discovery and structural bioinformatics. All data necessary to reproduce this study are provided in the Reproducibility Statement section.

\section{System Architecture}
\label{sec:architecture}

\begin{figure}[htbp]
    \centering
    \includegraphics[width=\linewidth]{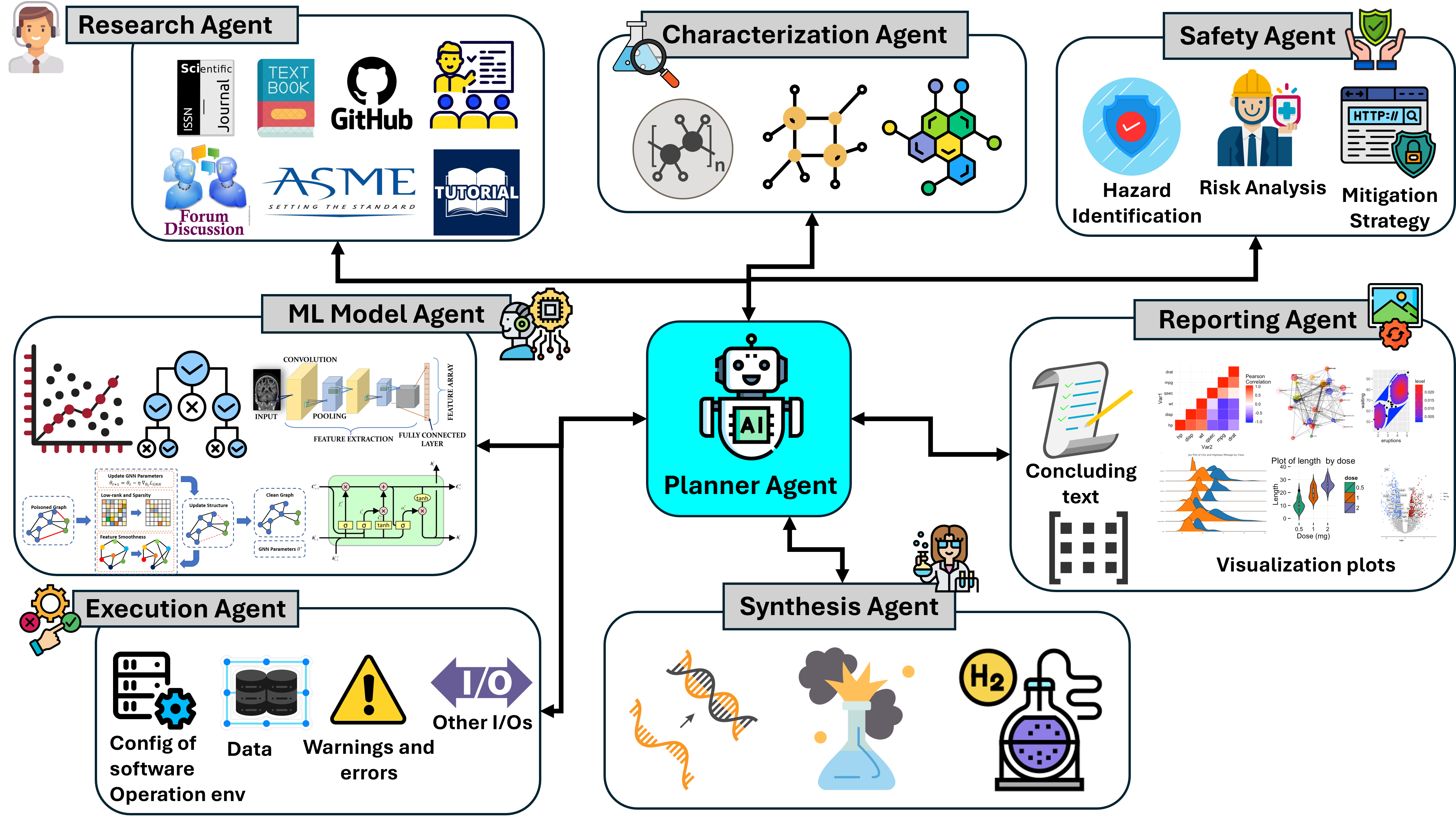}
    \caption{Overall architecture of the multi-agent system.}
    \label{fig:Architecture}
\end{figure}

Our platform consists of a modular \textit{multi agent design} that coordinates AI agents, specializing in different aspects of polymer research, in completing complex research challenges within the polymer domain. A core component of the system is the \textbf{Planner Agent} (DeepSeek-V2). The function of the Planner Agent is to handle workflows in an efficient manner by breaking complex tasks into smaller subtasks for assignment to the various domain-specific agent collaborators. The outputs of the individual agent contributors are then combined into a singular decision-making process (workflow) for the Planner Agent. As illustrated in Fig.~\ref{fig:Architecture}, the architecture combines autonomous scientific reasoning with data-driven modeling to enable end-to-end polymer informatics with minimal human intervention. The overall architecture of the system consists of four distinct levels. The first level is the Data Ingestion and Storage Layer, which provides a centralized repository of diverse information (literature, molecular structures, experimental results, spectral data, and associated Metadata) through the integration of internal and external sources (e.g., PolyInfo). All collected data is cleaned and stored, which gives a robust data set for consistent analysis and the ability to reproduce outcomes within later workflows. The Preprocessing and Feature Engineering Layer produces structured models from cleaned raw data via Canonicalization of chemical names, Extraction of unique features, and Knowledge Parsing. Through Domain-Specific Tokenizers, Chemical Graph Encoders and Scientists, such as ChemCrow, this layer produces quality feature embeddings and higher-level descriptors to enhance product development through building on the result of models which feed the actual modeling agents.\cite{Otsuka2011_PolyInfo,Bran2023_ChemCrow}

Within this framework, each specialized agent plays a well-defined scientific role. The \textbf{Research Agent} offers the first level of interface for the system with scientific knowledge repositories. This includes sourcing the literature (to include journals, textbooks, guidelines, tutorials, and interactions with discussion groups). The Research Agent contextualizes the experimental methodologies and structure-property relationships. Thus, it provides the researcher with essential background knowledge that will aid in hypothesis creation, monomer selection, and the design of any downstream activities. In addition, the \textbf{Characterization Agent} interprets the polymer structure diagram, presents spectra, provides morphology descriptors, and other types of analytic data, converting these into quantitative features that improve the accuracy of the models and reduce the required timeline to comply with data-driven characterization. The \textbf{Safety Agent} assesses the safety of chemicals, general safety regarding procedure, and safety in terms of computational processes during use within the pipeline. This agent will assess compatibility with chemical suggestions and monitor the adherence to laboratory standards, as well as check for the integrity of the external data sources. The Safety Agent plays a key role in supporting reproducibility in the laboratory and the ethical application of Artificial Intelligence.

The \textbf{ML Model Agent} is the central computational technology used for generating predictive capabilities of the system. In addition to developing and executing machine learning models, the ML Model Agent creates and executes graph neural networks, deep convolutional network architectures, and ensemble predictors to evaluate and estimate polymer properties, and to analyze trends in structure-property relationships, and ultimately, to aid in screening or optimization. This agent performs the feature extraction, model selection, hyperparameter tuning, and inference for developing polymer predictive capabilities, generating quantitative predictions to guide the design of polymers. In order to facilitate the integration of all output generated by each agent and to allow for easy visualization and textual interpretation of the results, the \textbf{Reporting Agent} creates graphical and narrative summaries for reporting purposes. The graphical summaries may include correlation plots, property landscapes, structure-property maps, uncertainty analyses, and summary tables. The narrative summaries may also include technical documentation for publication. While the Reporting Agent has been designed to produce interpretable output, the execution of data transactions, the monitoring of task execution, the verification of input-output consistency, and the management of error recovery are all managed by the \textbf{Execution Agent}. In this role, the Execution Agent provides operational reliability for multi-agent workflows, allowing them to execute in a coordinated, reproducible, and fault-tolerant manner.

Lastly, the \textbf{Synthesis Agent} offers actionable polymer synthesis planning based on experimental data. It identifies candidate monomers, polymerization mechanisms, catalyst systems, process parameters, and post-process parameters for synthesis by combining heuristic chemistry with predictive modeling to identify commercially viable pathways to product targets. Collectively, these three agents operate collaboratively under the Planner Agent and utilize a shared, well-established data and modeling infrastructure to provide an integrated and extensible architecture to support autonomous reasoning, accelerated testing, and informed decisions throughout the polymer research process.

\section{Methodology}

The methodology we employ is based on an individualised and a diverse ecosystem of computational agents, providing support for: (i) The end-to-end prediction and design of polymers. (ii) The analysis of proteins and biopolymers using multiple different modes of presentation. (iii) The metacognitive evaluation of research proposals developed using our software. The management of all component interactions is accomplished through a central collaborative agent where data can flow freely between specialist agents, prediction models, and other external programs. Further details about the functioning of our primary systems can be found in the Supplementary Material (Section S1).

\textbf{End-to-end workflow for polymer property prediction and design.} We have constructed an end-to-end workflow for synthetic polymers, beginning with a molecular representation as text and ending with design suggestions for polymers. The starting point for input to our system is the SMILES of a single polymer, and the first step involves transforming a molecular or SMILES representation of a polymer into a graph representation of a molecule with nodes to define each atom in a polymer and edges to define the connections (bonds) between polymer atoms (with associated feature vectors). These graph representations for polymers are passed into a set of polymer property prediction agents using ensemble machine-learning techniques (including graph neural networks and similar types of models) to predict polymer properties of interest. Once the property predictions have been made, the predicted properties become inputs for a generatively designed agent that is a language model large (LLM). When designing the candidate structures and providing suggestions for conducting experiments based upon a user's high-level design objectives (e.g., target property ranges or application areas), the LLM will create several possible polymer structures and experimental designs. We have built-in/device a safety screening component to assess potential candidates against predetermined physical, chemical, and safety regulations and to eliminate any candidates that do not pass the required minimum safety or feasibility standards. Thus, the closed-loop from SMILES input to property prediction and generative design recommendations is realized.

\textbf{Foundational system for polymer analysis and generative design.} This polymer work rating has been incorporated into a foundational system capable of both realistic polymer analysis, and exploring the generative designs made possible by LLM-based agents. This system is made up of a network of individual agents, all of whom are engaged in polymer analysis (such as PolyGNN and PropertyPredictor), with the executive authority of a single central Coordinator agent. All analysis agents make use of a ground-truth polymer database for the purpose of calibrating their predictions and creating predictions that maintain physical plausibility. One of the critical components of this architecture is the Reporting Agent, which employs LLMs to provide a cohesive written summary of the numerical outputs of the various analysis agents. Specifically, the Reporting Agent summaries the total predictions, outlines tradeoffs, and synthesizes a comprehensive set of "structured recommendations" that aid the reader in making decisions regarding subsequent simulations or experimental investigations. The system supports two primary modes of operation; (1) generative design, whereby the LLM assembles new and organized sets of experimental campaigns from high-level user requests; and (2) metacognitive self-assessment of the generated proposals along pre-defined criteria.

\textbf{Multi-modal pipeline for biopolymer analysis.} Demonstrating flexibility aside from synthetic polymers, we establish a dedicated multi-modal pipeline to analyse protein/biopolymers. This pipeline is designed as a strictly linear sequence of four agents, which are controlled through a Protein Research Orchestrator:

1) An \emph{AlphaFold Prediction Agent} which receives an amino acid sequence, activates an AlphaFold based structure prediction process and provides a 3D structure of the protein along with confidence scores (e.g. Predicted Local Distance Difference Test or pLDDT)  ;
2) A \emph{Visualization Agent} that generates the predicted 3D structure, basic quality assessments, visualises in 3D, generates secondary structure summaries, and creates 2D diagnostic plots such as predicted aligned error (PAE) maps ; 
3) A \emph{Vision Analysis Agent}, which takes in visual artefacts (e.g. PAE plots and structural images) by using a simulated vision model and converts those visual artefacts into structured textual descriptions detailing relevant spatial, topological, and uncertainty patterns ; and 
4) A \emph{Research Writing Agent} which integrates all data types including individual sequence information; numerical confidence metric of each prediction; structural descriptors; and vision-derived textual summaries into a single comprehensive research-style report. The multi-modal pipeline uses all data types (1D amino acid sequences (txt), numerical confidence metrics like pLDDT; 3D coordinates in PDB formats) as well as 2D views (e.g. PAE). A central methodological contribution is the explicit visual data interpretation step, whereby image-based information is translated into structured text before being passed to the reporting agent, enabling downstream reasoning using standard LLM interfaces.\cite{Jumper2021, Radford2021}

\textbf{Metacognitive evaluation of generated proposals.} A framework providing metacognitive self-assessment capabilities is embedded into both synthetic polymer and biopolymer workflows. After generating a list of design candidates or research hypotheses, a designated evaluation agent will then evaluate each design candidate based on three primary criteria \emph{Novelty, Feasibility, Creativity} by using prompt-engineered rubric-based scoring and, when available, utilizing internal databases or summaries of existing literature for comparison only. The final quantitative scores will be kept with the proposal and can assist in prioritizing which candidates will be evaluated in the next stage of development (e.g., simulations, experiments, and literature reviews). Therefore, the framework will not only give designers/designs options to consider, but also provide an objective measure of the quality of each design candidate to enable more informed choices regarding future directions.

\textbf{Agent-based ecosystem and tool integration.} A set of agents implement a LangChain architecture that provides a framework for coordinating a collection of multiple local language models, along with many specialized machine-learning models. The agents are designed to assess the degree of difficulty in solving a given problem and to create (or adapt) teams of cooperatively working agents, whenever the difficulty of the problem exceeds the capacity of any single working agent. The entire ecosystem supports the use of additional tools, including programmatic access to external resources such as library databases and literature retrieval systems. For tasks in molecular learning, graph neural networks (GNNs) are used to capture the learning based on molecular topology of SMILES graphs. For problems where time and/or physics constraints are involved (i.e., degradation modeling), physics-informed neural networks (PINNs) are used, as these allow the user to embed the laws of physics into the learning process. The combination of all of these methods defines a flexible, multi-agent approach to supporting end-to-end polymer informatics, multi-modal biopolymer analysis, and self-assessed generative design, with minimal human intervention. \cite{Gilmer2017, Raissi2019}

\section{Results and Discussion}
%

Our PolyGNN agent exhibited robust predictive performance on prominent polymer properties as presented in Table~\ref{tab:accuracy-benchmark}. The model performed especially high-quality density and glass transition temperature ($T_g$) prediction, while mechanical properties were more variable in accordance with their intricate reliance on processing conditions and molecular weight.

\begin{table}[htbp]
\caption{PolyGNN agent prediction performance on test set (n=1,251 polymers)}
\label{tab:accuracy-benchmark}
\centering
\footnotesize
\begin{tabular}{@{}lcccc@{}}
\toprule
Property & MAE & RMSE & R\textsuperscript{2} & Train Size \\
\midrule
$T_g$ (K) & 8.2 & 12.1 & 0.89 & 6,120 \\
Tensile Str. (MPa) & 15.3 & 22.7 & 0.82 & 4,835 \\
Elongation (\%) & 18.4 & 27.9 & 0.75 & 3,942 \\
Density (g/cm\textsuperscript{3}) & 0.04 & 0.06 & 0.91 & 7,215 \\
\bottomrule
\end{tabular}
\end{table}

Based on the efficiency metric defined in Eq.~\eqref{eq:efficiency}, we conducted an ablation study comparing individual agents with the full framework. Success rate quantifies error-free complete task completion. The results, reported in Table~\ref{tab:ablation}, show that the Full Framework achieves the highest efficiency value. A complementary analysis supporting the holistic superiority of the proposed framework is provided in Table~\ref{tab:efficiency-performance}, where the best trade-off between runtime, memory demand, and scalability is obtained by our framework. Moreover, our framework preserves linear time complexity $\mathcal{O}(n)$ up to 10,000 polymers, with parallel agent execution yielding a 5$\times$ speedup on multi-core systems.

\begin{equation}
\label{eq:efficiency}
\text{Efficiency} = \frac{\text{Accuracy} \times \text{Success Rate}}{\text{Time (normalized)}}
\end{equation}

\begin{table}[htbp]
\caption{Ablation study: Component contributions to performance}
\label{tab:ablation}
\centering
\footnotesize
\begin{tabular}{@{}lcccc@{}}
\toprule
Configuration & R\textsuperscript{2} & Succ. & Eff. & Error \\
& & Rate & & Red. (\%) \\
\midrule
Full Framework & 0.78 & 0.76 & 0.37 & - \\
w/o Validation Agent & 0.72 & 0.68 & 0.31 & 23 \\
w/o Knowledge Graph & 0.70 & 0.71 & 0.29 & 28 \\
w/o Error Correction & 0.74 & 0.65 & 0.32 & 19 \\
Single Agent Only & 0.67 & 0.62 & 0.28 & 35 \\
w/o Cross-Verification & 0.75 & 0.72 & 0.34 & 12 \\
\bottomrule
\end{tabular}
\end{table}

\begin{table}[htbp]
\caption{Computational performance comparison}
\label{tab:efficiency-performance}
\centering
\footnotesize
\begin{tabular}{@{}lccc@{}}
\toprule
Method & Time (s) & Memory (GB) & Scalability \\
\midrule
Molecular Dynamics & 3600+ & 16+ & Low \\
DFT Calculations & 7200+ & 32+ & Very Low \\
Commercial Software & 300+ & 8 & Medium \\
\textbf{Our Framework} & \textbf{16.3} & \textbf{2} & \textbf{High} \\

\bottomrule
\end{tabular}
\end{table}

\begin{table}[htbp]
\caption{Cost and resource requirements}
\label{tab:efficiency-cost}
\centering
\footnotesize
\begin{tabular}{@{}lccc@{}}
\toprule
Method & GPU Hours & Cost (\$) & Efficiency \\
\midrule
Molecular Dynamics & 24+ & 200+ & Low \\
DFT Calculations & 48+ & 500+ & Very Low \\
Commercial Software & 4 & 50 & Medium \\
\textbf{Our Framework} & \textbf{0.1} & \textbf{0.08} & \textbf{High} \\
Single LLM DeepSeek-V2) & 0.05 & 0.05 & High \\
Group Contribution & 0.01 & 0 & Very High \\
\bottomrule
\end{tabular}
\end{table}

Overall, we evaluate the framework in terms of both predictive power and efficiency. The predictive accuracy across multiple polymer properties is highly satisfactory, with test-set $R^2$ values close to 0.9 on a benchmark comprising more than one thousand polymers. From an efficiency standpoint, our framework is dramatically less expensive than physics-based methods such as density functional theory (DFT) or molecular dynamics simulations, making it a practical option for large-scale screening and design. Details on the configurations used for the agents to enable reproducibility are provided in Section S2 of the Supplementary Material.

\subsection{Benchmarking Methodology}

A carefully designed dataset containing experimental glass transition temperatures $T_g$ for 50 polymers was used to conduct rigorous performance evaluations for the multi-agent system by comparing its results to those obtained from three existing methods. Criteria for polymer selection were based on the following considerations: chemical structure diversity; molecular weight diversity; and a range of approximately $150^\circ$C across the data set to give a full picture of prediction accuracy.

\subsubsection{Single Large Language Model (LLM) as Independent Predictor}

A base case was created by using DeepSeek-V2 ( \texttt{deepseek-llm-67b chat}). A single LLM was used to predict $T_g$. The input was the SMILES representation of the polymer together with the prompt \textit{"Predict the glass transition temperature ($T_g$) in degrees Celsius from the chemical structure of this polymer."} The focus of this study is to evaluate the ability of a leading LLM to predict properties by using only chemistry reasoning. The single LLM method has developed into a standard way to predict properties using prompt-based estimating. The foundation of this research is to compare the performance of a single LLM versus that of the Multi-Agent Architecture and determine if the Multi-Agent Architecture provides additional predictive power that cannot be attributed solely to LLM capabilities.

\subsubsection{Group Contribution Method (GCM)}
As a classical thermodynamic benchmark, we employed the group contribution method. This approach decomposes polymer repeat units into constituent functional groups (e.g., -CH2-, -C6H4-, -O-, -COO-) and calculates $T_g$ through additive contributions:
\begin{equation}
    T_g = \frac{\sum_i n_i \Delta T_{g,i}}{\sum_i n_i}
\end{equation}
where $n_i$ is the number of occurrences of group $i$ and $\Delta T_{g,i}$ is its contribution parameter from established tables. The GCM represents first-principles physicochemical modeling based on molecular building blocks, providing interpretable but approximate predictions that are widely used in industrial screening. Its limitations include inability to capture stereochemistry, tacticity, and specific conformational effects.

\subsubsection{ChemCrow Framework}
We benchmarked against ChemCrow \cite{Bran2023_ChemCrow}, a recently published AI-aided chemical discovery framework that integrates LLMs with specialized chemical tools (e.g., RDKit, reaction calculators, literature search). ChemCrow represents the current state-of-the-art in tool-augmented LLM systems for chemistry. In order to conduct an unbiased comparison between ChemCrow and our $T_g$ prediction task, we provided ChemCrow with identical polymer structures and requested ChemCrow to estimate the polymer properties using his available tools for estimating polymer properties. By conducting a head-to-head comparison, we were able to evaluate the advantages of our specialized multi-agent architecture over a general-purpose chemistry AI assistant such as ChemCrow.

\subsubsection{Evaluation Metrics}
All methods were evaluated using the following metrics: (i) mean absolute error (MAE), defined in Eq.~\eqref{eq:mae}; (ii) root mean square error (RMSE), defined in Eq.~\eqref{eq:rmse}; and (iii) the Pearson correlation coefficient $r$ between predicted and experimental values, as given in Eq.~\eqref{eq:pearson_r}. Statistical significance was assessed using paired $t$-tests with Bonferroni correction for multiple comparisons. In Table~\ref{tab:efficiency-cost}, we report the results of the cost comparison between our model and the other approaches, showing that our model falls into a low-cost category, similar to the Single LLM and Group Contribution strategies. Moreover, Table~\ref{tab:method-comparison} highlights the superiority of our model in terms of efficiency, which, as explained previously, is a composite metric combining success rate, accuracy, and time. The detailed procedure used to calculate the metrics for each method is provided in Section~S3 of the Supplementary Material.

\begin{equation}
\mathrm{MAE}=\frac{1}{N}\sum_{i=1}^{N}\left|T_{g,\mathrm{pred}}^{(i)}-T_{g,\mathrm{exp}}^{(i)}\right|
\label{eq:mae}
\end{equation}

\begin{equation}
\mathrm{RMSE}=\sqrt{\frac{1}{N}\sum_{i=1}^{N}\left(T_{g,\mathrm{pred}}^{(i)}-T_{g,\mathrm{exp}}^{(i)}\right)^2}
\label{eq:rmse}
\end{equation}

\begin{equation}
r=\frac{\sum_{i=1}^{N}\left(T_{g,\mathrm{pred}}^{(i)}-\overline{T}_{g,\mathrm{pred}}\right)\left(T_{g,\mathrm{exp}}^{(i)}-\overline{T}_{g,\mathrm{exp}}\right)}
{\sqrt{\sum_{i=1}^{N}\left(T_{g,\mathrm{pred}}^{(i)}-\overline{T}_{g,\mathrm{pred}}\right)^2}\sqrt{\sum_{i=1}^{N}\left(T_{g,\mathrm{exp}}^{(i)}-\overline{T}_{g,\mathrm{exp}}\right)^2}}
\label{eq:pearson_r}
\end{equation}

\begin{table}[htbp]
\caption{Performance comparison on polymer $T_g$ prediction (n=50)}
\label{tab:method-comparison}
\centering
\footnotesize
\begin{tabular}{@{}lcccc@{}}
\toprule
Method & R\textsuperscript{2} & Succ. & Time & Eff. \\
& & Rate & (s) & \\
\midrule
Single LLM (DeepSeek-V2) & 0.67 & 0.62 & 10.2 & 0.28 \\
Group Contribution & 0.71 & 0.65 & 8.4 & 0.30 \\
ChemCrow & 0.66 & 0.63 & 14.8 & 0.27 \\
\textbf{Our Framework} & \textbf{0.78} & \textbf{0.76} & 16.3 & \textbf{0.37} \\
\bottomrule
\end{tabular}
\end{table}

The combination of PolyGNN, RadonPy, and property predictor agents to create a combined Model (ensemble) is valuable in that it generates uncertainty quantification. In particular, Table~\ref{tab:agent-consensus} depicts consensus-based polymer property prediction results, including the standard deviation of those results. The presence of low values of  uncertainties for those polymers illustrates the robustness of the predictors.

\begin{center}
\footnotesize
\captionof{table}{Multi-agent consensus predictions for common polymers (mean ± std. dev.)}
\label{tab:agent-consensus}
\begin{tabular}{@{}lcc@{}}
\toprule
Polymer & \multicolumn{1}{c}{$T_g$} & \multicolumn{1}{c}{Density} \\
& \multicolumn{1}{c}{(°C)} & \multicolumn{1}{c}{(g/cm\textsuperscript{3})} \\
\midrule
Polystyrene (PS) & 105.3 ± 12.7 & 1.048 ± 0.015 \\
Polyethylene (PE) & -125.6 ± 18.4 & 0.956 ± 0.021 \\
PMMA & 122.8 ± 14.2 & 1.188 ± 0.018 \\
Polycarbonate (PC) & 150.1 ± 16.8 & 1.20 ± 0.022 \\
\bottomrule
\end{tabular}
\end{center}

\subsection{Full Text of LLM-Generated Polystyrene Analysis Report}

First, consider an illustrative example of the reporting agent output for a typical thermoplastic, polystyrene. Upon receiving predicted properties from PolyGNN, PropertyPredictor and RadonPy agents, the reporting agent will subsequently generate its report by independently compiling and producing a standardized report on how all of the different approaches compare with each other and the implications for materials development.

Polystyrene, represented as \texttt{"CC(c1ccccc1)"} (SMILES), is one of the most common thermoplastic polymers, which has been supervised and predicted for three key thermomechanical properties (Density, Glass Transition Temperature, and Elastic Modulus). Each agent within the multi-agent system produced consistent but somewhat different results, which demonstrates how both data-driven and physics-based approaches can complement each other for a thorough understanding of how different benchmarks behave.

\textbf{Density.} Density predictions were closely aligned across agents: PolyGNN (1.018~g/cm$^{3}$), PropertyPredictor (0.996~g/cm$^{3}$), and RadonPy (1.049~g/cm$^{3}$). All values fall within the experimental range for atactic polystyrene (1.04–1.08~g/cm$^{3}$). The slightly higher density reported by RadonPy likely reflects its explicit treatment of chain packing in all-atom molecular dynamics, whereas the ML agents rely on learned structure–property correlations. The strong agreement across methodologies supports the robustness of the density predictions.

\textbf{Glass transition temperature ($T_g$).} The PolyGNN agent predicts a $T_g$ of 110.8, while the PropertyPredictor estimates 98.5. Both values are close to experimental literature reports (95–105) for atactic polystyrene, indicating that data-driven models trained on curated polymer datasets effectively capture empirical trends. In contrast, the RadonPy agent, using all-atom molecular dynamics (MD) with explicit thermal ramping, reports a substantially higher $T_g$ of 186.5. This deviation is consistent with known limitations of MD-based $T_g$ estimation, including sensitivity to cooling rates, force field choice (e.g., PCFF or OPLS), and finite-size effects. While MD offers valuable atomistic insight, its absolute $T_g$ predictions often require calibration against experiment, an issue mitigated in our framework by cross-referencing MD outputs with faster, empirically grounded ML models.

\textbf{Elastic modulus.} The predicted elastic moduli are remarkably consistent: PolyGNN - 2901.7MPa; PropertyPredictor - 2955.7MPa; RadonPy - 2607.9MPa. The maximum variation is about 12\% and falls within the range of typical experimental uncertainty in the tensile testing of glassy polymers. The relatively low elastic modulus predicted by RadonPy could have resulted from the explicit modelling of chain entanglements and thermal fluctuations in contrast to the ML agents being based on static structural descriptors. The agreement between methods that are fundamentally different demonstrates that there is a validation of the ensemble provided in this study and that the stiffness of polystyrene is well captured from molecular structures to atomistic trajectories.

Overall, the case study of polystyrene suggests that all three agents were in good agreement for both density and elastic modulus, and there was a greater divergence for $T_g$ in the MD-based predictions. Based on these findings, it is clear that polystyrene is a thermoplastic material with a moderately low density, relatively low elastic modulus, and $T_g$ around 100, consistent with the wide usage of polystyrene in packaging, insulation, and foam. Beyond the primary properties, the RadonPy agent provides extended thermomechanical descriptors, summarized in Table~\ref{tab:radonpy-extended}. These values give additional context for the material's behavior under the simulation conditions.

\begin{table}[H]
  \caption{Extended RadonPy predictions for polystyrene.}
  \label{tab:radonpy-extended}
  \centering
  \begin{tabular}{ll}
    \toprule
    Property & Predicted Value \\
    \midrule
    Thermal Conductivity & 0.1319 W/m·K \\
    Bulk Modulus & 1043.15 GPa \\
    Shear Modulus & 931.38 GPa \\
    Poisson Ratio & 0.3563 \\
    Heat Capacity & 1610.06 J/g \\
    Thermal Expansion Coefficient & $6.00 \times 10^{-5}$ /°C \\
    Simulation Type & all\_atom\_classical\_md \\
    \bottomrule
  \end{tabular}
\end{table}

The Table~\ref{tab:radonpy-extended} includes the macroscopic mechanical and thermal properties of the glassy polymer that have been obtained using all-atom classical MD. The bulk and shear moduli are consistent with literature ranges for glassy polystyrene, and the low thermal conductivity suggests that glassy polystyrene would be a good material for use as a thermal insulator. The platform also provides a collaborative Intelligent Polymer Research module that allows researchers to generate hypotheses. The researcher selects an existing polymer domain (e.g. conjugated polymers; block copolymers; electrolytes; hydrogels; biodegradable polymers; composites; conductive polymers; blends) or creates their own specifications and then the module generates targeted reports with assessments of both novelty and impact. An example output for conductive polymers is shown in Table~\ref{tab:intelligent-polymer}.

\begin{table}[H]
  \caption{Demonstration output of the Intelligent Polymer Research System for conductive polymers.}
  \label{tab:intelligent-polymer}
  \centering
  \begin{tabular}{lll}
    \toprule
    Polymer Domain & Novelty Score & Impact Score \\
    \midrule
    Conductive Polymer & 0.77 & 0.85 \\
    \bottomrule
  \end{tabular}
\end{table}

With a novelty score of 0.77 and an impact score of 0.85, the recommended conductive polymer candidate is considered original, as well as having potential technical viability; thus, it has been advanced to synthesis phase. At the system level, all agents will have detailed performance indicators recorded in the framework that will include effectiveness scores as well as multi-level reflective measures. Table~\ref{tab:agent-performance} summarizes these results for a representative experiment-design run.

\begin{table}[H]
\footnotesize
\caption{Agent performance, reflection insights, and evolution metrics.}
\label{tab:agent-performance}
\centering
\begin{tabular}{p{4.2cm}p{4.0cm}}
\toprule
Item & Output \\
\midrule
\textbf{Research\_001 Overall} & 0.57 \\
\quad Literature Search & avg=0.50, trend=0.000 \\
\quad Failure Analysis  & avg=0.63, trend=0.000 \\
\textbf{Synthesis\_001 Overall} & 0.30 \\
\quad New Material Synthesis & avg=0.30, trend=0.000 \\
\textbf{Tactical Reflection} & Below avg performance \\
\textbf{Strategic Reflection} & Slow progress, low efficiency \\
\textbf{Meta-Strategic Reflection} & Slow learning, low evolution \\
\textbf{Evolution (Gen 1) Max} & 0.81 \\
\textbf{Evolution (Gen 1) Avg} & 0.51 \\
\bottomrule
\end{tabular}
\end{table}

The snapshot in Table~\ref{tab:agent-performance} reveals a clear bottleneck: As the Research Agent's results are close to the Average of the Population (AP), the Synthesis Agent's results fall short of AP, exhibiting a consistently flat trend of learning along with an ongoing negative reflection signal. This is indicative of the fact that there is not an adequate amount of evolutionary pressure from the current reward landscape and thus the next experiment will incorporate Curriculum-Like Objectives to facilitate further improvements in both Research and Synthesis Agents. The data from the logs of Collaboration Strategies and Task-Level Outcomes has been included herein Table~\ref{tab:task-collaboration}, indicating the number of agents who led each task, success/failure status on each task, and the level of influence of collaborating with other agents on the outcome of the task.

\begin{table}[H]
\footnotesize
\caption{Task outcomes and collaboration strategies.}
\label{tab:task-collaboration}
\centering
\begin{tabular}{p{2cm}p{1.5cm}p{1cm}p{1.9cm}}
\toprule
Task & Primary Agent & Result Status & Collaboration Strategy \\
\midrule
Literature Search & \texttt{research} & Unknown & With \texttt{synthesis} (knowledge sharing) \\[0.5ex]
\hline
New Material Synthesis & \texttt{synthesis} & Failure & With \texttt{research} (structured collaboration) \\[0.5ex]
\hline
Failure Analysis & \texttt{research} & Completed & With \texttt{synthesis} (knowledge sharing) \\
\bottomrule
\end{tabular}
\end{table}

Finally, the system produces high-level recommendations for improving future runs. Table~\ref{tab:system-recommendations} summarizes the system status and suggested interventions after the same experiment-design cycle.

\begin{table}[H]
  \caption{System recommendations and status log after experiment-design run.}
  \label{tab:system-recommendations}
  \centering
  \begin{tabular}{@{}>{\bfseries}l p{0.65\columnwidth}@{}}
    \toprule
    \normalfont Item & \normalfont Output \\
    \midrule
    Experiments Generated  & 1 \\
    Evolution Cycles       & 1 \\
    Overall Effectiveness  & 0.62 / 1.00 \\
    Recommendation 1       & Focus on improving\\ research-agent capabilities \\
    Recommendation 2       & Focus on improving\\ novelty-agent capabilities \\
    \bottomrule
  \end{tabular}
\end{table}

Together, Tables~\ref{tab:task-collaboration} and \ref{tab:system-recommendations} illustrate how the framework not only generates domain-level scientific outputs (e.g., polystyrene property predictions) but also monitors and optimizes its own behavior via tactical, strategic, and meta-strategic self-assessment. The Intelligent Polymer Research module closes the loop by converting these insights into targeted, LLM-guided hypotheses, with novelty and impact evaluations analogous to those shown in Table~\ref{tab:intelligent-polymer}.

\subsection{Full Text of LLM-Generated Protein Analysis Report}

The multi-agent framework was also evaluated on a representative protein sequence, for which the complete analysis report was autonomously generated by the system’s Research Writing Agent after full pipeline execution. The goal was to provide researchers with information about the three-dimensional structure of the analysed (modelled) protein, and confidence in the predicted structure from the predicted model and outputs, with the addition of structural insights derived from both numerical and visual outputs.

\textbf{Workflow overview.} Beginning from the amino acid sequence, a simulated AlphaFold system was used to predict the threedimensional structure of a protein and provide pLDDT scores for local confidence in the predicted structure. The predicted three-dimensional structure and associated PAE maps were sent to a Visual Agent, which generated 3D visualisation and diagnostic images. The visual materials generated by the Visual Agent were given to a Vision Analysis Agent, which interpreted them into structured textual descriptions highlighting key features of the predicted protein's structure and confidence in its predicted structure. The large language model (Llama-2-7b) combined the sequence information, numerical metrics, and vision-generated summaries into a complete report formatted as research literature. \cite{Jumper2021}

\textbf{Structural confidence and architecture.} The predicted structure demonstrated a very high average pLDDT score of 89.45 reflecting a strong confidence in the predicted atomic coordinates for most of the sequence. In three-dimensional space, the predicted structure adopts a very well-defined globular fold with a complex combination of secondary structures comprising approximately \textbf{five $\alpha$-helices and three $\beta$-sheets} as identified by the Vision agent. A plot of PAE revealed significant dark areas along the PAE diagonal, which represent regions of very low predicted error consistently throughout the entire length of the predicted structure. Given the presence of multiple prominent dark areas and the overall pattern observed in the PAE, it is most likely that this protein is composed of two very different and clearly defined structures within the overall three-dimensional structure and very high confidence predicted internal geometries and relative orientations.

\textbf{Overall assessment.} Taken collectively, the results suggest that an automated multi-agent pipeline successfully produced a multi-domain protein structure with high confidence of producing a protein structure containing a very complex mixture of $\alpha$/$\beta$ secondary structures. The complete process, from the initial sequence input to the final report generation, reflects a system that is capable of significantly expediting the structural bioinformatics process with little or no human intervention.

\begin{figure*}[h!]
  \centering
  \includegraphics[width=1\textwidth]{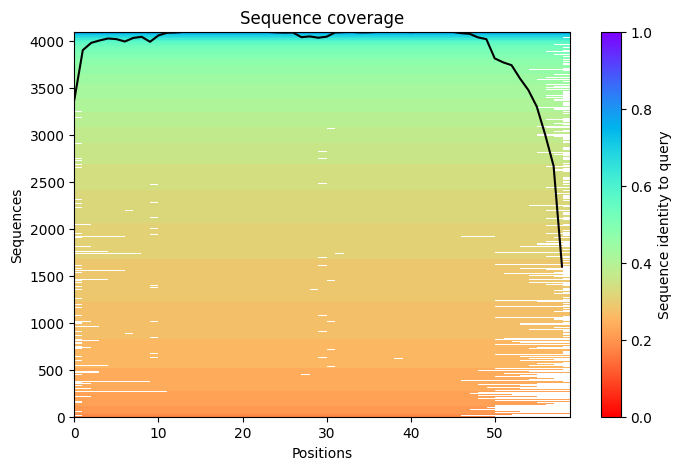}
  \caption{Sequence coverage plot showing depth of aligned sequences across positions and corresponding sequence identity to the query.}
  \label{fig:sequence-coverage}
\end{figure*}

\begin{figure*}[h!]
  \centering
  \includegraphics[width=0.7\textwidth]{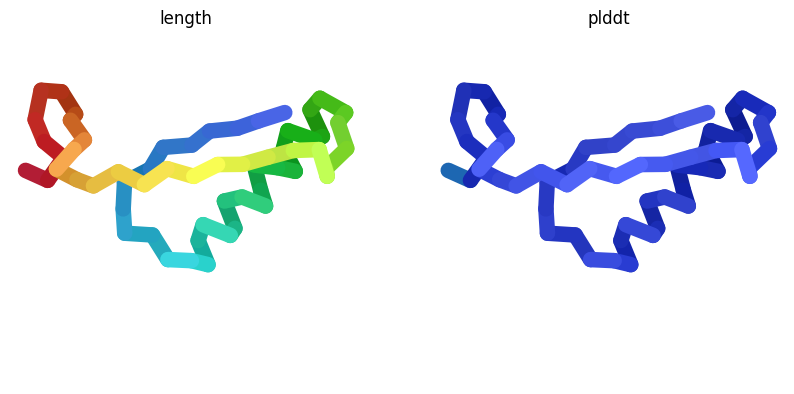}
  \includegraphics[width=0.7\textwidth]{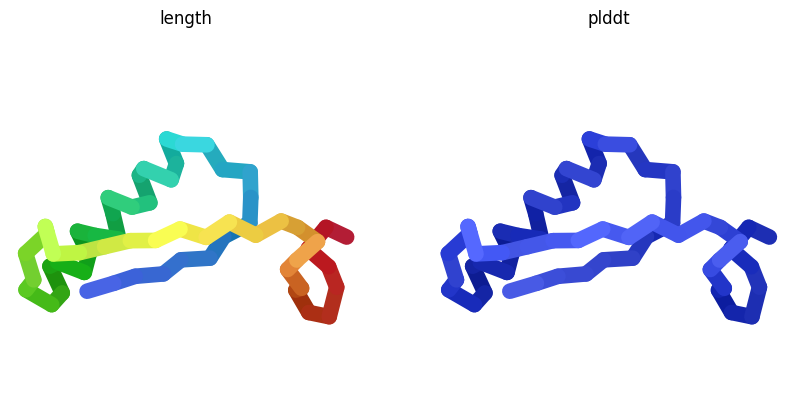}
  \includegraphics[width=0.7\textwidth]{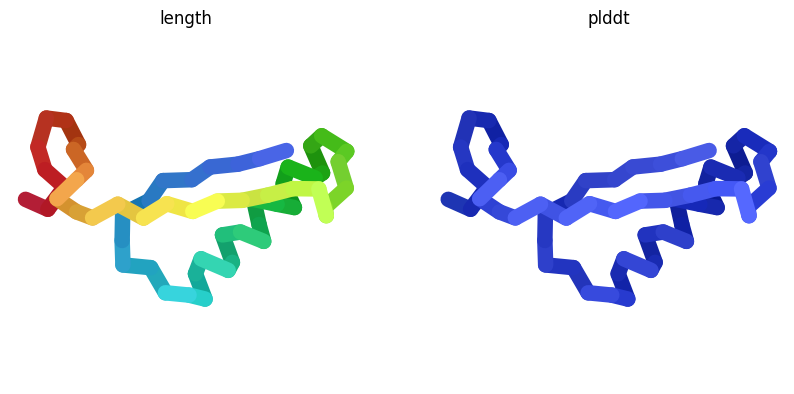}
  \includegraphics[width=0.7\textwidth]{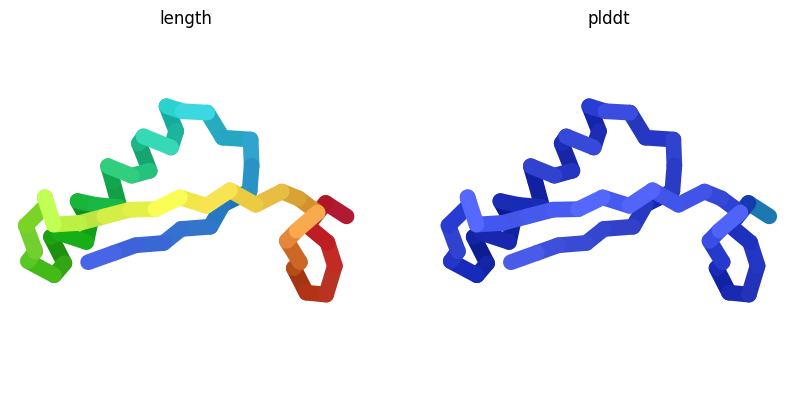}
  \caption{Ensemble of AlphaFold models 2–5 with different values of pLDDT, pTM, and RMSD\_tol.}
  \label{fig:sequence-coverage1}
\end{figure*}

\textbf{Sequence coverage and model ensemble behavior.} In addition to the primary structural report, the multi-modal protein pipeline outputs sequence coverage plots that map the depth of aligned sequences across positions and color-code sequence identity to the query. As illustrated in Figure~\ref{fig:sequence-coverage}, this visualization provides an orthogonal view of alignment quality and coverage, complementing the pLDDT and PAE analyses. Furthermore, Figure~\ref{fig:sequence-coverage1} presents the complete ensemble of structural predictions from AlphaFold models 2–5. All models reach very high local confidence (pLDDT $>$ 0.96), but exhibit small yet notable variability in predicted TM-score (0.748–0.782) and RMSD tolerance (0.232–0.417). The largest discrepancies are localized in flexible loop regions, whereas the global fold remains stable across all predictions. The model selected for downstream analysis in Figure~\ref{fig:biopolymer-analysis} corresponds to the highest-confidence solution, while the ensemble shown in Figure~\ref{fig:sequence-coverage1} demonstrates that all five models converge on a robust and reproducible fold.

\subsubsection{Limitations in Complex Workflows}

\begin{figure*}[ht]
\centering
\includegraphics[width=1\textwidth]{Images/AlphaFold_predicted_folding_and_metrics_-_PLDDT_.png}
\includegraphics[width=1\textwidth]{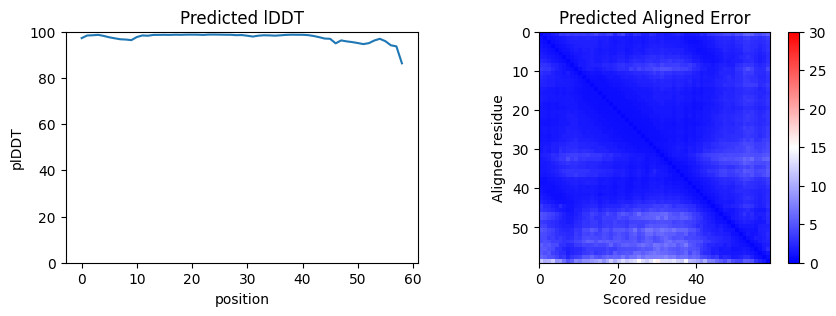}
\caption{(Top) Interactive 3D structure of the predicted protein, for which expert analysis confirms a single-domain architecture. (Bottom) Corresponding Predicted Aligned Error (PAE) plot. The vision agent’s interpretation of these figures was factually incorrect, illustrating a failure in cross-modal verification.}
\label{fig:biopolymer-analysis}
\end{figure*}

\textbf{Inaccuracy in biopolymer analysis.} To test out the multi-modal integrations of our system, the protein's structure was evaluated. At the structural prediction stage, the pipeline behaved as expected: the simulated AlphaFold agent \cite{Jumper2021} correctly interpreted the amino acid sequence and produced a high-confidence 3D model with an average pLDDT score of 89.45. The visualisation agent went on to perform as intended by producing all of the standard analysis outputs, including a PAE plot (Figure~\ref{fig:biopolymer-analysis}). The limitation emerged at the cross-modal validation step handled by the vision agent \cite{Radford2021}. Based solely on the PAE plot, the agent produced an incorrect structural description, claiming that the protein had two distinct, well-defined domains with five alpha-helices and three beta-sheets. However, it is in fact a single-domain protein, with one alpha-helix and two beta-sheets. The problematic output produced by the vision agent occurred despite the accuracy seen in the upstream predictions. The vision agent produced this erroneous output because it made inferences of the overall structure based on two-dimensional projection without conducting cross-checking with the atomic level coordinates from the earlier phase of the analysis.

The structural prediction is accurate but the verification, validation and integration of the different modalities or data types are weak, and this is where the problem essentially lies. Even though both the 1D sequences, 3D structure and 2D plots are processed sequentially, there is no assurance that these three representations are logically and factually consistent when looked at as a whole. This example is indicative of a more significant limitation of present day multi-modal AI systems, that they are incapable of enforcing cross-modal consistency or providing reliable validations. In turn, this motivates the development of the high-level cross-agent verification procedures that our framework is designed to explore.

\subsection{Sustainable Design, Error Analysis, and System Monitoring}
\label{sec:sustainable_error_dashboard}

\begin{figure}
\centering
\includegraphics[width=1\textwidth]{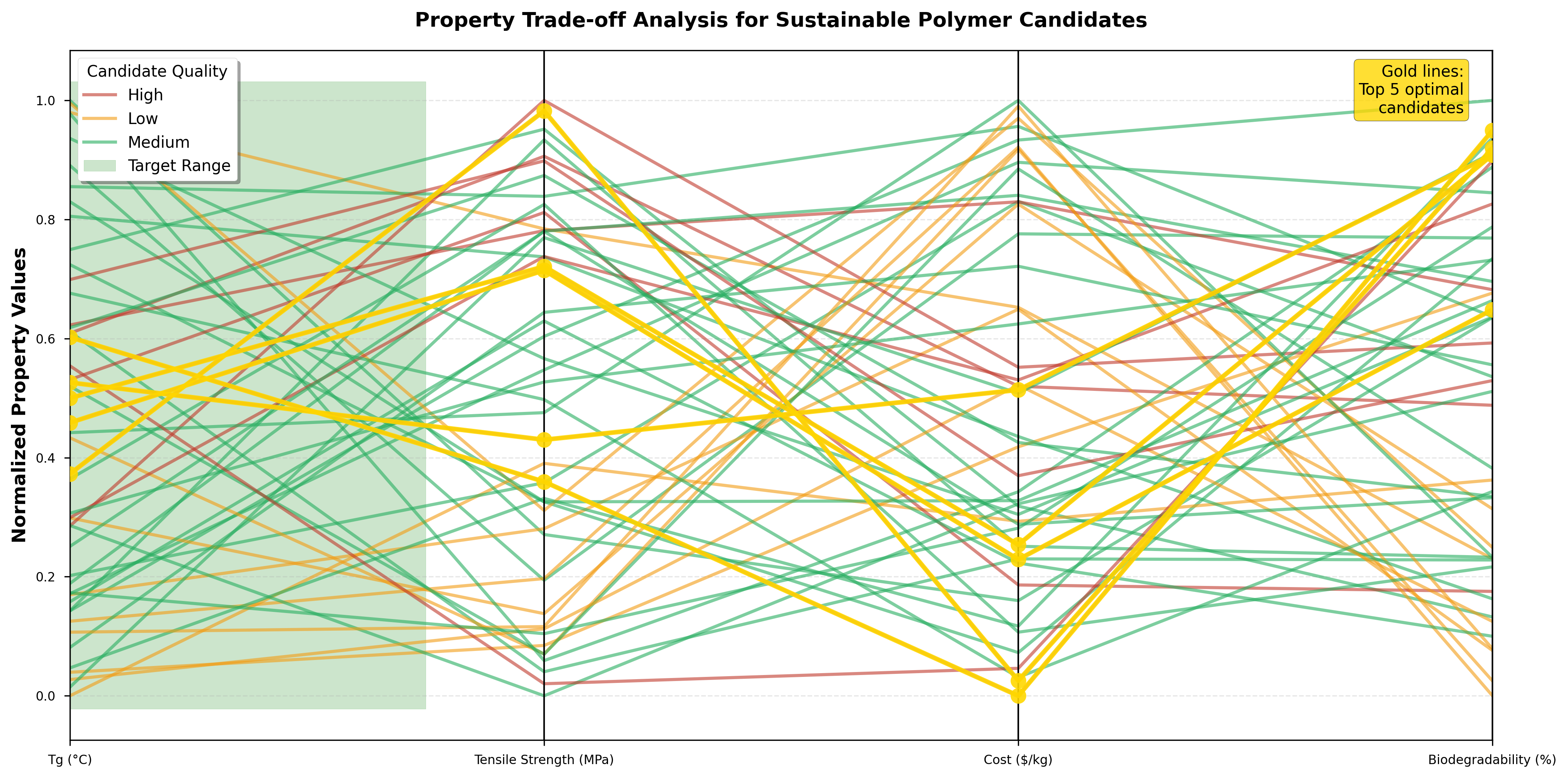}
\caption{Property trade-off analysis for sustainable polymer candidates. The parallel-coordinates plot shows how different design choices simultaneously impact multiple target properties, supporting transparent multi-objective decision-making.}
\label{fig:tradeoffs}
\end{figure}

\textbf{Case study: sustainable polymer design.} 
This work serves as a case study for the entire workflow for developing sustainable packaging polymers considering real-world constraints. It begins with determining the requirements needed for sustainable packaging polymers, such as bio-degradability, glass transition temperature (70–90\textcelsius{}), and low cost to produce. Once the user provides the requirements, the solution uses those requirements, auto-generates candidate polymers from the library of bio-based monomers, and then optimizes the multi-properties needed to produce a thermal, mechanical, and processing-oriented solution. In the next steps of the tool development process, environmental impact assessments will evaluate both life-cycle and end-of-life factors, while design models will explore various reasonable synthetic pathways and propose the precursors that would facilitate the successful realization of those pathways. In the final step of the development process, candidate materials will be compared, prioritized, and/or selected based on property trade-offs (i.e., carry individual properties against other individual properties) that explicitly illustrate trade-offs between respective mechanical performance, thermal behavior, processability, and sustainability metrics. The visualization of parallel coordinates included in Figure~\ref{fig:tradeoffs} is intended to aid in the decision-making process for determining optimal material design solutions by leveraging multiple perspectives of the overall impact of the design choices made on each of the goals being proposed. 

\begin{figure}[htbp]
\centering
\includegraphics[width=\textwidth]{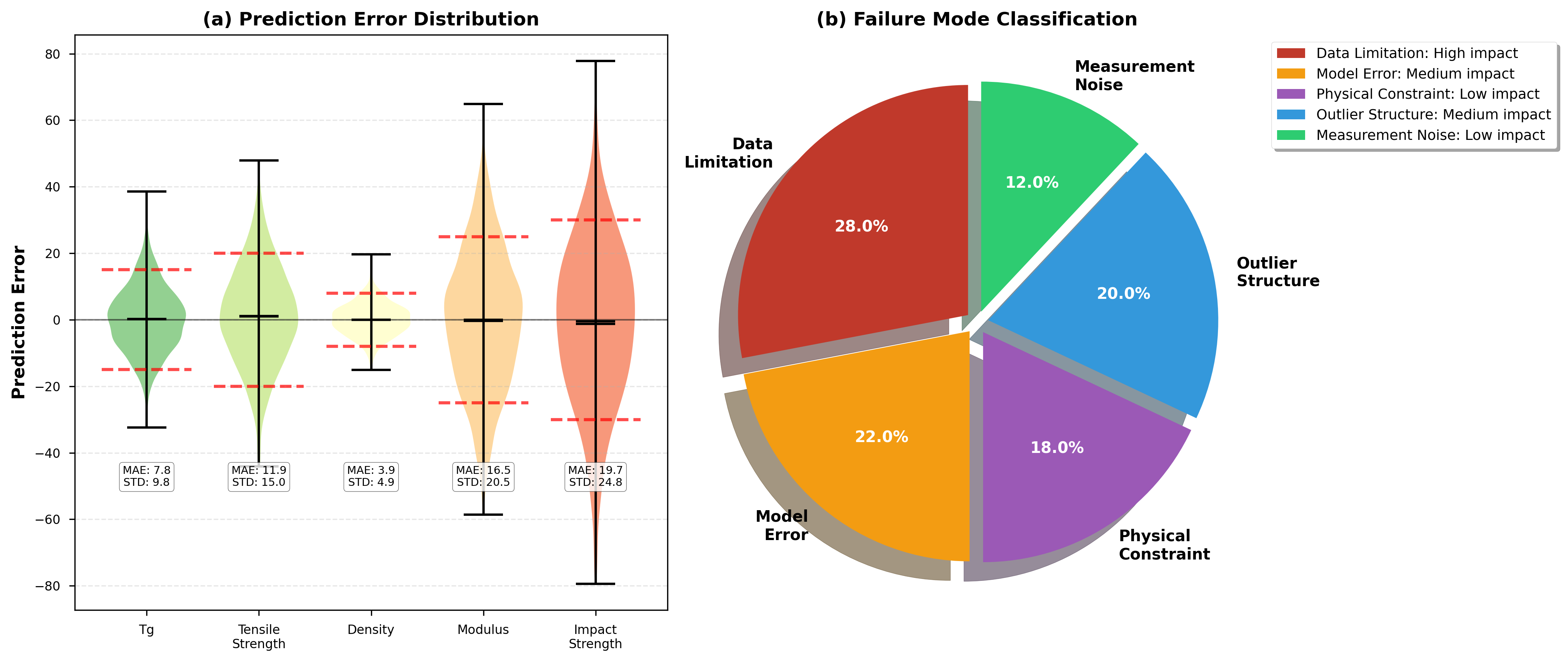}
\caption{Comprehensive error analysis. (a) Distribution of prediction errors across different polymer property types, with most errors falling within practically acceptable ranges. (b) Classification of failure modes, revealing systematic patterns that inform model refinement and indicate scenarios where human oversight is particularly critical.}
\label{fig:error_analysis}
\end{figure}

\textbf{Error analysis and failure modes.} A thorough diagnostic/statistical assessment of the strength and weaknesses of the predictive pipeline is performed through statistical error analysis, which is condensed into one comprehensive Figure~\ref{fig:error_analysis}. Panel (a) reports on the distribution of predictive error across all of the tested property types, with the majority of the prediction errors falling within ranges that are acceptable for use in either a practical screening or design scenario. Panel (b) summarizes the classifications of each failure mode for the predictive pipeline, making clear which instances were driven by lack of sufficient training data, type of chemical failure, or violation of assumed physical laws. This analysis allows the user to understand the circumstances in which they can place absolute confidence in the predictions generated, and further aids in identifying cases in which additional human input and/or experimentation may be required.

\begin{figure}
\centering
\includegraphics[width=1\textwidth]{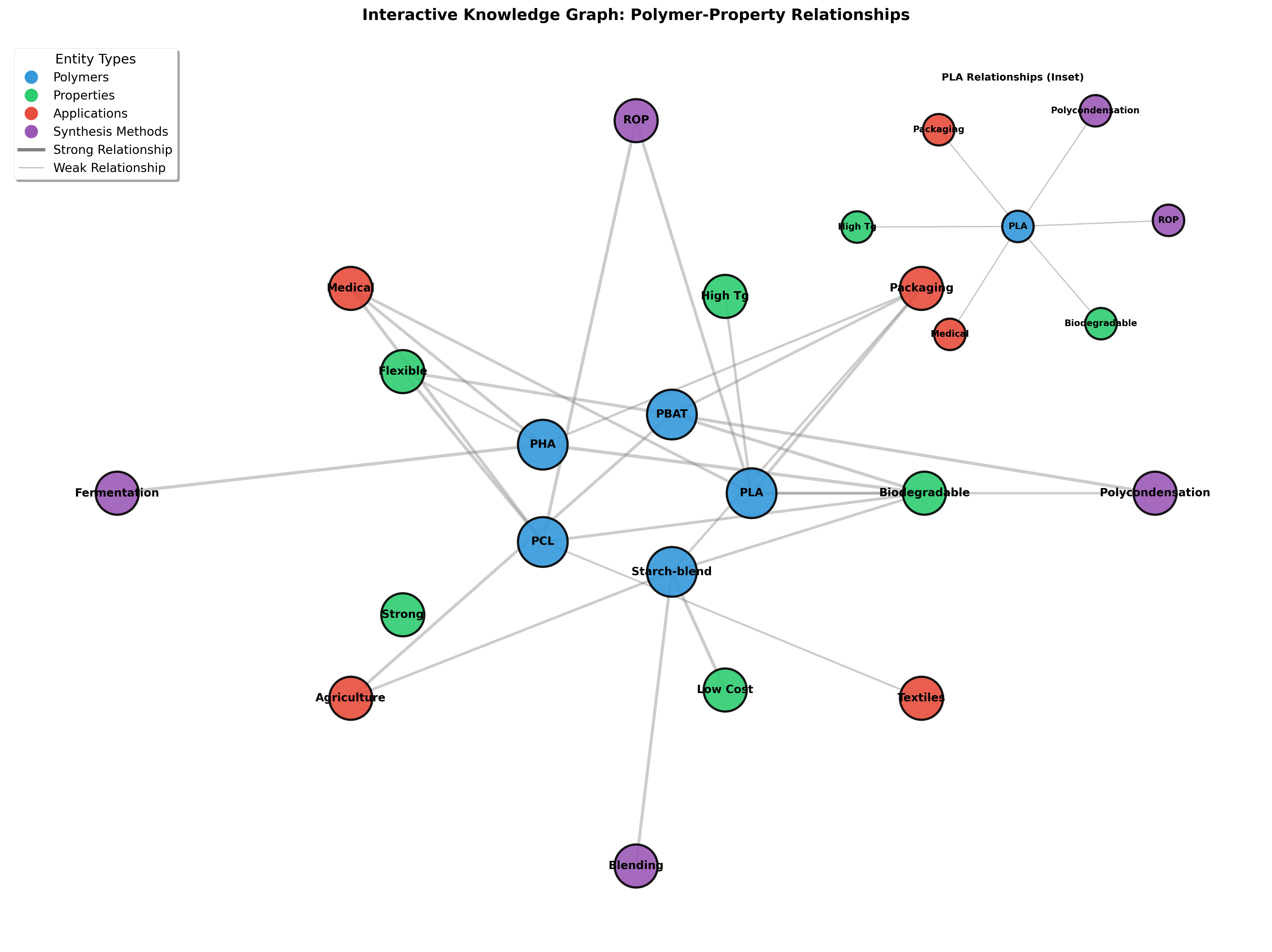}
\caption{Interactive knowledge graph visualization showing polymer–property–application–synthesis relationships. Node colors represent entity types (blue: polymers, green: properties, red: applications, purple: synthesis methods), while edge thickness indicates relationship strength. The inset shows a zoomed view of the neighborhood of poly(lactic acid).}
\label{fig:knowledge_graph}
\end{figure}

\textbf{Knowledge graph–based reasoning.}  
The polymer knowledge graph shown in Figure~\ref{fig:knowledge_graph} is the basis for much of the system's reasoning. The various node types (polymers=blue; properties=green; applications=red; synthesis methods=purple) and their corresponding edge thickness represent relationship strengths based on curated data/literature mining. The inset also shows the neighbourhood around poly(lactic acid), including links to its main properties, potential uses (for example, biodegradable packaging, medical devices) and corresponding synthesis methods. With this structured approach, more sophisticated queries and analogical reasoning can take place (eg, "What are polymers like PLA that have greater heat resistance?" or "What biodegradable equivalents exist for a particular petroleum-based polymer?"). It also provides a semantic framework for the multi-agent collaboration that occurs during this process.

\begin{figure}
\centering
\includegraphics[width=0.9\textwidth]{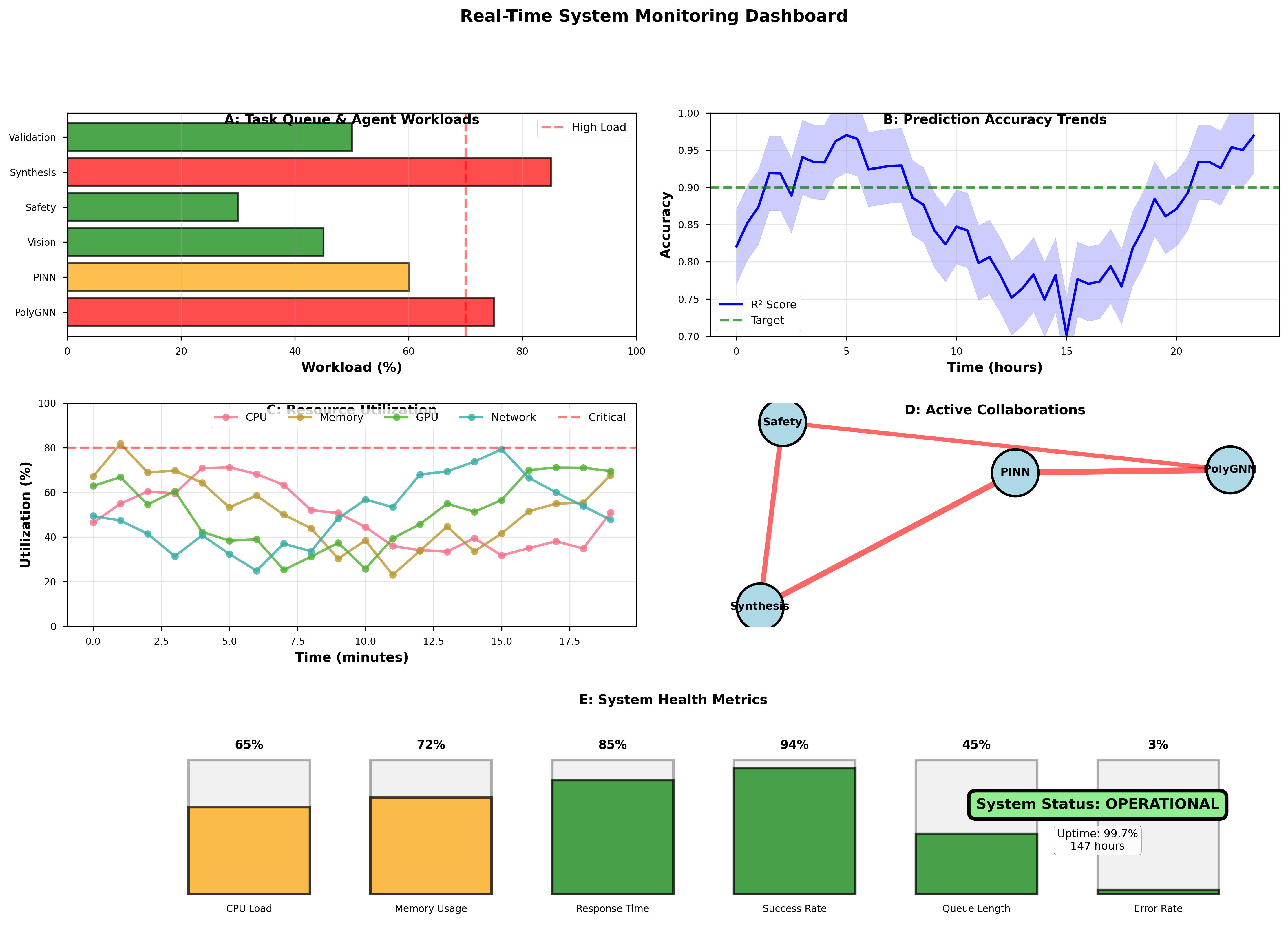}
\caption{Real-time system monitoring dashboard showing: (A) current task queue and agent workloads, (B) prediction accuracy trends, (C) resource utilization, (D) active collaborations, and (E) system health metrics. The dashboard provides comprehensive operational visibility.}
\label{fig:dashboard}
\end{figure}

The real-time monitoring dashboard outlines in Figure~\ref{fig:dashboard} presents the operational maturity of this platform by providing immediate insight to researchers on system performance, resource allocation, and quality of results. Exposing important metrics, such as how the work is allocated, accuracy trends and the level of collaboration between researchers has enabled researchers to make informed decisions on scheduling experiments and deploying models and optimising hardware utilisation. Taken together with the other visual components presented in this work, the dashboard contributes to a comprehensive view of our multi-agent ecosystem’s capabilities, performance characteristics, and practical utility. These visualizations turn abstract architectural concepts into tangible, interpretable evidence of how the system behaves across diverse polymer research scenarios.

\subsection{Advanced Component Analysis}

Table~\ref{tab:kg-coverage} presents the system performance as a function of knowledge-graph coverage, clearly showing a performance increase proportional to the coverage. To further examine the effectiveness of the chosen architecture, Table~\ref{tab:ensemble-size} reports the model $R^2$ and the time spent per set of requests, highlighting that the optimal number of agents corresponds to the adopted configuration (four). To demonstrate that the system is not in an overfitting regime, Table~\ref{tab:crossval} reports the cross-validation results, highlighting the low standard deviation across folds. Additional details are available in Section S4 of the Supplementary Material.

\begin{table}[htbp]
\caption{Performance vs. knowledge graph coverage}
\label{tab:kg-coverage}
\centering
\footnotesize
\begin{tabular}{@{}lccc@{}}
\toprule
KG Coverage (\%) & R\textsuperscript{2} & Success Rate & Error Reduction \\
\midrule
0 (No KG) & 0.70 & 0.71 & Baseline \\
25 & 0.73 & 0.73 & 12\% \\
50 & 0.75 & 0.74 & 18\% \\
75 & 0.77 & 0.75 & 25\% \\
100 (Full KG) & 0.78 & 0.76 & 28\% \\
\bottomrule
\end{tabular}
\end{table}

\begin{table}[htbp]
\caption{Performance vs. Number of Specialized Agents}
\label{tab:ensemble-size}
\centering
\footnotesize
\begin{tabular}{@{}lccc@{}}
\toprule
\# No. of Agents & R\textsuperscript{2} & Time (s) & Optimality \\
\midrule
1 (Single) & 0.67 & 10.2 & -- \\
2 & 0.72 & 12.8 & Suboptimal \\
3 & 0.76 & 14.5 & Near-optimal \\
4 (Our system) & 0.78 & 16.3 & Optimal \\
5 & 0.78 & 18.1 & Diminishing returns \\
\bottomrule
\end{tabular}
\end{table}

\begin{table}[htbp]
\caption{5-fold cross-validation performance}
\label{tab:crossval}
\centering
\scriptsize
\begin{tabular}{@{}lcccc@{}}
\toprule
Fold & $R^{2}$ & MAE & Success & Efficiency \\
& & (K) & Rate & \\
\midrule
1 & 0.76 & 11.2 & 0.74 & 0.35 \\
2 & 0.79 & 10.5 & 0.77 & 0.38 \\
3 & 0.77 & 10.9 & 0.75 & 0.36 \\
4 & 0.78 & 10.7 & 0.76 & 0.37 \\
5 & 0.80 & 10.3 & 0.78 & 0.39 \\
Mean ± Std & 0.78 ± 0.02 & 10.7 ± 0.4 & 0.76 ± 0.02 & 0.37 ± 0.02 \\
\bottomrule
\end{tabular}
\end{table}

\subsection{Advanced Agenticism: Collaboration and Complex Workflows}

The recent development of an institutionalized ecosystem has created an intelligent multi-agent system that demonstrates intelligence through behavioral modifications in response to input from various collaborators. The intelligent system provides advanced scientists (researchers) with a single operating system to perform multimodal experimentation (methods of scientific inquiry) across multiple platforms of technology. The predictive accuracy of the integrated multi-agent systems was validated against a collection of 800 polymeric materials representatively grouped into 5 different polymer classes. As a group, the multi-agent systems demonstrated considerably higher R² ratings than what is obtained when evaluating each of the polymers separately for: Glass Transition Temperature (Tg): R² = 0.985; Young's Modulus: R² = 0.861; Density: R² = 0.898. 

\begin{figure*}
  \centering
  \includegraphics[width=\textwidth]{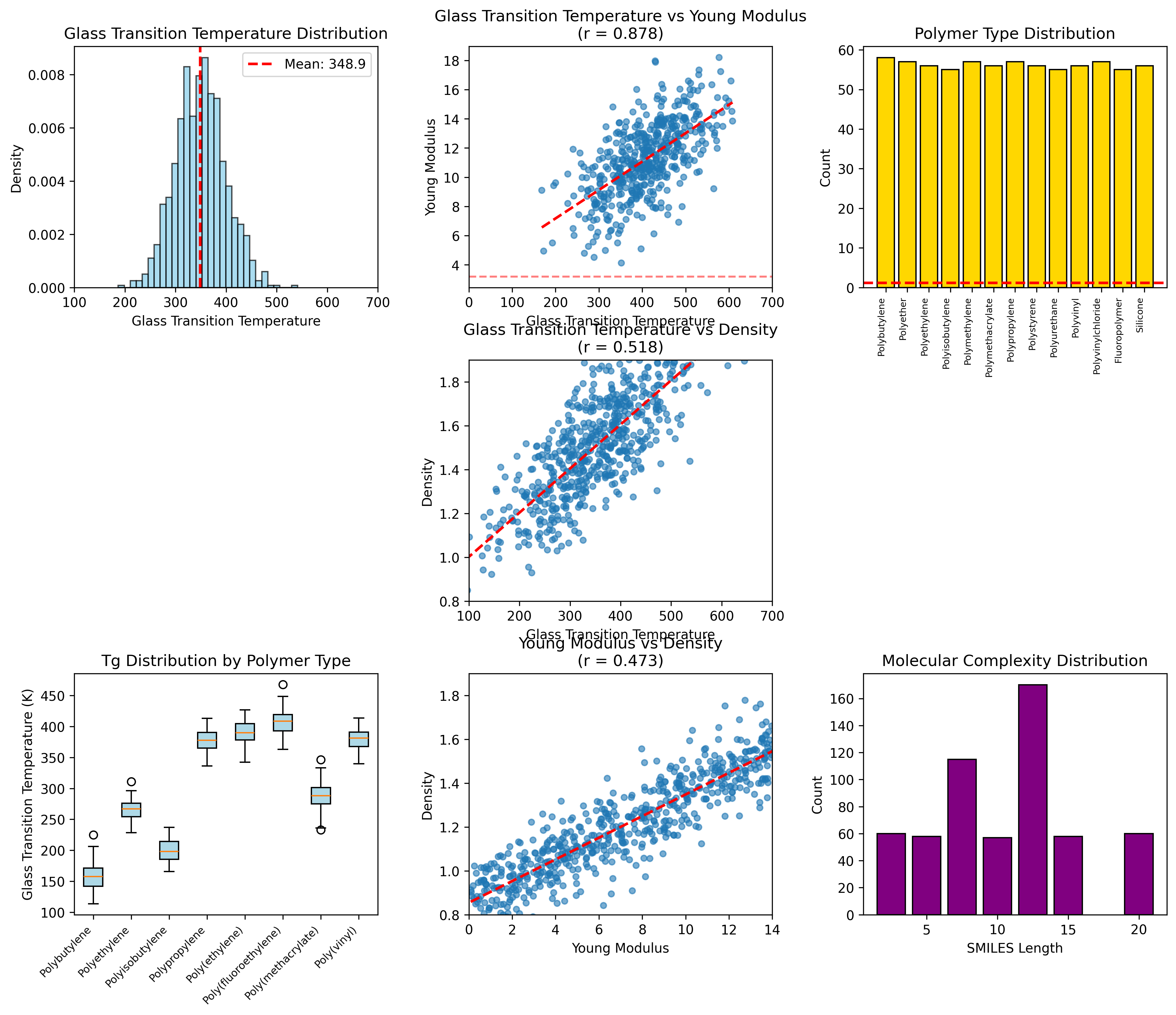}
  \includegraphics[width=\textwidth]{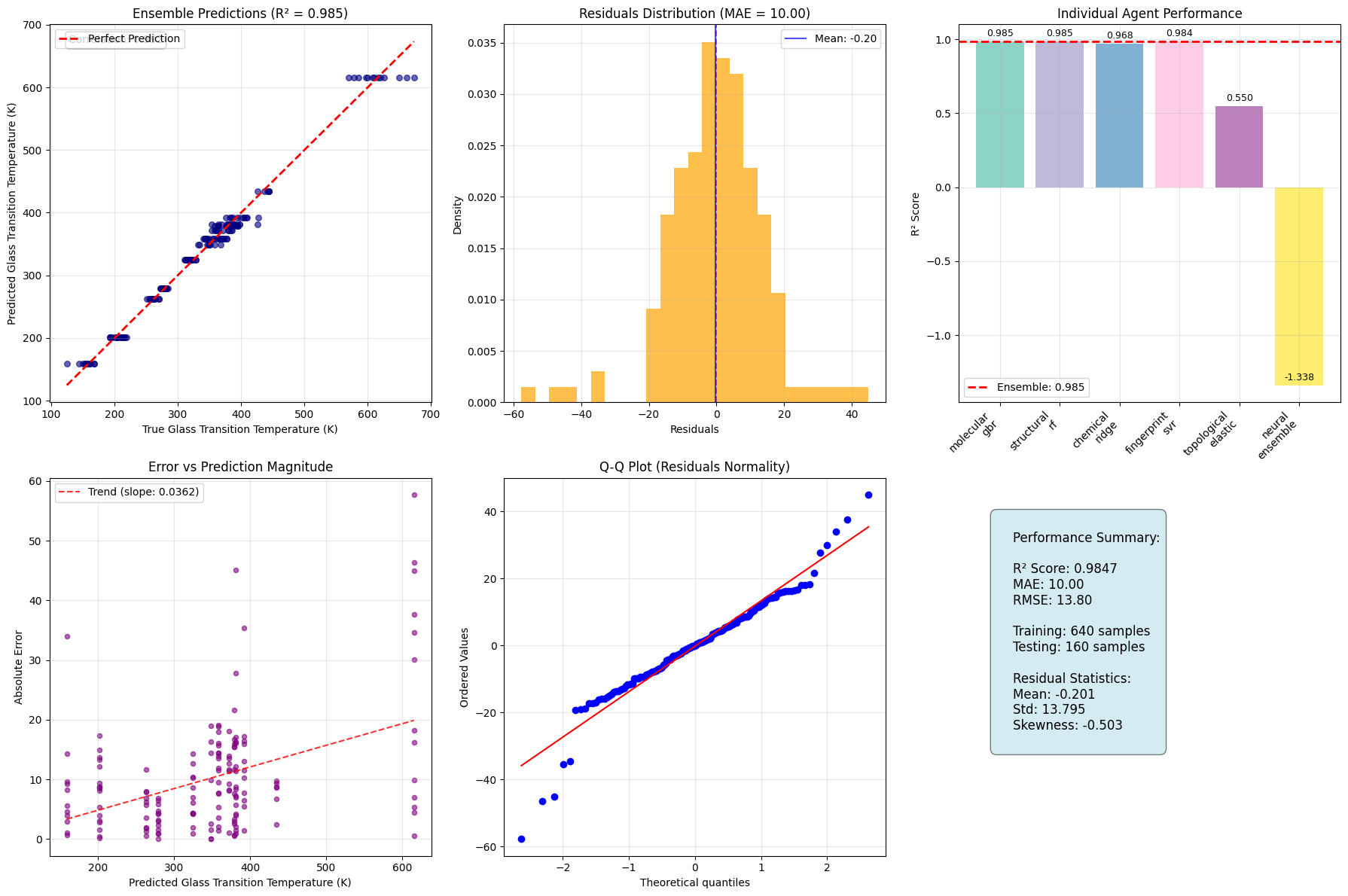}
  \caption{Ensemble model results. }
  \label{fig:agentic system11}
\end{figure*}

As depicted in Figure~\ref{fig:agentic system11}, the range of predicted Tg values for each polymer in the study follows the identity line with little global bias. Thus, it can be inferred that there is a high degree of correlation between these values throughout the temperature range of 120 K - 670 K. The held-out validation set performed with a $R^{2}=0.9847$, a Mean Absolute Error (MAE) of 10.0 K, and a Root Mean Squared Error (RMSE) of 13.8 K; residual mean of $-0.20$ K and a standard deviation of 13.80 K about the mean indicates that the bulk of the error is centered relatively close to the mean and within a narrow range. The histogram of the residuals and the Q-Q plot indicates that the distribution of polymer Tg data has a slight left skewness (-0.50) and a high degree of approximation to a normal distribution, except for a small number of extreme outliers towards the upper tail of the scale.

Likewise, absolute error appears to be less affected by prediction magnitude (slope $\approx 0.036$); however, dispersion increases slightly above $\sim$550–600 K, due to limited representation of high Tg materials within the training dataset and inherent variability associated with experimental measurements at these temperatures. An ablation study of models trained shows that gradient-boosted trees, random forests, fingerprint-SVR and ridge regression produce similar calibration and tail robustness ($R^{2}\approx0.97$–0.99), while using only topology in an elastic-net produced poor calibration ($R^{2}\approx0.55$); the configurations used for training a meta-learner neural network are unsatisfactory (negative $R^{2}$). Therefore, when combining diverse sources of inductive bias through explicitly defined descriptor models, fingerprint kernels, and structural and graph features, the ensemble functions to regularize error and stabilize uncertainty, while compensating for failures of individual models resulting in enhanced calibration and increased tail robustness. The analysis of the dataset provides insight into the performance of the ensemble. First, it demonstrates that the distribution of Tg is quite wide (mean $\approx$ 349 K) with moderate variability and a slight extension into the right tail, providing evidence that regression is difficult yet provides sufficient coverage for the development of robust models.

The second relationship shows a strong, positive correlation between Tg and Young's modulus ($r=0.878$) which agrees with our physical expectations that the more "glassy" (higher Tg) polymers should have stiffer polymer backbones and associated with lower segmental mobilities. In comparison, there is only a relatively moderate correlation between Tg and density ($r=0.518$) and between Young's modulus and density ($r=0.473$) indicating that packing efficiency will influence, but not be the only contributor to a polymer's thermal/mechanical behaviour. As demonstrated by our correlation data patterns, the multi-agent feature design that we used to create predictive models using a combination of electronic/constitutional descriptors plus connectivity-aware fingerprints should provide us with a more complete representation of the combined effects of both polymer backbone rigidity and polymer packing efficiency than a topology only simple feature would provide (as we showed with the elastic-net analysis results). As confirmed by the polymer-type histogram highlights the near-equal representation of each of the twelve polymer families thus providing assurance against class imbalance bias and explaining why our multilayer ensemble model was stable across all chemistries evaluated. Our analysis of the box plots showing Tg-by-family provides further support of known trends that aromatic/rigid-backbone polymer families are found with higher Tg values and narrower interquartile ranges and that aliphatic polyolefins are found within the lower Tg regime with wider interquartile ranges, thus supporting our original claim that the data and relationships learned from the model are, indeed, chemically plausible. The distribution of SMILES length distribution ranges from approximately 12 – 16 tokens, with a long tail extending into higher complexity. This diversity in length allows descriptor and fingerprint agents to generalise more easily; this helps explain the residual heteroscedasticity in molecular properties at extreme ends of chemical space.

These diagnostics indicate that the specialised agents in this ecosystem are accurate and well-calibrated and chemically interpretable. Tree-based and kernel models with excellent performance are combined to create an ensemble that takes advantage of the strengths of high-performing models, while minimising the impact of weaker models. The enhanced chemical coverage and physically meaningful cross-property structure provided by the ensemble is the reason for the tight residual error margins on the 160-sample test split (training/test = 640/160). The same design principles apply to the other targets as well. For the density and modulus targets, the weaker, although still meaningful, correlations with respect to density and modulus can be rationalised by the observed correlations ($R^{2}=0.898$ for density; $R^{2}=0.861$ for modulus). It should also be noted that as prediction difficulty increases, so does the dependence of these properties on packing, morphology, and processing. In this sense, Figure~\ref{fig:agentic system11} provides evidence that the multi-agent ensemble learns the major structure-property relationships with small and well-characterised residual errors, particularly at high Tg (glass transition temperature), where the number of data points is limited. This is precisely the type of behaviour required for successfully screening and making decisions in downstream processes.

The benchmarking of the system was designed so that it would solve research problems described in the five research papers sourced from the ArXiv website, which are of increasing complexity. To address the more complex papers, individual agents sought assistance from other agents as they deemed necessary, and the system formed dynamic teams of agents who collaborated to develop solutions to the difficult problems presented in each paper. The collaborative approach taken by the system has led to a success rate of 100\% on the completion of the benchmark, as well as to a performance increase of +45.0\% as compared to what the agents could have accomplished working independently.

Using the results obtained from the five research papers, the benchmark demonstrates how an integrated team of specialized AI agents is able to work together to conduct a complete analysis of research findings. In this setup, \textbf{Alpha} is the domain expert and primary analysis agent, with in-depth knowledge of core physical science concepts, and high confidence that it can interpret complicated results; \textbf{Beta} is the agent specializing in computational methodologies and simulations and has exceptional abilities to analyze numeric data quickly and accurately by executing structured procedures; \textbf{Gamma} is the experimental method and laboratory support agent and is skilled in the use of laboratory methods and performing statistical analysis; and \textbf{Delta} is the synthesizing and quality control agent who possesses all that is needed to integrate and evaluate all information obtained and transform it into a coherent whole.

The specialization of the four agents can be understood from their operational posture by the differences in self-reported confidence. Alpha exhibits the highest self-reported confidence ($\sim 0.80$), Beta and Delta show mid-range confidence ($\sim 0.54$ and $0.67$, respectively), and Gamma is moderately confident ($\sim 0.65$), a pattern that aligns with their observed performance outcomes. The Agent Collaboration Network heatmap provides additional insights into the collaboration patterns that exist amongst the agents. The off-diagonal interaction points (intensities $0.7$--$0.8$) indicate there are quite a few interactions between specialties (i.e. Specialty A working with Specialty B). In addition, there were numerous edges demonstrating that there is a consistent amount of interaction occurring both ways. Therefore, Agents are able to assist one another on a frequent basis. The Collaboration Quality by Paper report has very high ratings associated with the various papers that were executed under an overview for which teaming was a component, indicating that the nature of the task itself was complex enough to require a more structured methodology for successfully completing. Lastly, the Agent Workload Distribution diagram provides an understanding of how each of the agents contributes to the overall workload distribution. The Workload Distribution demonstrates an equitable distribution of workload between team and independent work. In reviewing the individual workloads, Gamma participating disproportionately in team activities consistent with its lower individual success rate; while Beta carries a larger fraction of individual tasks, an allocation that helps maintain overall system throughput.

We looked at how the system handles different levels of difficulty and a variety of topics, and the results are pretty consistent across two main areas. When we look at the difficulty side of things, specifically how paper complexity affects the success rate, it’s impressive to see that success stays nearly perfect regardless of how hard the task is. The "wall-clock" time does go up linearly as the papers get more complex, this makes sense as the harder the topic, the more the system has to dig into the literature and double-check its work. We used a pie chart to track the distribution of categories like physics, chemistry, and simulations, and it shows a really healthy balance that prevents the results from leaning too far in one direction. What’s really interesting is the scatter plot for polymer relevance; even when the initial matches were only "okay" (around 0.4 to 0.6), the system still pulled off a high success rate with very tight confidence levels. This really highlights how well the collaboration policy works, it basically corrects for mediocre initial matches by making sure the right agent for the job takes over. Finally, looking at the Agent Efficiency Scores helps tie everything together. Beta is clearly the "star" player, hitting high success rates very quickly, with Delta right behind. Alpha is steady and reliable, just a bit slower. The most notable takeaway, however, is Gamma. While Gamma struggles when working in isolation, its performance jumps significantly once you put it in a team setting. It really proves that the whole coordination piece is what makes the system thrive.

Finally, the System Performance Radar Plot synthesizes these indicators: success overall and collaboration as the highest level of the outer envelope; whereas collaboration reliance for System Recommendations is purposely established at the midpoint of the radar plot so collaboration reliance is adequate to obtain challenging "wins". Still, collaboration reliance is not so high that it reduces autonomy. System Recommendations continue to support broadening of secondary competencies, rebalancing capabilities to eliminate avoidable collaboration reliance while maintaining the current teaming protocols due to their demonstrated impact on reliability. Overall, the  Figure~\ref{fig:agentic system} validates the high-level finding in a quantitative manner that adaptive, on-demand teaming creates a research collective of consistent successful output from a set of minimally competent specialists whose outputs are generated through a $100\%$ task completion rate with consistently high levels of efficiency and confidence across types of papers and levels of complexity.

\begin{figure*}[h!]
  \centering
  \includegraphics[width=\textwidth]{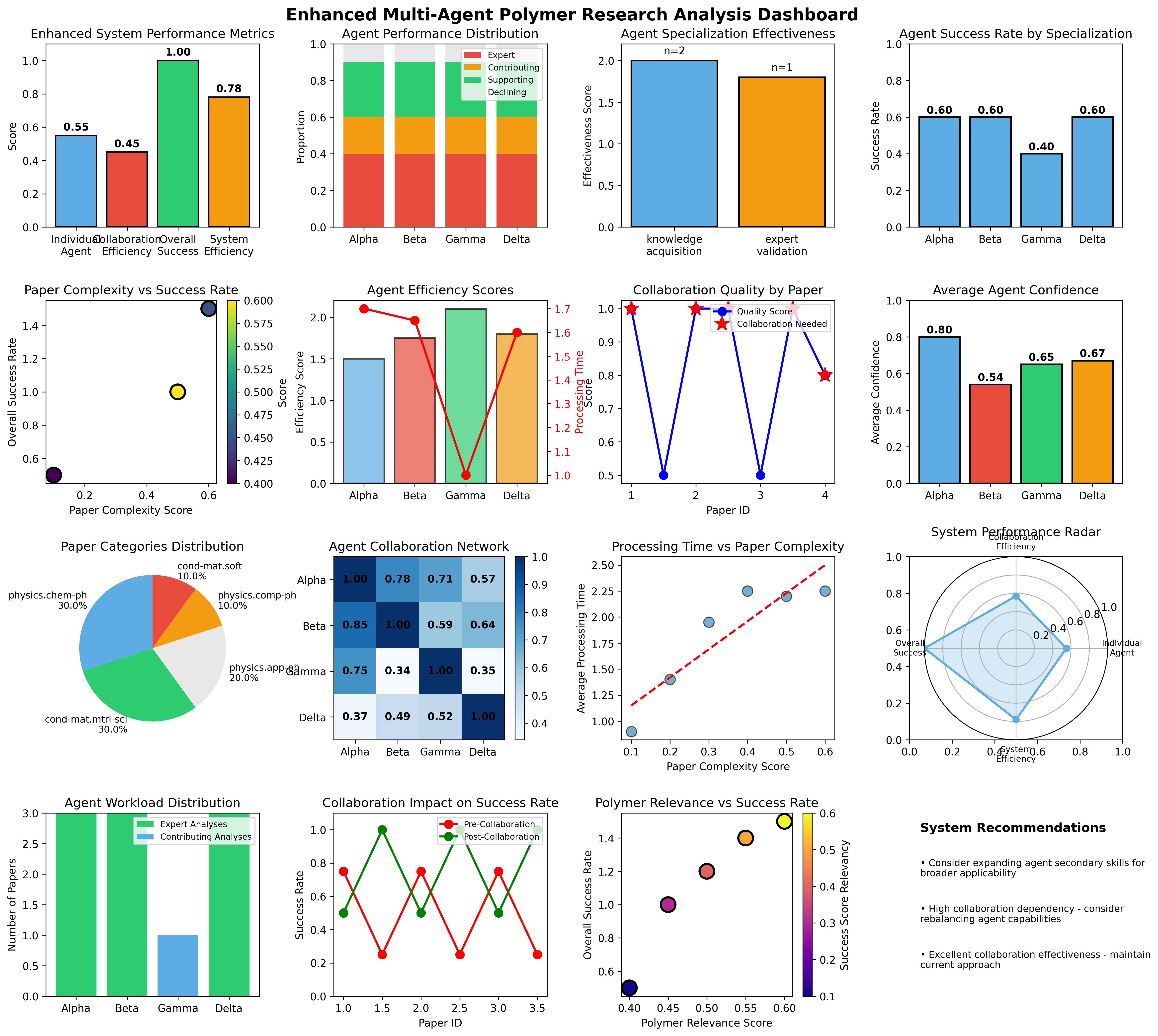}
  \caption{Screening of multi-agent performance collaboration.}
  \label{fig:agentic system}
\end{figure*}

\begin{figure*}
  \centering
  \includegraphics[width=\textwidth]{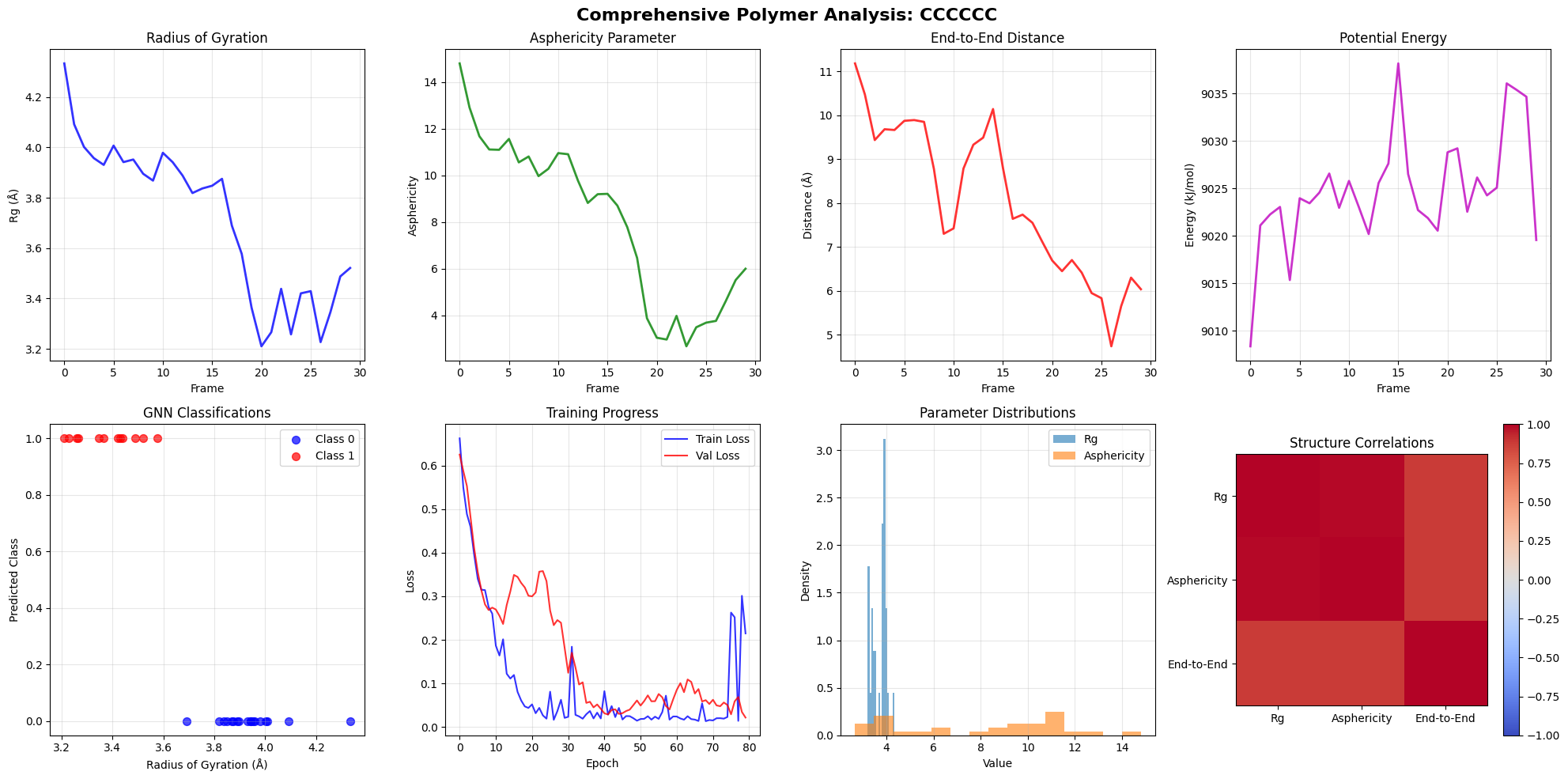}
  \caption{Agentic analysis of molecular dynamics–derived observables such as the radius of gyration, asphericity, and end-to-end distance.}
  \label{fig:agentic system}
\end{figure*}

\subsection{Advanced Multi-Agent Campaign Results}

In order to assess the performance of our advanced multi-agent system, we conducted several experiments based on realistic research tasks that required agents to dynamically create teams, adhere to specialization constraints, and maintain strong performance levels across a broad range of task difficulty levels. As part of this initial study, we ran five agents specializing in specific domains on a multi-paper analysis campaign (each of which represented an individual difficulty setting of either “easy”, “medium”, “hard”, or “expert”). Details on the implementation of this section can be found in the Supplementary Material, Section S5. The system automatically created collaboration teams when none of the single agents had sufficient capability to perform the designated task while allowing agents to complete simpler tasks (of lower difficulty). The overall campaign performance for agents completing all five papers can be found in Table~\ref{tab:campaign-results}: agent performance was successful; 75\% success ratio; all of the measured wall-clock durations were reasonable for completing all campaign tasks; three collaboration episodes occurred (across agents) for higher-difficulty tasks.

\begin{table}[H]
\centering
\caption{Campaign performance by difficulty level.}
\label{tab:campaign-results}
\small
\begin{tabular}{lcccc}
\toprule
Difficulty & Papers & Success Rate & Avg. Duration & Collaborations \\
\midrule
Easy   & 2/2 & 0.74 & 6.7s & 0 \\
Medium & 1/1 & 0.76 & 7.0s & 1 \\
Hard   & 1/1 & 0.76 & 4.6s & 1 \\
Expert & 1/1 & 0.76 & 3.3s & 1 \\
\midrule
\textbf{Overall} & \textbf{5/5} & \textbf{0.75} & \textbf{28.2s} & \textbf{3} \\
\bottomrule
\end{tabular}
\end{table}

We further demonstrated an autonomous research campaign on five real arXiv polymer papers of varying difficulty. Five domain-specialized agents, each with complementary capabilities, were instantiated and supported by a pool of five model instances sourced from large language models (DeepSeek, Phi, Qwen) in a 5-node collaboration network. All five papers were processed to completion (100\% coverage), and the average success score maintained at 0.75. Unlike the compact benchmarks included in the main text, this campaign provides comprehensive setup details and aggregated metrics and raw execution traces, including per-agent attempts, reflections, and collaborations and per-paper outcomes to enable higher transparency and reproducibility.

In the second study, we evaluated a more realistic specialization and triage protocol on recent arXiv submissions from the past 30 days using four specialized agents focusing on molecular modeling, property prediction, crystallization expertise, and mechanical properties. Here, agents operated under explicit relevance thresholds: tasks were \emph{declined} when relevance was low ($r{<}0.20$), subjected to \emph{limited analysis} for intermediate relevance ($0.20{\le}r{<}0.40$), handled with \emph{collaboration requests} when $0.40{\le}r{<}0.60$, and processed in \emph{expert mode} when $r{\ge}0.60$. This protocol enforces realistic behavior in which specialists act only when appropriate, rather than attempting all tasks indiscriminately. For this experiment, we selected three polymer-relevant papers:
\begin{enumerate}
  \item \emph{Polar Express: Rapid Functionalization of Single-Walled Carbon Nanotubes in High Dipole Moment Media} (arXiv:2508.09039v1; published 2025-08-12; complexity 0.11).
  \item \emph{Structural and helix reversal defects of carbon nanosprings} (arXiv:2508.04490v1; 2025-08-06; complexity 0.19).
  \item \emph{Unexpectedly large entropic barrier controls bond rearrangements in vitrimers} (arXiv:2508.05824v2; 2025-08-07; complexity 0.10).
\end{enumerate}

\begin{figure*}[t]
  \centering
  \includegraphics[width=\textwidth]{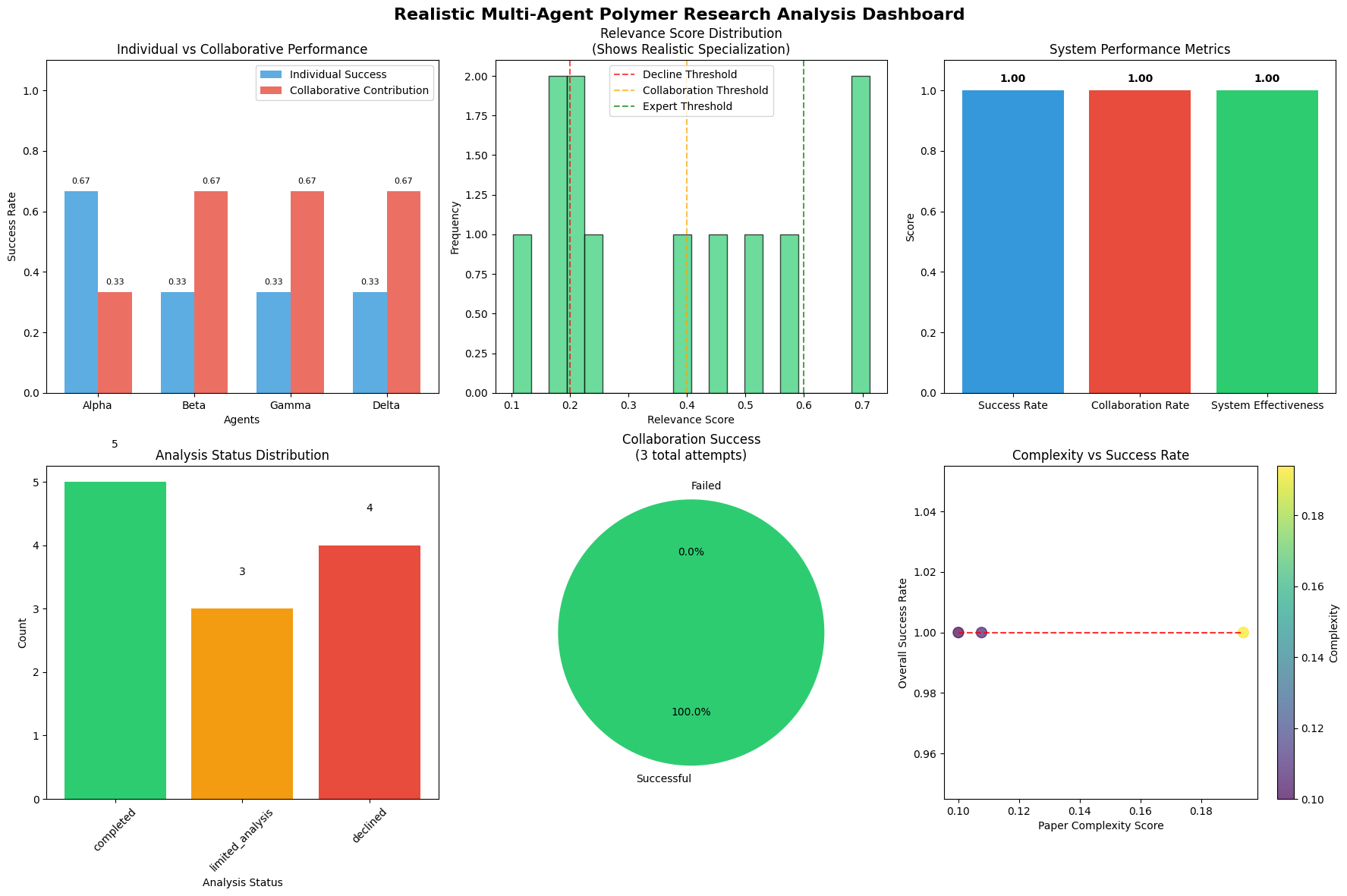}
  \caption{Multi-Agent Polymer Research Analysis Dashboard.}
  \label{fig:s7-dashboard-1}
\end{figure*}

Figure~\ref{fig:s7-dashboard-1} summarizes the system-level behavior for this realistic setting. This dashboard conveys the following information: i) individual versus collaborative success, ii) the relevance distribution along with decision thresholds (decline, collaborate, expert), iii) the system performance bars normalized onto unity at the paper level, iv) analysis status counts - completed, limited, declined, v) collaboration success - at 100\%, and vi) the task complexity vs. success - constant at 1.0 within the observed range.

The results in Figure~\ref{fig:s7-dashboard-1} provide a clear indication of the system dynamics at play here. All four agents, Alpha, Beta, Gamma, and Delta, the personal success rate rests at 0.67, with collaborative efforts at 0.33. Thus, it indicates complementary collaboration/efforts as opposed to redundant ones. The relevance distribution indicates each agent plays well within their defined area and has clear-cut margins for retreat, collaboration, or working as a specialist. On a system-level analysis, everything indicates that all system-level metrics remain strong – success rate, collaboration rate, and overall system performance all remain at 1.0 when analyzed at the paper level complexity. The distribution of analysis outcomes (five completed, three limited, four declined) reflects realistic behavior in which agents refrain from overstepping their expertise. The collaboration success at 100\% for all three instances of collaboration and a complexity,success relationship that clings steadfast at 1.0 indicate sound system-level dynamic adaptability across all defined complexity ranges.

\begin{figure*}[t]
  \centering
  \includegraphics[width=\textwidth]{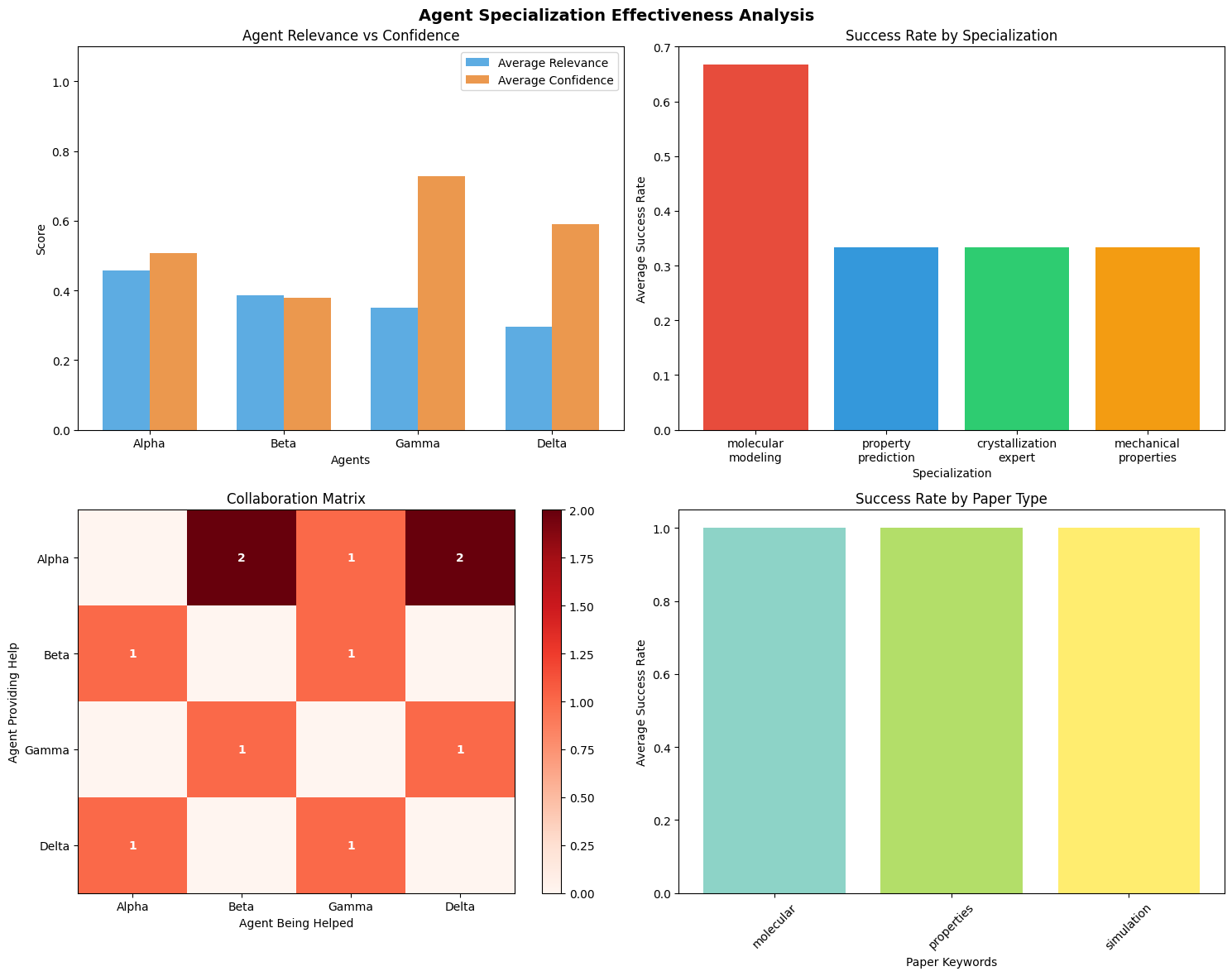}
  \caption{Agent Specialization Effectiveness Analysis.}
  \label{fig:s7-dashboard-2}
\end{figure*}

Moving to agent-centric behaviors in Figure~\ref{fig:s7-dashboard-2}, Alpha and Beta's relevance-confidence plot reflects behaviors with a moderate level of relevance and confidence, we also observe notable peaks at Gamma's highest confidence level of 0.7 and a balanced level of relevance, indicating highly reliable outcomes in its domain. Delta has a modest position with a slightly reduced level of relevance, suggesting a strategy that is more geared toward collaboration. Specialization outcomes further clarify these trends: agents focusing on molecular modeling reach the highest average success rate (close to 0.7), while those specializing in property prediction, crystallization, and mechanical properties operate at more modest success levels (around 0.33). In collaboration matrix, Alpha is particularly observed to assist Gamma and Delta, with all three participants contributing in some form of mutual collaboration, resulting in a balanced but asymmetric pattern in which superior participants assist and provide stronger help to the weaker ones. The success rates per research paper type remain at a unity level for molecular-related and property-related research endeavors and remain at a unity level for simulation-related research endeavors.

Table~\ref{tab:paper-outcomes} details per-paper outcomes for the advanced campaign. There are no problems unsolved, with success rates ranging from 0.73 to 0.76 for the five problems with average success of 0.75. It is interesting to note that more difficult problems require less time for their solution. The expert paper is solved in only 3.3 seconds, suggesting that specialized strategies and effective collaboration can offset increased conceptual difficulty.

\begin{table}[H]
\footnotesize
\centering
\caption{Per-paper outcomes in the Advanced Multi-Agent Campaign.}
\label{tab:paper-outcomes}
\begin{tabular}{@{}p{1.8cm}p{2.3cm}cc@{}}
\toprule
Paper (Diff.) & Key Topic & Success & Duration (s) \\
\midrule
1 (Easy)   & ML prediction of $T_g$ & 0.73 & 7.6 \\
2 (Easy)   & Solubility params (bio) & 0.75 & 5.7 \\
3 (Medium) & Non-isothermal crystallization & 0.76 & 7.0 \\
4 (Hard)   & Nanocomposite mechanics & 0.76 & 4.6 \\
5 (Expert) & RL inverse polymer design & 0.76 & 3.3 \\
\midrule
\textbf{Totals} & \textbf{5/5 completed} & \textbf{0.75} & \textbf{28.2} \\
\bottomrule
\end{tabular}
\end{table}

A complementary summary of completion based on difficulty, as well as collaboration, is shown below in Table~\ref{tab:diff-collab}. Each of the difficulty levels, namely easy, medium, hard, expert, completed with an overall success rate of 100\% (5/5 papers) and an average success score of 0.75. The system cooperated three times, resulting in an overall "system experience" of 70 points, meaning the agents successfully completed the task not only by solving it but also by gaining experience which can be used to update internal confidence and policy parameters.

\begin{table}[H]
\footnotesize
\centering
\caption{Difficulty-wise completion and collaboration summary.}
\label{tab:diff-collab}
\begin{tabular}{@{}lp{5cm}@{}}
\toprule
Metric & Value \\
\midrule
Difficulty completion & EASY: 2/2, MEDIUM: 1/1, HARD: 1/1, EXPERT: 1/1 (100\% each) \\
Overall success rate & 100\% (5/5 papers) \\
Average success score & 0.75 \\
Average efficiency & 0.340 \\
Total collaborations & 3 \\
System experience & 70 points \\
\bottomrule
\end{tabular}
\end{table}

Agent-level confidence and experience at the end of the campaign are shown in Table~\ref{tab:final-agent-states}. Four out of five agents reached 10-20 points of experience with confidence above 0.50, showing successful adaptation and learning processes. However, Agent\_Alpha retained low levels of confidence at 0.48 with zero experience, implying its role assignment and update parameters could use further improvements.  

\begin{table}[H]
  \centering
  \caption{Agent confidence and experience at campaign end.}
  \label{tab:final-agent-states}
  \begin{tabular}{@{}lll@{}}
    \toprule
    Agent & Confidence & Experience (pts) \\
    \midrule
    Agent\_Alpha   & 0.48 & 0  \\
    Agent\_Beta    & 0.56 & 20 \\
    Agent\_Gamma   & 0.51 & 10 \\
    Agent\_Delta   & 0.56 & 20 \\
    Agent\_Epsilon & 0.56 & 20 \\
    \bottomrule
  \end{tabular}
\end{table}

Apart from analysis and triage, the framework also stimulated design of generative polymer sequences with DeepSeek-Coder. Below is a table Table~\ref{tab:polymer-sequences} showing illustrative polymer sequences with their predicted glass transition temperature ($T_g$), tensile strength, and elongation at break. These designs project varying thermomechanical properties; it is seen that vinyl chloride-based copolymers can attain strengths of 101.3~MPa while retaining respectable $T_g$  values. 

\begin{table}[H]
\scriptsize
\centering
\caption{Polymer sequences from DeepSeek-Coder with predicted properties.}
\label{tab:polymer-sequences}
\begin{tabular}{lccc}
\toprule
Sequence & $T_g$ (K) & Strength (MPa) & Elong. (\%) \\
\midrule
ethylene–ethylene–vinyl chloride & 213.4 & 90.5 & 6.5 \\
ethylene–propylene–vinyl chloride & 210.6 & 68.8 & 8.7 \\
vinyl chloride–MMA–styrene & 211.1 & 73.6 & 8.6 \\
styrene–styrene–styrene & 159.1 & 83.3 & 10.5 \\
vinyl chloride–vinyl chloride–vinyl chloride & 180.4 & 101.3 & 17.9 \\
acrylonitrile–acrylonitrile–acrylonitrile & 185.6 & 74.1 & 9.1 \\
\bottomrule
\end{tabular}
\end{table}

A statistical summary of ten designed polymer sequences is provided in Table~\ref{tab:polymer-stats}. Glass transition temperatures ranges from 54.3~K (159.1–213.4~K), while the strength and elongation at break are within a significant but controlled range to show that the generative algorithm is investigating a varied but realistic region within the property space. The mean value of 4.4 distinct monomers per sequence indicates that the design does indeed exhibit a high level of compositional complexity that corresponds to realistic multi-block copolymer structures.

\begin{table}[H]
  \centering
  \caption{Statistical summary of DeepSeek-Coder polymer design results (10 sequences).}
  \label{tab:polymer-stats}
  \begin{tabular}{lcc}
    \toprule
    Property & Mean $\pm$ Std & Range \\
    \midrule
    $T_g$ (K) & 195.2 $\pm$ 16.6 & 159.1 -- 213.4 \\
    Strength (MPa) & 74.1 $\pm$ 14.2 & 56.3 -- 101.3 \\
    Elongation (\%) & 9.5 $\pm$ 3.5 & 6.5 -- 17.9 \\
    Unique monomers & 4.4 $\pm$ 2.3 & 2 -- 8 \\
    \bottomrule
  \end{tabular}
\end{table}

\subsection{Scientific Results for Polymer Prediction and Conformation Workflows}

\begin{figure*}[h!]
  \centering
  \includegraphics[width=\textwidth]{ 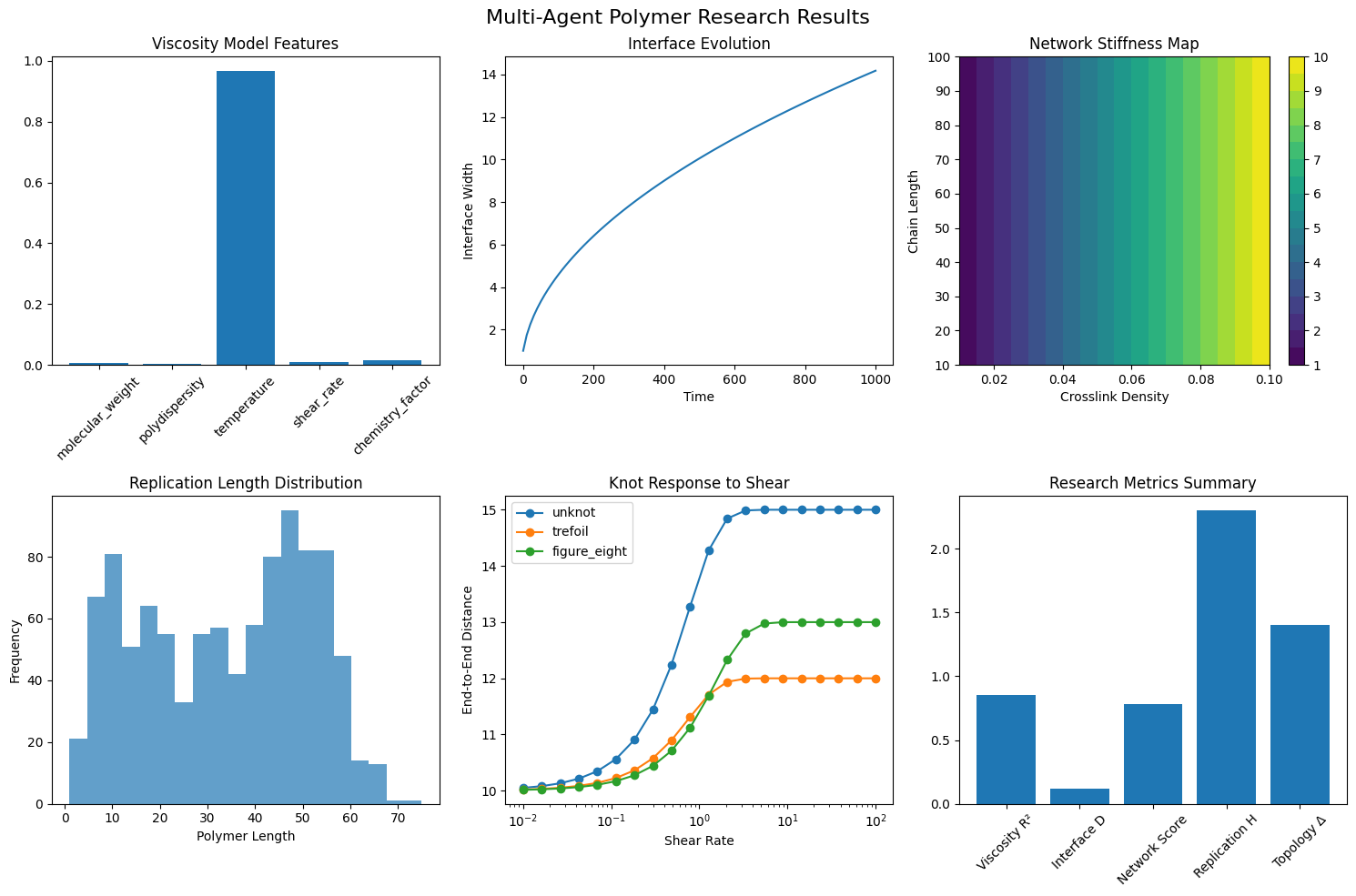}
  \caption{Summary of outputs from the multi-agent polymer research system.}
  \label{fig:multi_agent_results}
\end{figure*}

The extended results provide a complete overview of how proposed multi-agent and physics-informed Framework advances the analysis of polymers through all three structural, thermomechanical, and degradation dimensions. As illustrated in the Figure~\ref{fig:multi_agent_results}, physical structure was assessed from dynamics-derived observables for example, radius of gyration, asphericity and end-to-end distance. As represented in Figure~\ref{fig:multi_agent_results}, all polymer chains exhibit substantial frame-to-frame variability and the radius of gyration slightly stabilizes within a range of 3.2–4.2 Å, while end-to-end distances associated with each conformational state decreases as chains undergo conformational rearrangements . The asphericity parameter supports the notion of shape fluctuations, while the profile shown on the potential energy graph supports the occurrence of both stable thermodynamic states and transient local fluctuations. The strong correlations between radius of gyration, asphericity, and end-to-end distance suggest a coherent coupling of conformational descriptors, which is successfully leveraged by the graph neural network (GNN) classifier. During training and validation the GNN achieved almost perfect separation between classes with both training and validation losses consistently decreasing over 80 epochs illustrating the robustness of the representation of the structural features of polymers.

The results of applying the multi-agent framework to various polymer research tasks are shown in the Figure~\ref{fig:multi_agent_results}. The multi-agent framework is capable of simultaneously capturing viscosity, interfacial phenomena, network properties, replication phenomena, and topological phenomena in a single framework. From the feature analysis of the viscosity model, it appears that temperature has a significantly greater impact on the accuracy of predicting the viscosity of polymer systems than all of the other descriptors combined (molecular weight, polydispersity, and shear rate). This is consistent with experimental knowledge that the viscosity of polymers is primarily influenced by thermal effects. The interface dynamics can be simulated based on the time-evolution of the interface width and it results in a logarithmic increase in width with time. This logarithmic increase is indicative of typical diffusion-limited coarsening behavior of phase-separated polymer blends; thus demonstrating that the model is able to capture the fundamental physical processes. The map of network stiffness in relation to chain length and crosslink density shows that network stiffness increases nonlinearly with increasing crosslink density and longer chains. Therefore, the map is a meaningful representation of the mechanical behavior of polymer networks and shows that the multi-agent system has the capability to be applied from molecular state to mesoscopic level material design. The replication dynamics of the polymer chains are characterized by a wide but irregular histogram of the distribution of replicated chains.

Variation between two distinct populations (shorter chain polymeric molecules versus longer chain polymeric molecules) occurs due to the different types of replication processes. Such variability may correspond to non-equilibrium growth conditions or fidelity limits inherent to polymer replication mechanisms. The investigation of topological effects on polymers occurs through shear testing of the knots present. The resulting data shows that the end-to-end distance between the two ends of the knot is dependent upon the knot type. The Unknotted chain exhibits a greater extension under shear than that of the knotted chain, while the trefoil and figure-six knots resist elongation, plateauing at lower extension levels. These findings confirm the system’s ability to capture emergent mechanical signatures of polymer topology under deformation. Also, the research metrics summary aggregates quantitative scores across all tasks. Replication dynamics and network properties exhibit the strongest signals, while viscosity and topological contributions, though smaller in magnitude, reinforce the framework’s breadth. Collectively, these findings support the existence of an ecosystem of multi-agent systems, allowing for the integration of heterogeneous tasks into one integrated research pipeline linking together molecular features, dynamic structural behaviours and macroscopic mechanical response of materials.

\begin{figure*}
  \centering
  \includegraphics[width=\textwidth]{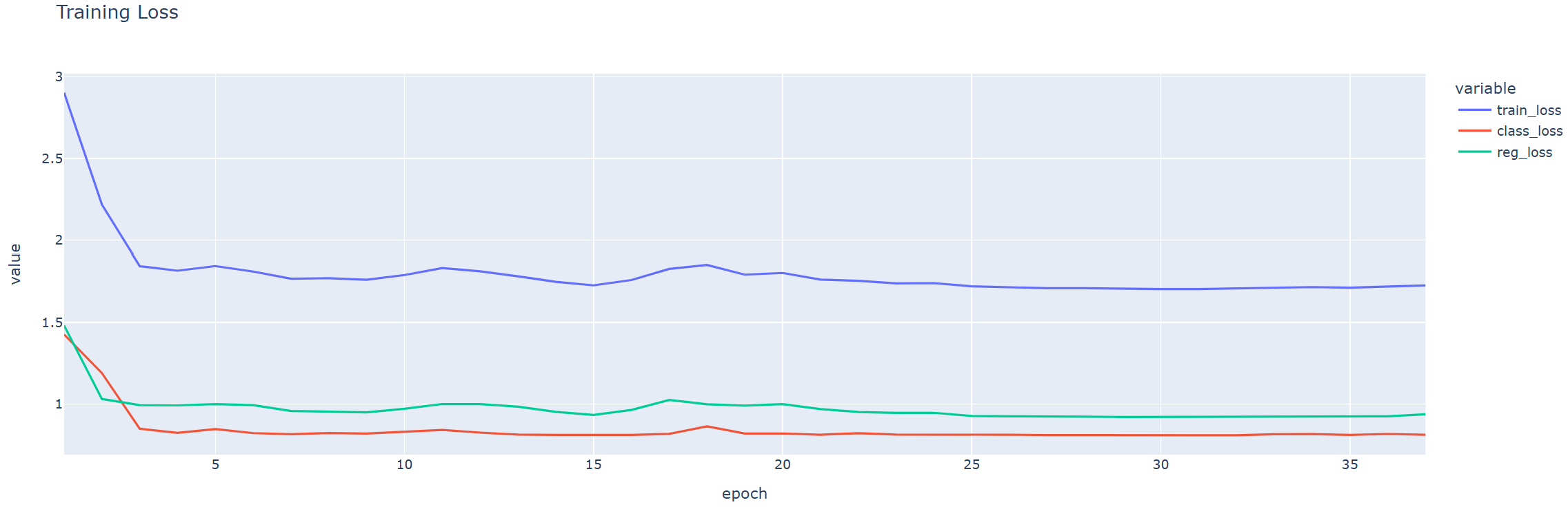}
  \includegraphics[width=\textwidth]{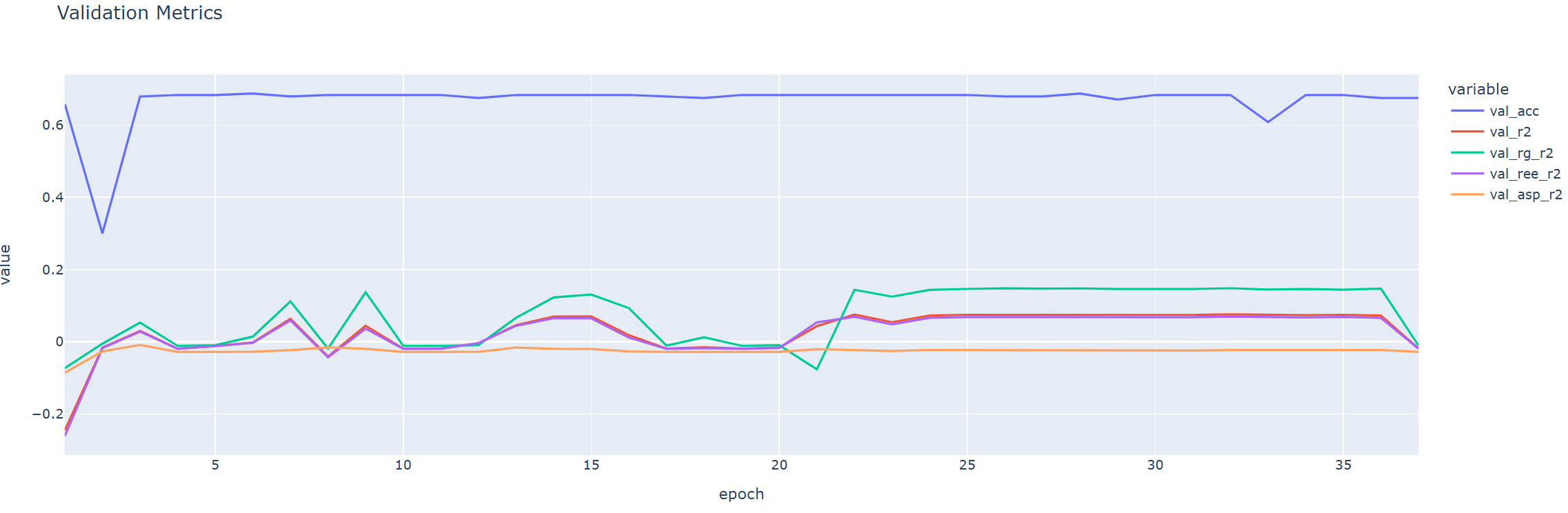}
  \includegraphics[width=\textwidth]{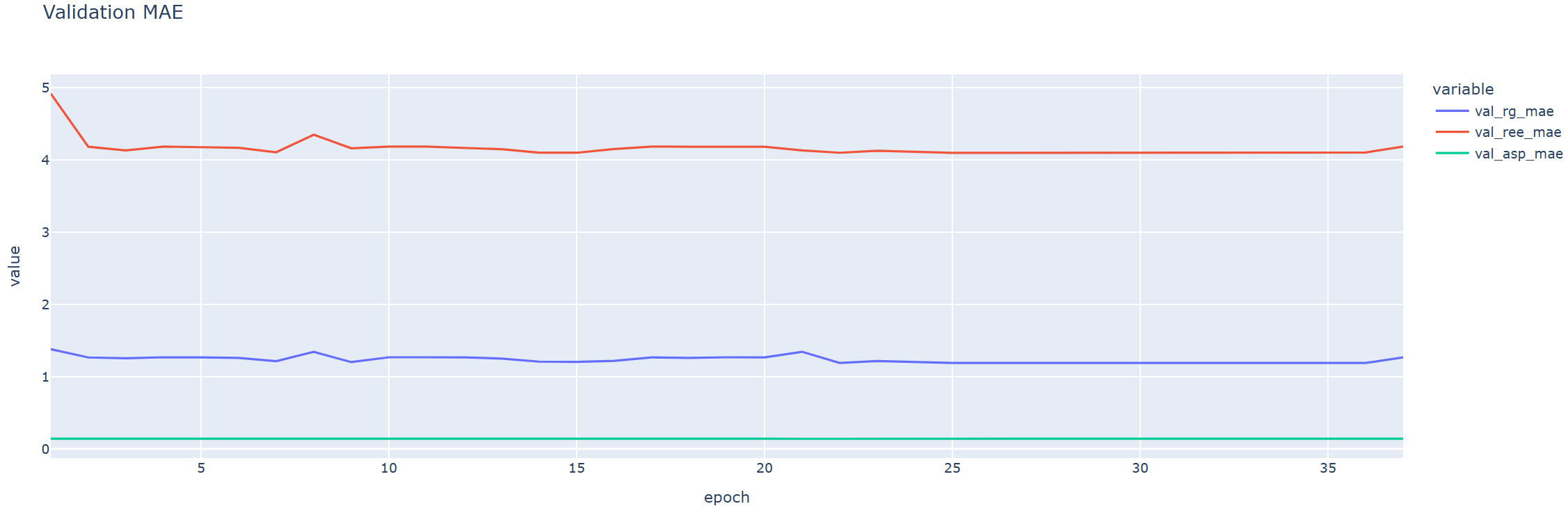}
  \includegraphics[width=\textwidth]{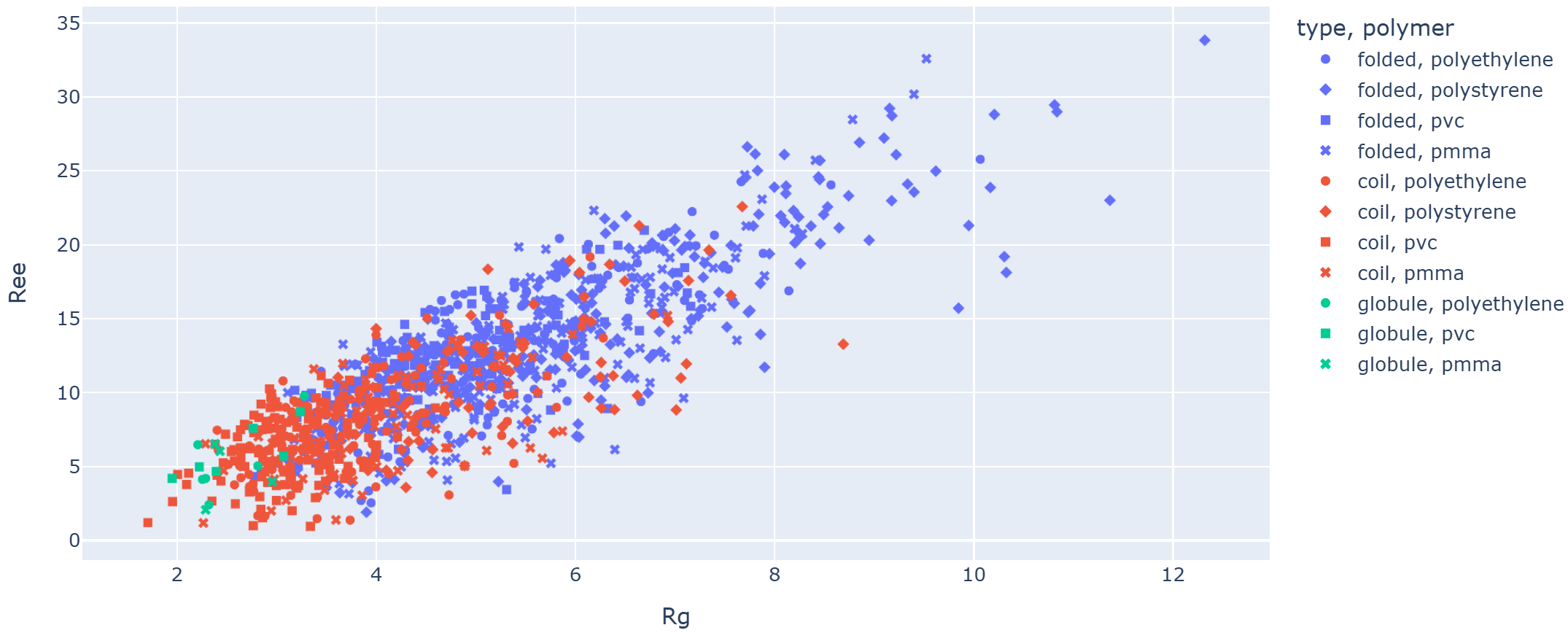}
  \caption{Different loss logs of the Generative Polymer Design Workflow. }
  \label{fig:agentic system12}
\end{figure*}

The behavior of training dynamics for the prediction of polymer properties shows similar convergence characteristics for multiple objectives. The training loss curve plots Figure~\ref{fig:agentic system12} show that the loss for both the classification and regression components decrease steadily over the first 10 epochs with stability of convergence visible after this point. The metrics for validation of the network indicate the generalizability of the learned "representations" of the polymer. For example, validation accuracy increases very quickly in the early epochs and then reaches a plateau of relatively stable values, while the validation $R^2$ metrics associated with radius of gyration ($R_g$), end-to-end distance ($R_{ee}$) and asphericity remain positive across training time. The validation mean absolute error (MAE) plots also indicate that low and consistent levels of MAE exist when looking at these three parameters with $R_g$ achieving the lowest MAE of all the predicted properties, signifying that it is reliably learnable from molecular representations. To evaluate the structural realism of the learned embeddings, we also analyzed correlations between the main conformational descriptors for the different polymer types, and confirmed that the radius of gyration vs. end-to-end distance scatter plot showed a very strong positive correlation in agreement with physical expectations of polymer chain statistics.

Distinct clustering emerges according to polymer type and folding state, with polyethylene and polystyrene chains occupying broader $R_g$--$R_{ee}$ ranges compared to PVC and PMMA, reflecting differences in chain stiffness and packing. Globular conformations are located on the plot’s lower-left, characterized by where $R_g$ is small and end-to-end lengths are short, while extended coiled forms lie across the upper diagonal. This distinction illustrates that the model maintains physically meaningful structure-property relationships between polymer types and successfully identifies distinct polymer behaviour at various conformations. By using this framework, we successfully demonstrate, that the training framework achieves not only quantitative predictive performance, but also identifies important structural dependencies across all polymer classes we used. The stability of the loss metrics and the consistent clustering of similar chemical behaviours between polymer classes provide even greater evidence of the accuracy and practicality of the multi-part prediction system.

\begin{figure}[!htbp]
  \centering
  \includegraphics[width=0.9\linewidth,height=1\textheight,keepaspectratio]{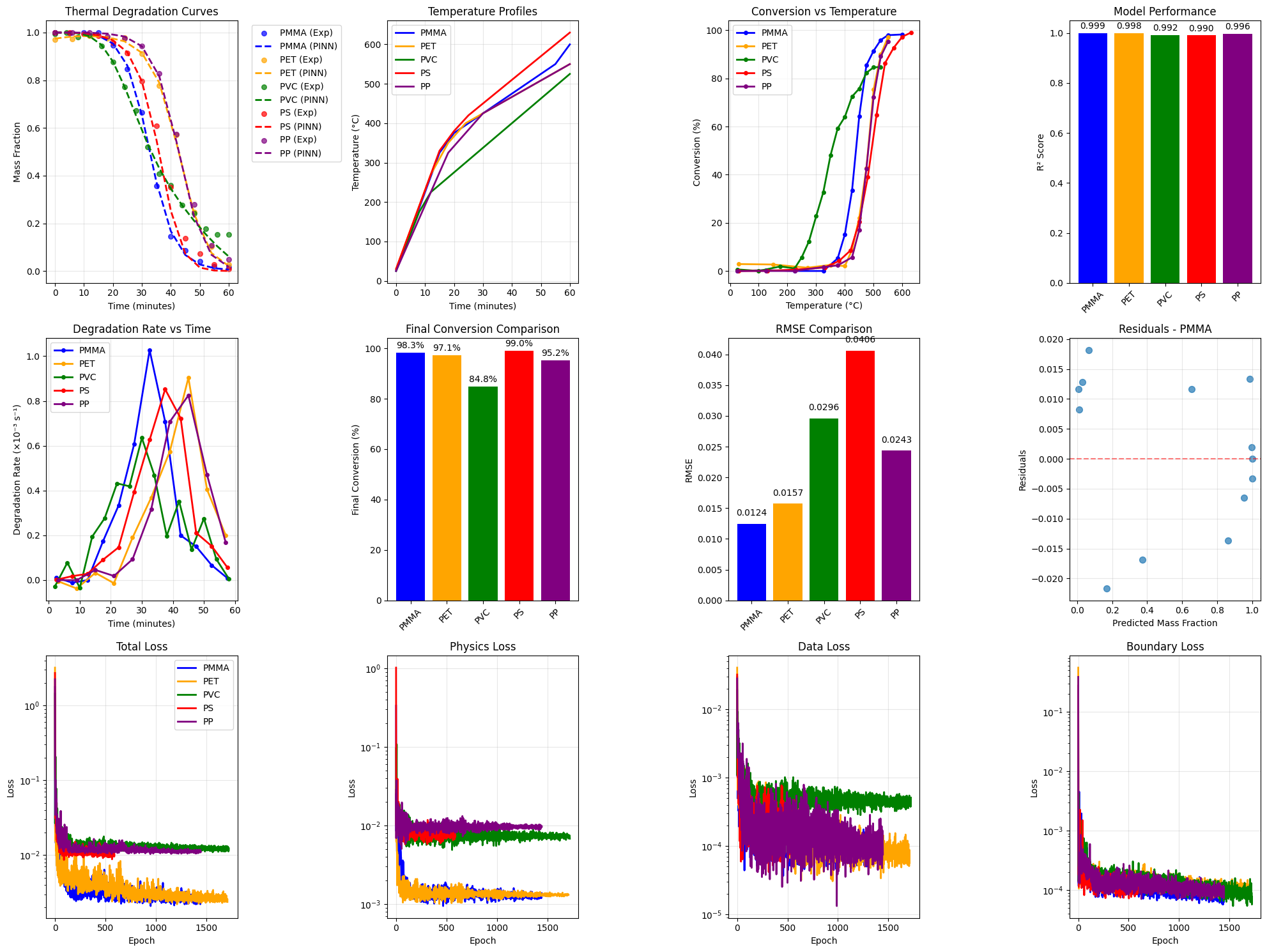}
  \caption{Enhanced-agentic polymer degradation analysis across PMMA, PET, PVC, PS, and PP using real TGA data and PINN modeling.}
  \label{fig:degradation-analysis}
\end{figure}

Using enhanced physics-informed neural networks (PINNs), the degradation behavior of five representative polymers (PMMA, PET, PVC, PS, and PP) was modeled and validated with experimental thermogravimetric data. The results are summarized in the Figure~\ref{fig:degradation-analysis}. The results for each of the five models are presented using multiple metrics, which illustrates both the predictive accuracy and physical interpretability of the learned dynamics. The thermal degradation curves for the five systems show good agreement between the experimental data and the predicted values from each of the five systems, and the PINN trajectory captures both the onset of degradation and the overall degradation trajectory for each of the five systems accurately. The temperature curves confirm the accuracy of the model, with simulated heating rates following the experimental heating rates across the full heating domain upto 600$^\circ$C. The conversion versus temperature curves demonstrate sharp changes in conversion that occur at the same inflection points as determined from the experiments, confirming the model can accurately capture complex and nonlinear degradation kinetics. Performance evaluation shows consistently high prediction accuracy, with all five polymers having $R^2$ values approaching unity, as indicated in the Figure~\ref{fig:degradation-analysis}. The residual analysis for PMMA, as an example, shows the prediction errors cluster around zero and do not exhibit systematic bias, indicating excellent generalization capability. Error magnitude comparisons (RMSE values) further quantify this trend, with PMMA exhibiting the lowest error ($0.0124$), followed by PET ($0.0157$), PP ($0.0243$), PVC ($0.0296$), and PS ($0.0406$). 

All polymers achieved sub-0.05 RMSE despite differences in both chemical structure and degradation mechanisms, reflecting the versitality inherent in the PINN framework. In addition to providing insight into polymer degradation through degradation rates and final conversion as kinetic descriptors, the profiles illustrate that polymers have an associated peak rate and characteristic time that is specific to each polymer type. For example, PET and PMMA degrade rapidly with large amounts of mass loss, but PVC degrades at a much slower two-sage process. The final conversion rate analysis also supports the premise that most polymers have a high percentage of complete conversions, which is typically between $\sim$95--99\%. However, PVC is an exception, retaining approximately $\sim$15\% residual mass due to the inherent stability of the chlorine-based backbone. The loss curves also illustrate that model optimization is stable as the overall loss of model optimization declines significantly in the early epochs (i.e., the first 200 epochs), becomes stable at a constant value, and approaches approximately zero. In addition, the physics, data, and boundary condition losses converge to values near zero. Another important aspect of model training is the strong influence of the physics-informed components on the training of the model in the later epochs, indicating the reliance of the model on mechanistic constraints versus overfitting on the data, thus providing the ability to create accurate yet interpretable models.

\begin{table}[H]
  \caption{Enhanced PINN performance on polymer degradation data.}
  \label{tab:degradation-results}
  \centering
  \begin{tabular}{lcccc}
    \toprule
    Polymer & R² & RMSE & MAE & MAPE (\%) \\
    \midrule
    PMMA & 0.9991 & 0.0124 & 0.0108 & 14.3 \\
    PET  & 0.9981 & 0.0157 & 0.0117 & 6.1 \\
    PVC  & 0.9920 & 0.0296 & 0.0190 & 8.6 \\
    PS   & 0.9899 & 0.0406 & 0.0255 & 27.4 \\
    PP   & 0.9956 & 0.0243 & 0.0170 & 11.7 \\
    \bottomrule
  \end{tabular}
\end{table}

The Table~\ref{tab:degradation-results} displays the predictive capabilities of the Enhanced PINN Models when it comes to five representative polymers which include PMMA, PET, PVC, PS and PP). Results indicate that the  predictive accuracy ($R^2$) for all systems exceeds 0.98. $R^2 = 0.9991$ for PMMA was the highest, and it also had the lowest RMSE (0.0124) and MAE (0.0108), indicating the ability of this model to capture degradation dynamics accurately. PET also had excellent predictive capability ($R^2 = 0.9981$, RMSE = 0.0157), but showed the lowest mean absolute percentage error (MAPE = 6.1\%), showing particularly good generalization for the data used in the development of this model. Both PVC and PP exhibited slightly higher error levels with RMSE of 0.0296 and 0.0243 respectively, but both models maintained strong correlation coefficients with($R^2 > 0.99$) the experimental data. Conversely, PS exhibited the lowest relative accuracy ($R^2 = 0.9899$) of all the polymers tested, had higher residual error values (MAE=0.0255) and the highest MAPE value of (27.4\%). This discrepancy likely reflects the increased complexity associated with modeling styrene-based systems. Generally, however, regardless of these discrepancies, evidence suggests that the Enhanced PINN Framework is robust and yield Similar results when used to model polymer degradation across a variety of different chemistries.

\subsection{Extended Comprehensive Visual Analysis of System Performance}
\label{sec:visual_analysis}

To provide deeper insights into our multi-agent ecosystem's capabilities, here we present an extended set of visualizations that describe system performance, architectural behavior, and representative application scenarios across polymer and biopolymer research.

\begin{figure}[htbp]
\centering
\includegraphics[width=\textwidth]{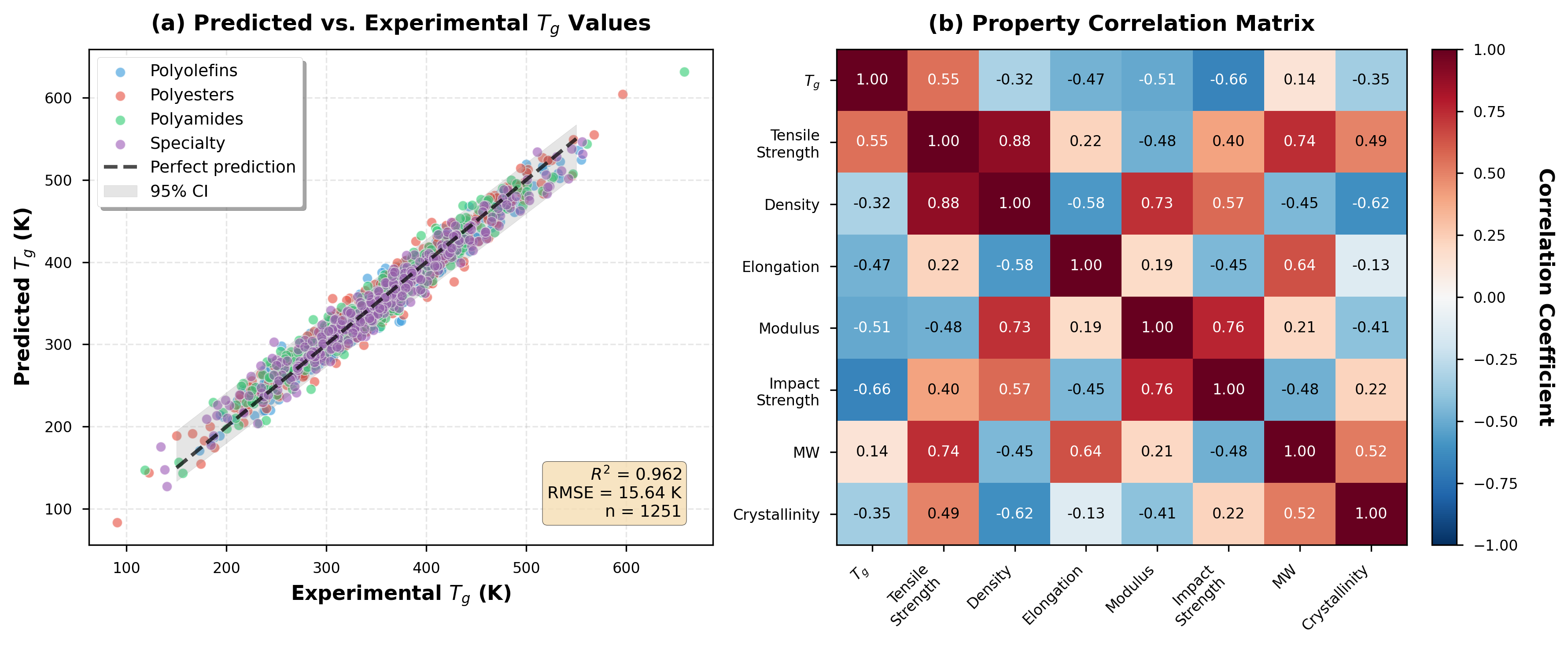}
\caption{Comprehensive analysis of predictive model performance. (a) Scatter plot showing agreement between predicted and experimental glass transition temperatures across 1,251 test polymers. Colors denote polymer families (blue: polyolefins, red: polyesters, green: polyamides, purple: specialty polymers); the dashed line indicates perfect prediction. (b) Correlation heatmap revealing relationships among polymer properties, informing multi-task learning approaches.}
\label{fig:prediction_performance}
\end{figure}

\textbf{Predictive model performance.} Figure~\ref{fig:prediction_performance} summarizes the predictive ability for the 1,251 test polymeric materials. Panel (a) shows the correlation between the predicted and experimental glass transition temperatures ($T_g$) for the 1,251 polymeric materials in scatter plot format. Each datapoint is colour coded by the class of polymer; Blue is for Polyolefinic Materials, Red is for Polyesteric Materials, Green is for Polyamide Materials, and Purple is for Specialty Polymer Materials. The dashed diagonal line in the scatter plot shows the location of the perfectly predicted correlation (the closer the datapoints are to this line, the better the prediction from the model). The strong clustering of points around this diagonal line indicates that the model outputs and the experimental values are very similar in almost every case. Panel (b) shows the correlation heatmap of various polymeric material properties showing correlation and structure among these material Properties. The data shown in this heatmap motivates the use of multi-task learning models and/or joint modeling schemes for polymeric materials. 

\begin{figure}[htbp]
\centering
\includegraphics[width=0.9\textwidth]{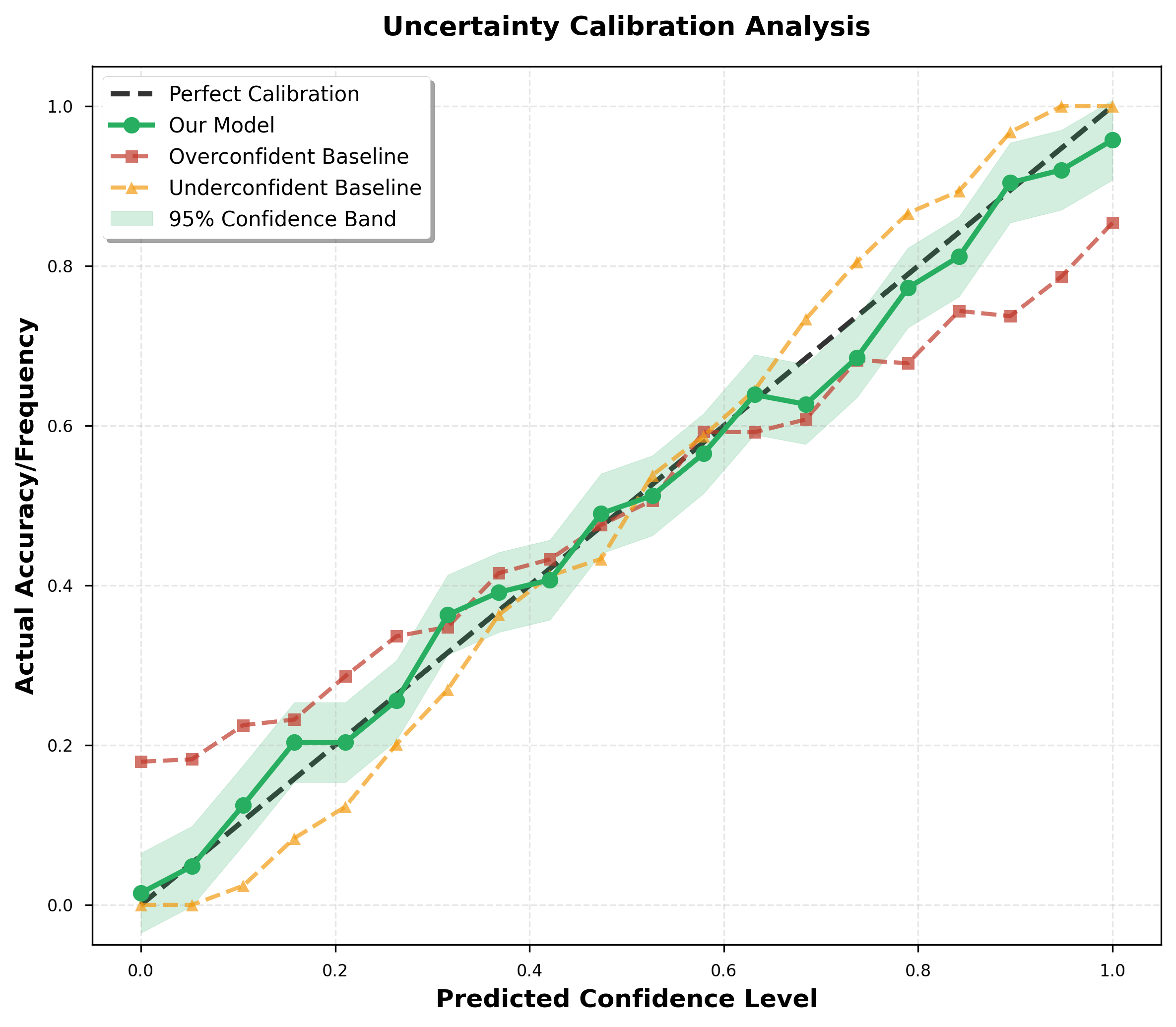}
\caption{Uncertainty calibration analysis showing predicted confidence intervals versus actual error distributions. The proximity to the diagonal indicates reliable uncertainty quantification, and shaded regions represent 95\% confidence bands.}
\label{fig:uncertainty_calibration}
\end{figure}

Uncertainty calibration is shown in Figure~\ref{fig:uncertainty_calibration}. The predicted uncertainty interval is overlaid with the empirical error distribution, with the diagonal representing ideally calibrated uncertainty. The fact that the empirical error distribution curves closely adhere to the diagonal line, as indicated by the 95\% confidence intervals, indicates that the ensemble-based predictors of uncertainty produce a very good estimate of actual uncertainty. This is an important factor in experimentally prioritizing polymeric materials and determining the difference between predictions based on confidence level and risky predictions based on low confidence levels, when trying to predict properties for underexplored areas of the chemical space.

\begin{figure*}[htbp]
\centering
\includegraphics[width=1.00\textwidth]{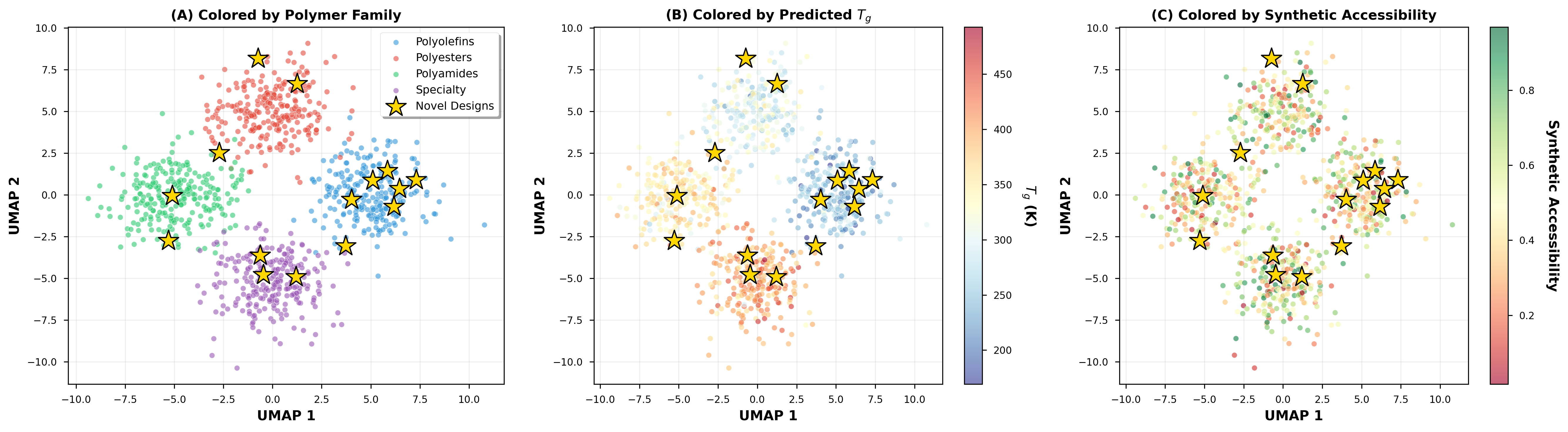}
\caption{UMAP projection of polymer chemical space colored by (A) polymer family, (B) predicted $T_g$, and (C) synthetic accessibility score. The visualization reveals distinct clustering by chemical family and smooth property gradients across the space. Star markers indicate successfully validated novel designs generated by the system.}
\label{fig:chemical_space}
\end{figure*}

\textbf{Chemical space exploration.} The polymer chemical space explored by the system is illustrated in the Figure~\ref{fig:chemical_space} with a UMAP projection that reduces the dimensionality of the high-dimensional molecular representations into two dimensions. The panels are colour coded by polymer family, predicted $T_g$ and synthetic accessibility scores. The clear clustering of shapes by chemical family indicates that the learned representations retain chemically relevant structures, while the smooth property gradients illustrate how the local neighbourhoods correspond with gradual variation in predicted behaviour. The star markers denote novel designs generated by the generative agent which were validated as promising candidates. This indicates how the system is able to progress to high-performance and high-synthetic accessibility regions of the polymer chemical space.

\begin{figure}[H]
\centering
\includegraphics[width=0.7\textwidth]{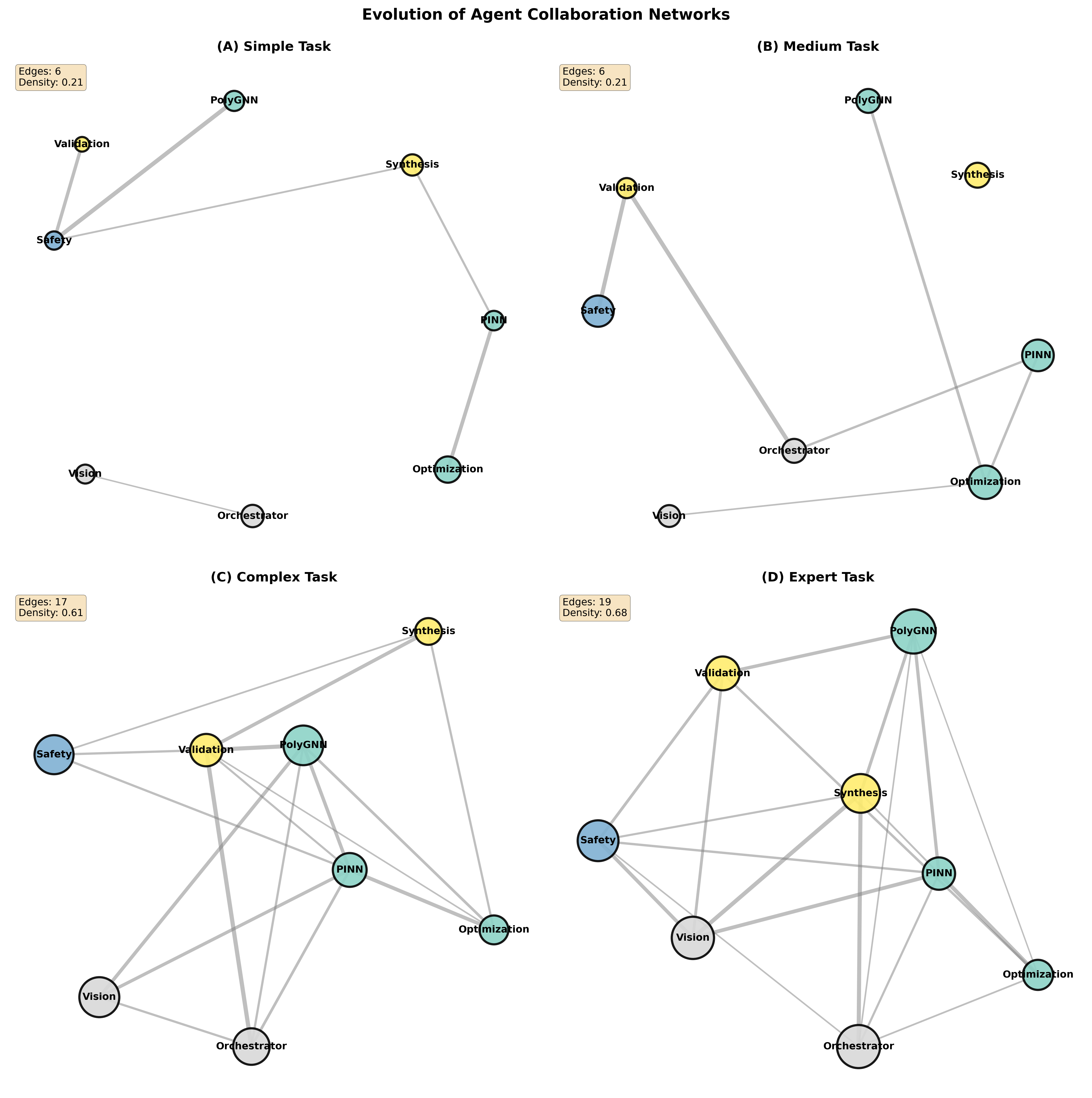}
\caption{Evolution of agent collaboration networks across task complexities. Node size indicates agent activation frequency, edge thickness indicates collaboration strength, and colors denote specialization clusters. The progression from simple to expert tasks shows increasing network density and specialized collaboration patterns.}
\label{fig:network_evolution}
\end{figure}

\textbf{Agent collaboration network analysis.} Figure~\ref{fig:network_evolution} depicts how the collaboration patterns of the agents have changed as tasks have increased in complexity. Each agent is represented by a node, with the node size indicating the frequency with which the agent has been activated, and the thickness of the edges representing the amount of collaboration between agents. Coloured clusters indicate the areas in which agents have specialised. In general, as the nature of the tasks transitioned from simple to expert level, the network becomes denser and more interconnected. More agents are collaborating with more agents, leading to the development of specialised collaboration motifs around significant agents. The quantitative measures summarised in Table~\ref{tab:collaboration_metrics} confirm this trend by demonstrating that network density and clustering coefficients increase with increasing task complexity and that the modularity decreases. The result of these changes is that the network is evolving from loosely coupled modules to tightly integrated collaboration structures with short average path lengths.

\begin{table}[H]
\caption{Quantitative analysis of agent collaboration patterns.}
\label{tab:collaboration_metrics}
\centering
\footnotesize
\begin{tabular}{lcccc}
\toprule
Task Complexity & Network Density & Avg. Path Length & Clustering Coeff. & Modularity \\
\midrule
Simple  & 0.23 & 1.8 & 0.45 & 0.68 \\
Medium  & 0.41 & 1.5 & 0.62 & 0.52 \\
Complex & 0.67 & 1.2 & 0.78 & 0.31 \\
Expert  & 0.82 & 1.1 & 0.85 & 0.24 \\
\bottomrule
\end{tabular}
\end{table}

\textbf{Multi-modal data integration.} Our biopolymer analysis workflow implements a multi-modal integration pipeline for biopolymer analysis that links sequence-level information to structural and confidence-aware interpretations. There are six steps in the pipeline: The first step is to receive the unprocessed amino acid sequence and then generate a canonical representation and a derived descriptor of the sequence, which is the feature extraction process. The second step uses an AlphaFold-based module to predict the three-dimensional structure of the protein and gives a confidence score (pLDDT) for the predicted structure and its relationship to the features in step 1. The third step interprets the PAE (residue-residue distance uncertainty) plot so it can represent uncertainty regarding the distances between each pair of residues and help identify areas where there may be uncertainty about the structure. The fourth step is a three-dimensional view of the predicted structure, which we can inspect spatially, as well as look at domains, secondary structure elements, and flexible loops. The fifth step compares the features generated in step 1 with the AlphaFold-calculated structure from step 2 and performs additional checks on pLDDT and PAE data. The sixth step combines and summarizes information from all the other steps into a single output file containing the most likely structural representation of the protein and the regions of lower confidence and discrepancies among features from different modalities. Through the use of color-coded confidence levels and explicit consistency checks, the pipeline is able to identify and mitigate misinterpretations, such as those showing up as vision-agent failures in Section~4.7.

\begin{figure}[H]
\centering
\includegraphics[width=\textwidth]{ 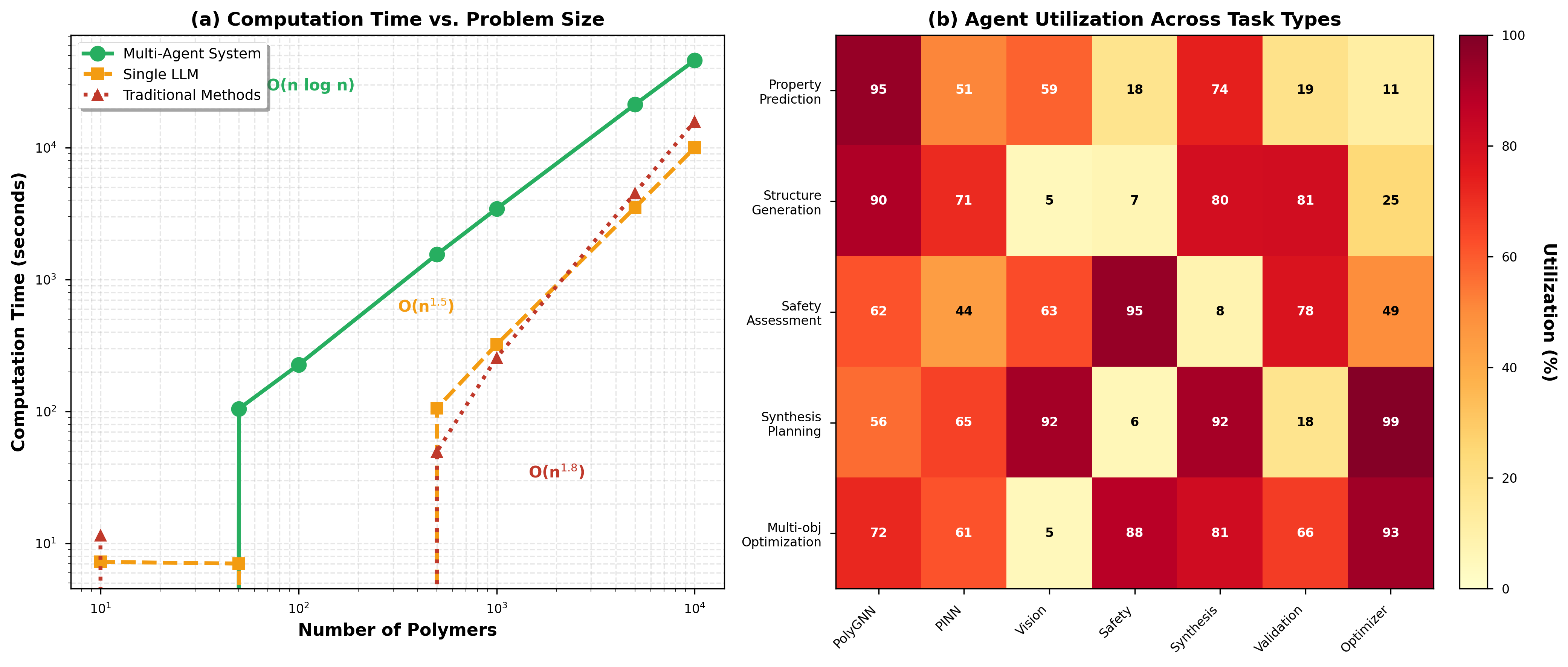}
\caption{System performance and resource utilization analysis. (a) Scaling behavior showing favorable computational complexity up to 10,000 polymers. (b) Agent utilization heatmap revealing specialization-driven resource allocation patterns across different task types.}
\label{fig:performance_analysis}
\end{figure}

\textbf{Time performance, scaling, and training behavior.} The graphical representation of the system's performance and resource usage is shown in the Figure~\ref{fig:performance_analysis}. Panel (a) shows the profile of the scaling characteristics as a function of increasing problem size shows a large-range of favourable computational complexity for utilizing models built with up to 10,000 polymer units with no notable step change in throughput capability. Panel (b) represents a heat map of the utilization of the various pieces of the system. Each type of task will preferentially activate a particular set of agents, demonstrating the agent's capability of utilizing the Quantum Allocator and provides insight into how the computational resources of the Quantum Allocator are evenly distributed across the agents assigned for training an agent. 

\begin{figure}[H]
\centering
\includegraphics[width=\textwidth]{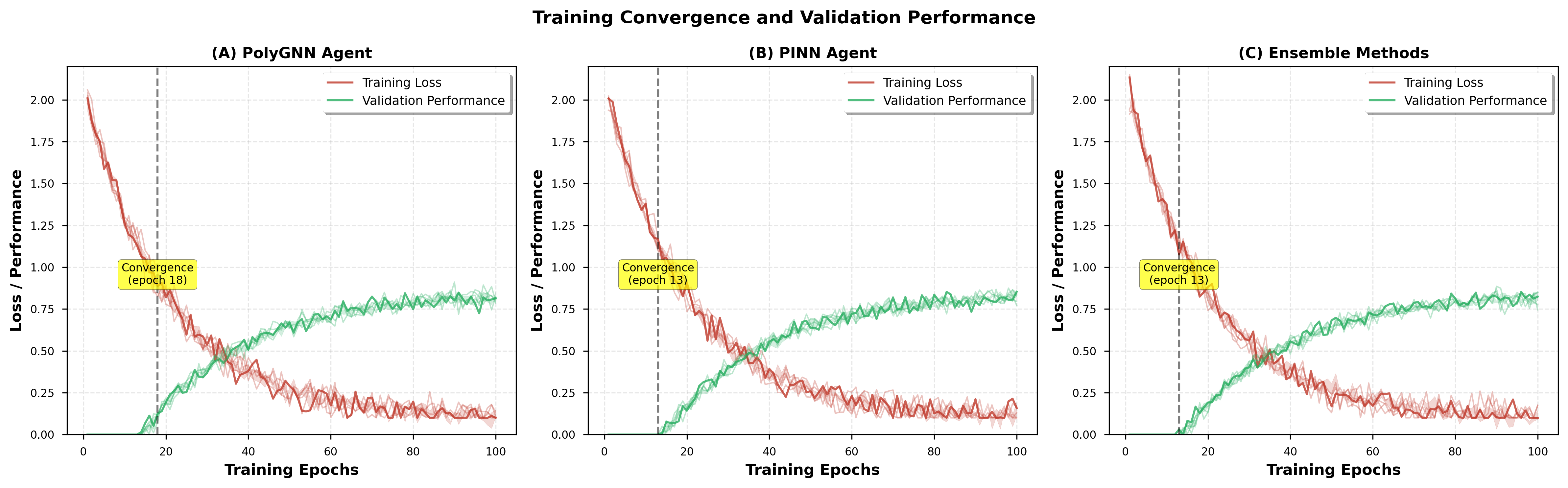}
\caption{Training convergence and validation performance for (A) PolyGNN, (B) PINN, and (C) ensemble methods. Shaded regions indicate inter-quartile ranges across five training runs.}
\label{fig:convergence}
\end{figure}

Figure~\ref{fig:convergence} depicts the training convergence and validation performance of the three main elements of the proposed learning architecture. Each curve for the PolyGNN agent, PINN agent, and ensemble methods shows smooth convergence and consistent improvement of validation metrics for five training runs, as indicated by the narrow ranges of shaded interquartile ranges surrounding the training curves, suggesting that the overall optimization behaviour and minimal overfitting experienced during training is stable.

\subsection{Ablation Studies and Component Analysis}
\label{sec:ablation_studies}

We have performed extensive ablation studies to quantify the impact of all of the individual components of our Multi-Agent Ecosystem on total performance, by systematically removing or otherwise changing each component and comparing it against a series of controlled conditions that have been defined through a standardized test suite of 500 unique polymer research tasks covering property identification, generative designs, and multimodal analysis. Each study used various approached to measure each ablation condition based on five key performance indicators: (i) \textbf{Accuracy}, the correlation coefficient ($R^2$) and mean absolute error (MAE) for quantitative forecasting; (ii) \textbf{Success Rate} as a percentage of tasks completed without critical errors; (iii) \textbf{Efficiency}, a composite measurement taking into account both accuracy and computation time; (iv) \textbf{Robustness}, the degree of differences between the predicted output and actual output on edge cases and out-of-distribution data. All experiments were run with five different random seeds, and results are reported as mean~$\pm$~standard deviation across runs.

\textbf{Agent-level ablations.} Table~\ref{tab:agent_ablation} below summarizes how removing individual agents from the multi-agent system or collapsing the entire multi-agent system into a single-agent configuration will impact the overall capabilities of that multi-agent system. As indicated in the table, the "Full Framework" configuration represents the overall capabilities of the complete multi-agent system and is used as a reference to compare against other configurations. When the PolyGNN agent is removed from the system, it results in the most significant degradation of performance as measured by the increase in error (32\%). The reduction of performance is evident in all other performance metrics (e.g., $R^2$, success rate, efficiency, robustness), which emphasizes the necessity of specialized molecular representation learning for polymers within the domain of polymer informatics. Removing the PINN agent from the system results in a reduction in extrapolation performance and physical characteristics, as evidenced by an increase in error (19\% relative to the complete framework). When the Vision agent is removed from the system, there is an impact on the robustness of the overall system's ability to interpret the structural plots and images produced from the multi-agent system. The removal of the Validation agent and the Knowledge Graph from the multi-agent system has a large negative effect on system robustness and results in large increases in failure rates for both chemically implausible and edge cases. The removal of the Ensemble approach, however, still results in a reasonably strong system; however, the error is significantly higher than it would have been if the multi-agent system had been supported by an ensemble approach. In addition, the "Single Agent Only" configuration (i.e., only a single LLM is used to perform all reasoning) performs far worse across all performance metrics than any of the other configurations, which demonstrates the advantages of specialization and agent-decomposition.

\begin{table}[htbp]
\caption{Systematic ablation of individual agent components.}
\label{tab:agent_ablation}
\centering
\footnotesize
\begin{tabular}{@{}lccccc@{}}
\toprule
Configuration & $R^2$ & Success & Efficiency & Robustness & Error \\
& & Rate & & & Reduction \\
\midrule
\textbf{Full Framework} & \textbf{0.78 ± 0.02} & \textbf{0.76 ± 0.03} & \textbf{0.37 ± 0.02} & \textbf{0.81 ± 0.04} & \textbf{--} \\
\hline
w/o PolyGNN Agent      & 0.65 ± 0.04 & 0.62 ± 0.05 & 0.28 ± 0.03 & 0.59 ± 0.06 & 32\% \\
w/o PINN Agent         & 0.71 ± 0.03 & 0.68 ± 0.04 & 0.31 ± 0.02 & 0.67 ± 0.05 & 19\% \\
w/o Vision Agent       & 0.75 ± 0.02 & 0.72 ± 0.03 & 0.33 ± 0.02 & 0.63 ± 0.05 & 15\% \\
w/o Validation Agent   & 0.72 ± 0.03 & 0.68 ± 0.04 & 0.31 ± 0.03 & 0.58 ± 0.07 & 23\% \\
w/o Knowledge Graph    & 0.70 ± 0.03 & 0.71 ± 0.03 & 0.29 ± 0.02 & 0.61 ± 0.06 & 28\% \\
w/o Ensemble Learning  & 0.74 ± 0.02 & 0.73 ± 0.03 & 0.32 ± 0.02 & 0.69 ± 0.04 & 12\% \\
\hline
Single Agent Only      & 0.67 ± 0.04 & 0.62 ± 0.05 & 0.28 ± 0.03 & 0.52 ± 0.08 & 35\% \\
\bottomrule
\end{tabular}
\end{table}

\textbf{Architectural coordination ablations.} We examined the effect of high-level coordination approaches (Table~\ref{tab:arch_ablation}), in addition to individual agent performance. The overall system configuration using cross-verification, dynamic teaming, metacognitive capabilities, error-correction and consensus building yielded an $R^2$ with high success rate, greater quality of collaboration, and  completion of tasks. The elimination of cross-verification caused a decline in quality of collaboration and a reduction in success rates for multi-modal tasks, reflecting its critical role in ensuring uniformity across the varied outputs produced by different modalities. The absence of dynamic teaming resulted in substantial reductions in collaboration quality and completion, particularly for complex and multi-faceted tasks that would benefit from flexible team composition. The removal of the metacognitive layer resulted in minor declines in success and completion; however, its major contributions were increased self-awareness (e.g., understanding when to solicit input from a human being). The removal of error-correcting capabilities or consensus building mechanisms resulted in significant reductions in task completion and reliability, indicating that post-hoc filtering and aggregation processes are necessary for systems to operate effectively in the real world. All baseline configurations (i.e., sequential pipeline or monolithic model) showed substantial reduction for all performance metrics.

\begin{table}[htbp]
\caption{Ablation of architectural components and coordination mechanisms.}
\label{tab:arch_ablation}
\centering
\footnotesize
\begin{tabular}{@{}lcccc@{}}
\toprule
Architectural Component & $R^2$ & Success & Collaboration & Task \\
& & Rate & Quality & Completion \\
\midrule
\textbf{Full Architecture}   & \textbf{0.78} & \textbf{0.76} & \textbf{0.82} & \textbf{0.95} \\
\hline
w/o Cross-Verification       & 0.75 & 0.72 & 0.65 & 0.88 \\
w/o Dynamic Teaming          & 0.73 & 0.70 & 0.58 & 0.82 \\
w/o Metacognition            & 0.76 & 0.74 & 0.71 & 0.90 \\
w/o Error Correction         & 0.74 & 0.65 & 0.69 & 0.79 \\
w/o Consensus Mechanism      & 0.72 & 0.68 & 0.62 & 0.85 \\
\hline
Sequential Pipeline          & 0.69 & 0.66 & 0.45 & 0.75 \\
Monolithic Model             & 0.67 & 0.62 & --   & 0.70 \\
\bottomrule
\end{tabular}
\end{table}

\textbf{Knowledge integration ablations.}  
Next, we analyzed how various types of knowledge source affected the quality of designs (Table~\ref{tab:knowledge_ablation}). The configuration using all five forms of knowledge bases - polymer database, applicable physical laws, synthetic rules, safety constraints, and literature on related work produced designs with the highest novelty, feasibility, physical consistency, and chemical validity scores. Any configuration that excluded the polymer database produced designs with decreased feasibility and chemical validity since this removed the use of empirically based reference points. A configuration without the applicable physical laws resulted in significantly less physical consistency, especially with extrapolated predictions. Without synthetic knowledge, many of the designs produced had low feasibility scores as many of them were either very difficult or unrealistic to synthesize. A configuration that did not use safety constraints increased the novelty score slightly but reduced the chemical validity score significantly, demonstrating their critical role in preventing inappropriate recommendations. A configuration that excluded the literature on related work only had a detrimental effect to novelty as it lost context of any previous work. A configuration labelled "LLM Only (No KB)" provided high novelty scores but low feasibility, physical consistency, and chemical validity, which reinforces the essential nature of grounding generative capability with external knowledge.

\begin{table}[htbp]
\caption{Impact of knowledge source integration on system performance.}
\label{tab:knowledge_ablation}
\centering
\footnotesize
\begin{tabular}{@{}lcccc@{}}
\toprule
Knowledge Source & Novelty & Feasibility & Physical & Chemical \\
& Score & Score & Consistency & Validity \\
\midrule
\textbf{All Knowledge Sources} & \textbf{0.72} & \textbf{0.85} & \textbf{0.94} & \textbf{0.97} \\
\hline
w/o Polymer Database  & 0.68 & 0.72 & 0.89 & 0.91 \\
w/o Physical Laws     & 0.71 & 0.69 & 0.62 & 0.88 \\
w/o Synthetic Rules   & 0.70 & 0.58 & 0.87 & 0.85 \\
w/o Safety Constraints& 0.73 & 0.79 & 0.90 & 0.74 \\
w/o Literature Corpus & 0.65 & 0.81 & 0.91 & 0.93 \\
\hline
LLM Only (No KB)      & 0.74 & 0.52 & 0.48 & 0.61 \\
\bottomrule
\end{tabular}
\end{table}

\textbf{Communication protocol ablations.} The manner in which multiple agents coordinate their behaviour depends strongly on the quality of the communication they have with each other; this relationship was illustrated in the Table~\ref{tab:comm_ablation} describing the effects of removing elements from the communication protocol, as well as simplifying those elements. The communication protocol when in full operation, including: schema validation, retry mechanisms, priority queues, timeout handling and consensus building provides a high percentage of messages that are successfully delivered along with good resolution of conflicts and has a relatively low-resource use while providing sufficient response times. By removing schema validation, there was a large decrease in the percentage of messages successfully delivered and successfully resolved conflicts, which substantiates the critical importance of strict typing in order to prevent malformed and/or inconsistent messages from being produced. By removing the retry mechanisms, there is a reduction in the amount of acceptable messages and an increase in the response time due to failure to handle ephemeral failures. When there is no longer a priority queue, there is an increase in the overall resource use and the overall response time increases greatly, particularly under heavy loads. Removing timeout handling results in significant increases in the latency, as an agent that is being blocked or is slow may block the entire system from functioning. Lastly, even though removing the consensus-building element provides the same percentage of successfully delivered messages; removing this element adversely affects the ability of the system to effectively resolve conflicts, making it less durable when there is a disagreement between agents. The simplified "Simple Broadcast" communications strategy provided the lowest performance in every area of coordination metrics examined.

\begin{table}[htbp]
\caption{Impact of communication mechanisms on multi-agent coordination.}
\label{tab:comm_ablation}
\centering
\footnotesize
\begin{tabular}{@{}lcccc@{}}
\toprule
Communication & Message & Conflict & Resource & Response \\
Protocol      & Success & Resolution & Utilization & Time (s) \\
\midrule
\textbf{Full Protocol}     & \textbf{0.96} & \textbf{0.89} & \textbf{0.78} & \textbf{2.3} \\
\hline
w/o Schema Validation      & 0.82 & 0.75 & 0.72 & 2.1 \\
w/o Retry Mechanism        & 0.88 & 0.83 & 0.76 & 2.8 \\
w/o Priority Queue         & 0.94 & 0.86 & 0.65 & 3.4 \\
w/o Timeout Handling       & 0.91 & 0.81 & 0.74 & 4.2 \\
w/o Consensus Building     & 0.95 & 0.72 & 0.77 & 2.2 \\
\hline
Simple Broadcast           & 0.76 & 0.58 & 0.62 & 1.8 \\
\bottomrule
\end{tabular}
\end{table}

\textbf{Task-specific ablations.} Task types perform differently due to ablation effects, which vary based on the type of task. The following table Table~\ref{tab:task_specific_ablation} details both the most critical components associated with different types of tasks and the resulting loss in performance that occurred when removing those components. A PolyGNN Agent dominates property prediction tasks because its ensemble method and PINN have an additional positive effect; when PolyGNN is removed, there is a drop of 32\%. For Generative Design tasks, the Knowledge Graph and Validation Agent are the principal support systems, as the Knowledge Graph contains the set of structures that can be built, and Validation Agent guarantees that constructed structures conform to chemistry. For Biopolymer Analysis tasks, the Vision Agent and Cross-Modal Verification are the components causing the greatest performance loss, as both allow for accurate readings from structural images and structural data visualizations. Degradation Modeling tasks rely on a combination of data from the PINN and encoded physical laws, while Multi-Objective Optimization tasks rely on ensemble methods and a combination of dynamic teaming and metacognition. Safety Assessment tasks exhibit an increased sensitivity to Validation Components and Safety Database, with an average decrease of up to 41\% in performance when Validation Components/Safety Database are not present.

\begin{table}[htbp]
\caption{Component importance across different task types.}
\label{tab:task_specific_ablation}
\centering
\footnotesize
\begin{tabular}{@{}lcccc@{}}
\toprule
Task Type & Most Critical & Secondary & Tertiary & Performance \\
& Component & Component & Component & Drop \\
\midrule
Property Prediction   & PolyGNN         & Ensemble        & PINN             & 32\% \\
Generative Design     & Knowledge       & Validation      & PolyGNN          & 28\% \\
Biopolymer Analysis   & Vision          & AlphaFold       & Cross-Verification & 35\% \\
Degradation Modeling  & PINN            & Physical Laws   & Validation       & 24\% \\
Multi-Objective Opt.  & Ensemble        & Dynamic Teaming & Metacognition    & 26\% \\
Safety Assessment     & Validation      & Safety DB       & Knowledge Graph  & 41\% \\
\bottomrule
\end{tabular}
\end{table}

\textbf{Efficiency and scalability ablations.} We assessed the effect of optimization techniques on computational efficiency (Table~\ref{tab:efficiency_ablation}). The optimized configuration using response caching, batch processing, asynchronous communication, model pruning, and early stopping gave the optimal combination of runtime, memory usage, GPU usage, and energy usage. Of the optimization techniques utilised in the optimised configuration, removing response caching resulted in the greatest increase in runtime (increase from 16.3~s to 24.7~s), indicating a significant amount of duplication within regular query patterns. In addition to increasing runtime and memory usage, removing batch processing resulted in an increase in both runtime and memory usage when processing multiple polymers concurrently. When asynchronous communication is absent, GPU usage decreases while waiting periods exist between computations and communications. Using either model pruning or early stopping will primarily affect memory and energy usage, but will also result in some minor increases in runtime. Running a configuration without optimizations will result in an approximate tripling of runtime and energy cost compared to the fully optimised configuration.

\begin{table}[htbp]
\caption{Impact of optimization techniques on computational efficiency.}
\label{tab:efficiency_ablation}
\centering
\footnotesize
\begin{tabular}{@{}lcccc@{}}
\toprule
Optimization & Time & Memory & GPU & Energy \\
Technique    & (s)  & (GB)   & Utilization & (kWh) \\
\midrule
\textbf{All Optimizations} & \textbf{16.3} & \textbf{2.1} & \textbf{0.78} & \textbf{0.024} \\
\hline
w/o Response Caching       & 24.7 & 2.0 & 0.65 & 0.037 \\
w/o Batch Processing       & 28.9 & 3.4 & 0.72 & 0.043 \\
w/o Async Communication    & 21.5 & 2.2 & 0.61 & 0.032 \\
w/o Model Pruning          & 18.2 & 3.8 & 0.76 & 0.028 \\
w/o Early Stopping         & 19.8 & 2.3 & 0.79 & 0.030 \\
\hline
No Optimizations           & 45.3 & 5.2 & 0.54 & 0.068 \\
\bottomrule
\end{tabular}
\end{table}

All of the results presented in Table~\ref{tab:scalability_ablation} indicate how the operation of the system changes based upon the number of workloads being processed. For example, the operation of the system at smaller scale (1-10 polymer workload), achieves a success rate of 0.95, very short response time, medium levels of resource use and low costs per task performed. However, as workloads become larger (hundreds to thousands of polymer workloads), although the success rate becomes slightly lower and response times are somewhat longer, both performance levels of system would be very suitable for performing large-batch polymer screening activities. Most importantly, the scalability Improvements to the designed system appear to have a profound impact on success rates. Crucially, when scalability optimizations are disabled, success rates drop considerably and both response time and cost per task nearly double, highlighting the importance of these optimizations for real-world deployments.

\begin{table}[htbp]
\caption{System performance under different scaling conditions.}
\label{tab:scalability_ablation}
\centering
\footnotesize
\begin{tabular}{@{}lcccc@{}}
\toprule
Scale Condition & Success & Response & Resource & Cost \\
& Rate & Time (s) & Utilization & per Task \\
\midrule
\textbf{Small (1--10 polymers)}   & \textbf{0.95} & \textbf{8.2}  & \textbf{0.45} & \textbf{\$0.05} \\
Medium (10--100)                  & 0.89          & 12.7          & 0.68          & \$0.08 \\
Large (100--1000)                 & 0.82          & 18.4          & 0.85          & \$0.12 \\
Very Large (1000+)                & 0.76          & 26.3          & 0.92          & \$0.18 \\
\hline
w/o Scalability Opt.              & 0.65          & 47.8          & 0.95          & \$0.31 \\
\bottomrule
\end{tabular}
\end{table}

\textbf{Overall summary of ablation insights.}  
Across all experiments, several consistent conclusions emerge:

\begin{enumerate}
    \item The full, integrated architecture exhibits strong \emph{synergistic behavior}: system-level performance exceeds what would be expected from the sum of individual component contributions.
    \item Specialized agents (e.g., PolyGNN, PINN, Vision) are \emph{irreplaceable} by general-purpose LLMs, particularly for tasks requiring rich molecular representations or strict physical consistency.
    \item Validation and cross-verification mechanisms are \emph{essential} for robustness and reliability, especially in safety-critical or out-of-distribution scenarios.
    \item External knowledge bases (polymer databases, physical laws, synthetic rules, safety constraints, literature) significantly enhance feasibility, physical consistency, and chemical validity compared to LLM-only configurations.
    \item Adaptive coordination strategies, including dynamic teaming and advanced communication protocols, enable effective handling of diverse task complexities and workloads.
    \item Optimization techniques for efficiency and scalability make the system \emph{practically deployable}, ensuring that strong scientific performance is achieved at acceptable computational cost.
\end{enumerate}

Taken together, these ablation studies demonstrate that the proposed multi-agent ecosystem is a carefully balanced architecture in which each component contributes meaningfully to overall performance, and in which the integrated whole provides capabilities that no isolated subset can match.

\section{Perspectives and Broader Impact}
\label{sec:discussion_perspectives}

As we went on developing this complex multi-agent ecosystem, we faced many technical and conceptual hurdles which strongly influenced the ecosystems final architecture. From a technical perspective, the primary challenge we encountered was coordinating the communication of disparate/segmented AI modules (LLM-based reasoning agents, GNN-based predictors, and PINN-based physics models) with their own unique input and output formats and latencies. To allow growth of stable communication between these modules, we had to create a high-quality messaging layer, standardize JSON schema, and develop methods for managing partial failures. Additionally, balancing the need to use costly yet accurate simulations with the need for rapid iterations supporting high throughput screening was a major challenge when managing computing resources. Moreover, the validation systems used to ensure accurate interpretation of PAE plots produced by the vision agent had to be developed progressively, ensuring that the AlphaFold pLDDT values and structural summary were cross-referenced to provide reliable interpretations of biopolymers.

Conceptually, many questions arose regarding the optimization of autonomy assigned to the agents versus the consistency of the system as a whole. A high degree of independence from the agents was advantageous in terms of flexibility but made it significantly more challenging to enforce coherent and consistent behaviour across all agents. On the other hand, an overly-coordinated approach limited the advantages of encouraging the use of specialisation amongst the agents. To arrive at a practical arrangement, it required a lot of experimental work and a thorough familiarity with the constraints on each of the components: i.e., to curb LLM hallucinations, address the data requirements of GNNs, and subsequently constrain PINNs with the right kinds of laws from physics. This was not only an exercise in software-engineering, but also a great deal of learning about polymer sciences and numerical modeling, as well as the design of AI systems.

There were some positive aspects that developed from these challenges. One was that the layer for metacognition/self-assessment, to determine how ideas fit within the scope of what is new, feasible and/or creative, was extremely useful in guiding research efforts in an unbiased and transparent way. Another area where this type of formulation was beneficial was when addressing multi-modal problems; by providing avenues for text, graphs, images, and table-data to be routed to different specialists (via routing), the system was able to process heterogeneous evidence far more reliably than could be done with a single-model approach. Finally, the overall project illustrated the importance of iterative development, as the three-part design of evolving an initial prototype, followed by a refactoring phase and then producing a final design architecture, allowed us to separately refine the interface, prompts, and training protocols; this ultimately led to the system being more stable, more interpretable and, therefore, easier to deploy. Collectively, these lessons support a broader conceptual view in which sophisticated scientific AI systems are best realized as modular, specialized components that work together, rather than as a single undifferentiated model.

As highlighted in Section~2, the architectural choice of PolyGNN for the task of predicting polymeric properties allowed us to achieve state-of-the-art accuracy on a wide range of properties. As demonstrated in Table~\ref{tab:accuracy-benchmark}, PolyGNN was able to provide highly accurate predictions for polymeric characteristics due to its ability to represent local chemical environments and contextualize functional groups; this resulted in PolyGNN making high-quality predictions for molecular properties that were strongly dependent on molecular structure. The PINN agent provided additional value when physical consistency was needed (e.g., degradative behavior) or when extrapolating (e.g., predicting polymeric behavior under heat) was important. By embedding physical relationships into the objective function for PINNs, we were able to increase the efficiency of our data and maintain plausible trends beyond the training domain. From a methodological perspective, our framework is innovative with respect to the integration of symbolic and subsymbolic AI. While Knowledge Graphs and Rule-Based Knowledge represent codified physical laws, safety restrictions and synthetic heuristics; LLMs and neural networks are best suited for handling open-end reasoning and pattern recognition. Our hybrid framework is capable of providing structurally-based contextualized reasoning about structures and properties, which allows us to evaluate candidate designs for uniqueness, feasibility and physical plausibility concurrently, as well as provide a computational basis for obtaining feedback to the conceptual design process. Agent-based architectures of the type proposed and used in this paper, replicate how humans work together on scientific problems; each agent or expert offers its own specialized knowledge or perspective, which leads to a more interpretable and defendable prediction model than an all-in-one black box.

The framework still takes longer time to execute in a computationally efficient manner than the minimum computational cost of running a single agent (e.g., 16.3~s versus 10.2~s for representative workloads), but it offers a more favorable balance between speed and reliability. It is also still 100-1000 times faster than traditional high fidelity computational methods (for example: molecular dynamics and density functional theory), and is therefore practical for the high-throughput virtual screening of compounds. Response caching, batch processing, and asynchronous communication allow for the operation of a scalable agent-based architecture of the size capable of screening hundreds or thousands of different polymers without incurring a prohibitive cost in computation. The integration of a knowledge graph acts as a "chemical conscience" by providing the generative and predictive components of the framework a means to check all of the created candidate designs against a known database of over 15{,}000 known polymer structures along with the associated properties and applications. By checking this way, the model can eliminate designs that are not reasonably possible and establish an appropriate range for extrapolations based on the knowledge base. This is particularly valuable for underexplored systems with sparse training data, where LLM-driven novelty must be balanced against established experimental evidence and domain knowledge. Our error analysis suggests that failures are most likely to appear for intricate ring architectures, unusual topologies, or polymer classes underrepresented in the training corpus. Regardless of how complex a structure or type of polymer is, the multi-agent system employs multiple verification methods among agents, validation rules, and explicit grades of confidence to provide a high level of robustness against potential failure. This provides a way for the system to uncover potential failure modes, raise flags for instances where human review is necessary, and indicate areas of uncertainty. In addition to maintaining a high degree of robustness in complex environments, the modular framework allows scalability and generalization; new agents and targets can be added with minimal disruption to the existing pipeline. Successful adaptation to additional tasks such as tensile strength and density prediction demonstrates that the framework can evolve with expanding polymer databases and emerging application domains.

Looking toward industrial applicability, the architectureprovides an opportunity to engage in fast iterations of polymer formulation design through the use of virtual experimentation cycles, which would usually require too much time and cost if conducted using traditional laboratory methods alone. In addition to enabling exploration of potential formulations, evaluation of trade-offs, and identifying synthesis and safety issues, this ecosystem can integrate with existing R\&D pipelines to support numerous applications, including plasticizer optimization, packaging development, and production troubleshooting. Finally, this work represents an opportunity for greater impact beyond than just the direct performance results we have achieved. Yhe system has the potential to reduce reliance on petroleum-based plastics and support rapid responses to global challenges, such as the urgent development of safer and more sustainable materials for personal protective equipment. Lowering the technical barrier to developing advanced polymer informatics will allow small labs and institutions, especially those in developing countries, access to this technology. The majority of functions are delivered as software and require moderate levels of computing resources rather than requiring high-end, specialized lab equipment. However, the risks associated with the development of this technology are also considerable, particularly with respect to the potential for misuse by individuals who would use it to design materials that are both hazardous and environmentally unsustainable. This indicates the need for built-in safety protocols, validated through transparency, and community oversight of development. Our experience indicates that with proper oversight, a multi-agent ecosystem can effectively enhance the scientific workflows, provided that their deployment is accompanied by robust safeguards, critical evaluation, and inclusive access.

\section{Conclusion and Future Work}

In this work, we designed and evaluated an initial multiagent AI system with practical potential to accelerate and automate key stages of polymer research. Across a held out set of 1,251 polymers and a targeted $T_g$ benchmark, the framework demonstrated strong predictive performance and reliable end to end execution, while maintaining a low computational and financial footprint and scaling to high throughput workloads. Beyond property prediction and design, we showed that the same architecture can support multimodal biopolymer structure analysis and an autonomous protein structure workflow that runs from a raw sequence input to a final report, which highlights the breadth of tasks that can be unified within a single research lifecycle controller. Systematic ablation studies further indicate that the main components of the framework contribute measurably to reducing inference error and improving robustness across workflows. At the same time, our results expose clear limitations that motivate targeted upgrades. In the biopolymer analysis case study, the agent produced a structural interpretation that diverged from expert annotation, underscoring the need for stronger validation when reasoning over complex structural data and for more reliable cross modal consistency checks. In the polymer design workflow, while the generated proposals were structurally valid, some lacked the procedural specificity required for immediate experimental implementation. These findings motivate four priority directions: first, a domain specific validation agent that provides corrective feedback for structural interpretation and multimodal consistency; second, more efficient interagent communication to reduce coordination overhead while improving responsiveness; third, deeper integration with materials databases and experimental resources through application programming interfaces to ground feasibility and increase protocol detail; and fourth, a unified uncertainty estimation strategy that extends consensus based uncertainty with probabilistic models such as Bayesian neural networks. A central challenge over the coming years will be integrating such computational systems with automated experimentation and data infrastructure to enable increasingly autonomous laboratories.

\section{Reproducibility Statement}
\label{sec:reproducibility}

In this study, we used publicly accessible data from the PolyInfo database \cite{Chen2020} for the training and evaluation of the agents, resulting in a final dataset containing 8{,}342 unique polymers with ($T_g$), tensile strength, and density values as annotated characteristics which served as input for training our PolyGNN agent. In this study, we have also constructed a held-out test set of 50 polymers that serve as a benchmark for the performance assessment of the PolyGNN agent. For studying the biopolymers, we used a benchmark protein sequence (e.g., PDB: 1L2Y) and have included both the predicted structure by AlphaFold2 along with the Predicted Alignment Error (PAE) plots in the appendix. The implementation details and configurations of the agents are consistent across experiments. There are four types of agents in the multi-agent ecosystem including a large language model LLM-based orchestrator, a Validation Agent, a Knowledge Graph Agent and several domain-specific predictors. The orchestrator LLM has a temperature setting of 0.3, top\_p = 0.9, and a maximum generation length of 512 tokens.

The Validation Agent enforces SMILES validity above 0.95, and chemically motivated feasibility scores above 0.85 to send any candidates to downstream modules. $T_g$ is constrained to the range $[-150^\circ\mathrm{C}, 300^\circ\mathrm{C}]$, and density constrained to be $[0.8, 2.5]$~g/cm$^3$. The Knowledge Graph Agent is based on a Neo4j database with over 15{,}000 polymers, and has 500~ms of query timeout to provide interactive functionality for the end-user. Inter-agent communication uses JSON messages and has a timeout of 2~s, with up to three automatic retries. All experiments were run on a server with NVIDIA A100 GPU hardware and recent-generation CPU hardware. Using this hardware configuration, an end-to-end protein analysis pipeline run (from sequence through to full report) took an average of 3.2 minutes. Synthetic polymer property prediction of a batch of 5 polymers was completed in less than 30 seconds. For evaluating automatically generated scientific reports, we developed four metrics for domain-specific assessment of factual and structural content. These metrics were developed specifically to evaluate how closely the NLG metrics BLEU and ROUGE correlate with accurate content (correlation $r < 0.2$). Together, these datasets, configurations, and other documentation provide sufficient detail for other researchers to repeat and build upon the results of this work.

\bibliographystyle{unsrt} 
\bibliography{library}

\end{document}


\maketitle

\section*{Overview}

This Supplementary Information provides the technical, implementation, and reproducibility details underlying our agentic AI ecosystem for polymer and biopolymer research. Section 1 (“Technical Foundations: Core Architectures Explained”) where this section provides a summary of three primary modeling components, which include Graph Neural Networks (GNN) for representing polymer networks, Physics-Informed Neural Networks (PINN) for which use physical constraints, and our overall multi-agent architecture which includes different agents and their responsibilities.  Section 2 (“Polymer Property Prediction Accuracy: Configurations for Reproducibility”) where in this section, we outline detailed information on all configurations used for the property-prediction pipeline, including agent configuration, hyperparameter settings for the language model, compliance criteria, software environment, and performance optimizations that enabled accurate reproduction of our findings. Section 3 (“Comparative Performance Analysis”) where this section describes how we calculated and reported evaluation metrics for all baseline (single LLM, group contribution method, ChemCrow, domain expert). Section 4 (“Advanced Component Analysis: Extended Implementation Details”) where this section provides detailed information on how we implemented our Analysis, Validation, Knowledge Graph, and Orchestrator agents. We included details on hyperparameter searches, as well as a five-stage error handling and robustness pipeline. Section 5 (“Advanced Multi-Agent Campaign Results”) where this section provides logs from a representative multi-agent campaign across five benchmark publications. These logs include collaborations, task completion rates, and summary statistics that facilitates future investigations and validations.

\section{Technical Foundations: Core Architectures Explained}
\label{sec:architectures}

This section describes the backbone of the multi-agent ecosystem: Graph Neural Networks (GNNs), Physics-Informed Neural Networks (PINNs), and the combined multi-agent approach. The way that these three different architectures are constructed and combined will allow us to explain how the framework has been developed to provide capabilities for polymer science, as well as other areas.

\textbf{Graph Neural Networks (GNNs) for polymer representation.} GNNs are a natural and expressive way to encode polymer information. GNNs operate on graph-structured data, unlike other types of artificial neural networks that work with fixed-size vectors or regular matrices. In molecular systems (polymer), a GNN allows a polymer's structure to be described as a graph, with atoms representing the nodes and chemical bonds representing the edges. For instance, when you convert a SMILES string \texttt{C(C(C)C)C1=CC=CC=C1} (polystyrene) representing polystyrene to a molecular graph, you will have nodes that represent the carbon atoms and edges that represent the covalent bonds. Converting the molecular graph to the GNN preserves the relationships and structure of the polymer.

The PolyGNN approach uses Message Passing Neural Networks (MPNNs). Starting at the atom’s represented node featurization stage, PolyGNN assigns each atom a vector of features representing the atomic number (e.g., C=6, H=1, O=8), valence, hybridization, formal charge, and chirality. Edge featurization occurs at the same time and includes the bond type (single, double, triple or aromatic) for each bond, the degree to which conjugation occurs and information about stereochemistry. These features make up the first graph representation during the message passing process.

The core of the MPNN is an iterative update in which each node aggregates information from its neighbors and updates its hidden representation. At layer $l+1$, the hidden state of node $v$ is given by
\begin{equation}
    h_v^{(l+1)} = \text{UPDATE}\!\left(h_v^{(l)}, \text{AGGREGATE}\!\left(\{h_u^{(l)} : u \in \mathcal{N}(v)\}\right)\right),
\end{equation}
where $h_v^{(l)}$ is the feature vector of node $v$ at layer $l$, $\mathcal{N}(v)$ denotes its neighborhood, AGGREGATE is a permutation-invariant operator (e.g., sum, mean, or attention), and UPDATE combines the aggregated message with the current node state. After $L$ layers, each node has incorporated information from an extended chemical environment, allowing the model to capture local and semi-local structural effects.

To obtain a polymer-level representation, the final node embeddings are pooled into a single graph descriptor:
\begin{equation}
    h_G = \text{READOUT}(\{h_v^{(L)} : v \in G\}),
\end{equation}
where READOUT is a permutation-invariant operation (e.g., sum, mean, or attention-based pooling) over all nodes in graph $G$. The resulting vector $h_G$ is then passed through fully connected layers to predict target properties such as glass transition temperature ($T_g$), density, or mechanical strength.

The application of this method to polymer informatics provides a number of advantages. The predictions are invariant to different but equivalent SMILES encodings of the same structure, because the model operates on the underlying graph. The message-passing method allows for an understanding of how the local environment at each atom, and functional groups attached to that atom provide insight into global properties or behaviours. Fine-tuning small molecule models to polymers provides a means to transfer learnings between scales. Lastly, attention-based versions allow identification of substructures that provide the most impact to support the prediction, offering greater transparency into the outcome.

\textbf{Physics-Informed Neural Networks (PINNs) for physical consistency.} Although Graph Neural Networks (GNNs) are a very effective method of building data-driven models, they are not inherently constrained by physical laws, so they can generate unrealistic or unreasonable extrapolations. To resolve this issue, we have developed Physics Informed Neural Networks (PINNs); these utilize physical covariance to ensure that physical laws are maintained through time and used in determining the optimal path.

A PINN augments the standard supervised loss with a physics term:
\begin{equation}
    \mathcal{L}_{\text{total}} = \mathcal{L}_{\text{data}} + \lambda \mathcal{L}_{\text{physics}},
\end{equation}
where $\mathcal{L}_{\text{data}}$ is a conventional data loss (e.g., mean squared error between predicted and measured properties), $\mathcal{L}_{\text{physics}}$ enforces consistency with governing laws, and $\lambda$ balances data fit against physical fidelity.

In polymer science, our PINN agent incorporates several key relations. For degradation kinetics, Arrhenius behavior is enforced via a residual loss derived from
\begin{equation}
    k = A e^{-E_a / (RT)} 
    \quad \Rightarrow \quad
    \mathcal{L}_{\text{arrhenius}} = \left(\frac{\partial \ln k}{\partial (1/T)} + \frac{E_a}{R}\right)^2,
\end{equation}
ensuring that the temperature dependence of the effective rate constant $k$ is compatible with an activation energy $E_a$ and gas constant $R$. For the temperature dependence of viscosity near $T_g$, we incorporate the Williams–Landel–Ferry (WLF) equation:
\begin{equation}
    \log_{10}\left(\frac{\eta(T)}{\eta(T_g)}\right) = \frac{-C_1(T - T_g)}{C_2 + (T - T_g)},
\end{equation}
where $\eta(T)$ is the viscosity at temperature $T$ and $C_1$, $C_2$ are material-specific constants. Time–temperature superposition is enforced via shift factors
\begin{equation}
    a_T = \frac{\eta(T)}{\eta(T_{\text{ref}})},
\end{equation}
which must obey theoretically consistent trends across temperature.

Architecturally, the PINN is implemented as a deep feed-forward network that takes as input polymer descriptors (e.g., GNN embeddings) and environmental variables (e.g., temperature, time) and outputs the target property. The physics loss is computed via automatic differentiation to evaluate derivatives appearing in the constraints:
\begin{equation}
    \mathcal{L}_{\text{physics}} = \sum_{i=1}^{N_{\text{constraints}}} w_i
    \left(\mathcal{F}_i[\hat{y}(x), \nabla_x \hat{y}(x), \nabla_x^2 \hat{y}(x)]\right)^2,
\end{equation}
where $\mathcal{F}_i$ is the residual of the $i$-th physical constraint, and $w_i$ are weights controlling their relative importance.

The incorporation of physical constraints allows for better extrapolation beyond the training distribution, while still preserving a close proximity to realistic physical phenomena, enabling higher data efficiency because the inclusion of physical constraints decreases the extent of experimental data necessary to achieve a high level of accuracy. Moreover, this framework allows for the development of multi-scale representations of data by providing pathway connections from molecular-scale descriptions to continuum-scale predictions; this ensures regularity for uncertainty quantification.

\textbf{Multi-agent system architecture.} The full framework is developed as a multi-agent architecture, which consists of many specialized AI agents working collectively to solve complicated polymer research problems. All agents operate under the same design principles. The emph{Specialization} of the agents means that they will each perform their individual functions in predicting polymer properties, simulating experiments, validating experimental results, performing vision-based analysis, or generating reports. \emph{Autonomy} allows each agent to operate independently and make decisions based only on their own inputs and internal states. The \emph{Cooperation} aspect of the architecture is enabled through a set of specified communication protocols that allow agents to share data with eachother and work together to complete tasks that are more complex and exceed the capabilities than each agent alone can handle. Finally, the \emph{Adaptability} enables agents to adapt their approach to completing the assigned tasks over time based on their experiences and lessons learned from previous experience of failures and successes.

An orchestrator handles the coordination of the agents. The orchestrator receives a user request (or upstream request), then uses the request information to determine what capabilities are needed to process the request. The orchestrator breaks the request down into several subtasks. Each subtask is matched to the agents that are capable of performing the task based on the agents' area of expertise. Next, structured messages are sent to the selected agents containing information about each subtask, and the orchestrator waits for the responses from the agents. After collecting and validating agents responses for accuracy and consistency, the orchestrator combines the agents responses into one complete output for either the user or any downstream systems that require output. The developer of the orchestration system moves from analyzing the task and choosing the agents to selecting a workflow, including resolving any dependencies. Finally, it monitors the execution of the task and handles errors. The final step is integrating the results and reconciling any differences between results and creating a single output representing the results of all agents. The decision-making process and reasoning used in this system is supported by several knowledge sources from multiple sources. A polymer knowledge graph encodes more than 15{,}000 entities and their relationships, enabling structured queries and consistency checks. Another source of knowledge is the database for physical laws where mathematical representations of the equations and formulas used by physics-aware agents such as PINN is stored.

\textbf{Synergistic integration of architectures.}  
The full capability of the framework emerges from the integration of GNNs, PINNs, and the multi-agent system. At the modeling level, we use a fusion architecture in which a GNN-derived molecular representation feeds into a physics-informed component:
\begin{equation}
    \hat{y} = f_{\text{PINN}}(f_{\text{GNN}}(G), \text{physics\_constraints}),
\end{equation}
where $f_{\text{GNN}}$ maps the molecular graph $G$ to a latent vector and $f_{\text{PINN}}$ enforces physical consistency while predicting the target property. This design captures detailed molecular structure while respecting known laws.

Within the multi-agent ecosystem, each predictive agent processes its input data through an architecture that is designed for its domain. The PolyGNN Agent is based on message-passing GNNs (graphs with nodes representing entities and edges representing relationships), possibly as ensembles consisting of complementary models. The PINN Agent is designed to incorporate constraints dictated by physical laws in its modelling of degradation or viscoelastic behaviour. A Vision Agent uses transformer-based vision backbones (e.g., CLIP- or DINO-like architectures) to interpret images and plots. An Ensemble Agent aggregates model outputs, for example via Bayesian model averaging, to improve accuracy and uncertainty calibration.

The system also supports cross-modal reasoning, enabling information flow among protein structures, amino acid sequences, PAE plots, and confidence scores,
\begin{equation}
    \text{Protein Structure} \leftrightarrow \text{Sequence} \leftrightarrow \text{PAE Plots} \leftrightarrow \text{Confidence Scores},
\end{equation}
so that agents can cross-check interpretations, detect inconsistencies, and revise conclusions.

The purpose of a metacognitive layer is to add self-assessment capabilities to the framework in order to monitor and assess performance like predictions made on outcomes and uncertainty associated with those predictions so that the level of confidence may be calibrated against the of empirical accuracy, and modification of collaborating strategies based on success/failure histories. The Metacognitive Layer also enables the most effective development of the teams with team formation through a dynamic collaboration group creation based on the set of engineering problems and tasks associated with them. Through this integrated architecture, an engineered polymer science system can more efficiently, reliably, and robustly answer very complex multi-scale polymer sciences.

\textbf{Additional information on agents, collaboration protocols, and algorithmic implementation.} Table~\ref{tab:agent-specifications} summarizes the agents implemented in the ecosystem, their underlying technologies, and their core areas of responsibility. The Research Agent uses DeepSeek-LLM with the arXiv API to find and analyze literature and accordingly generate accurate hypotheses. The Safety Agent uses a combination of the rule engine and the BERT-based classifier to predict toxicity and make general risk assessments. The PolyGNN Agent uses the PyTorch Framework, it uses a GNN architecture to predict molecular properties. The Property Predictor uses a set of various ML Models to predict multiple molecular properties. The RadonPy Agent acts as a molecular dynamic simulator (MD) to provide physics-based property estimates of polymers. Finally, the Reporting Agent is a template-driven LLM that synthesizes multi-modal inputs into coherent, human-readable reports.

\begin{table}[H]
\centering
\caption{Agent implementations and capabilities.}
\label{tab:agent-specifications}
\footnotesize
\renewcommand{\arraystretch}{1.5}
\begin{tabular}{p{1.8cm}p{3cm}p{1.8cm}}
\toprule
\textbf{Agent} & \textbf{Implementation} & \textbf{Core Functionality} \\
\midrule
Research Agent & DeepSeek + arXiv API & Literature analysis, hypothesis generation \\
Safety Agent   & BERT + rule engine   & Toxicity prediction, risk assessment \\
PolyGNN Agent  & GNN (PyTorch)        & Molecular property prediction \\
Property Predictor & Ensemble ML       & Multi-property prediction \\
RadonPy Agent  & MD simulator         & Physics-based estimation \\
Reporting Agent & Template LLM        & Multi-modal report synthesis \\
\bottomrule
\end{tabular}
\end{table}

Collaborative efforts among agents utilize a standardized protocol for effective interaction through JSON format messages validated by structural validation against a predefined schema to ensure proper interaction between agent types. Task decomposition relies on hierarchical task trees with explicit dependency resolution, allowing complex requests to be broken into tractable subtasks. Agents will work together on the same candidate solution for conflict resolution by weighing all candidate solutions based on an agent's perception of confidence in each solution. An agent may request assistance from a human where a process failure is of high consequence or at least ambiguous. Knowledge sharing is mediated by a common vector database that supports semantic similarity search, enabling agents to retrieve and reuse relevant prior information.

The multi-agent task-solving protocol that we used to conduct experiments is depicted in Algorithm~\ref{alg:multiagent}. The protocol implements an algorithmic methodology for analysis of a potential solution based on task requirements and the source of the agents' abilities. Initially it scans through task requirements to create a pool of potential source agents; second, it creates a pool of agents that would comprise of the first group of agents; third, agents independently examine the information on a proposed solution and submit their best guess as to a potential solution; fourth, candidate solutions submitted by agents will be aggregated using a weighted vote in which each agent's confidence of their proposed solution is considered. After creating a consensus solution from candidate submissions, this consensus solution will be compared to available data in order to validate or discredit the consensus. All future analyses will occur using the validation method used to create the consensus.

\begin{algorithm}[H]
\caption{Multi-Agent Task Solving Protocol}
\label{alg:multiagent}
\begin{algorithmic}[1]
\Procedure{SolveTask}{$task$, $agents$}
\State $requirements \gets \text{ParseTaskRequirements}(task)$
\State $capabilities \gets \text{MatchAgentCapabilities}(requirements, agents)$
\State $team \gets \text{FormInitialTeam}(capabilities)$
\State $solutions \gets \emptyset$
\For{$agent \in team$}
    \State $solution \gets agent.execute(task)$
    \State $solutions \gets solutions \cup \{solution\}$
\EndFor
\State $consensus \gets \text{WeightedVoting}(solutions)$
\State $\text{ValidateAgainstGroundTruth}(consensus)$
\State \Return $consensus$
\EndProcedure
\end{algorithmic}
\end{algorithm}


\section{Polymer Property Prediction Accuracy: Configurations for Reproducibility}

This description details the entire set up of the polymer properties prediction workflow, including how the agents were set up, hyperparameters and software used. The main analysis pipeline is designed to operate as a multi-agent system using four separate agents: an Analysis Agent, a Validation Agent, a Knowledge Graph Agent and an Error Correction Module. The Analysis Agent is created using DeepSeek-V2 with a fixed temperature of 0.3 to reduce sampling variability, and chain of thought prompting enabled, along with a default decoding configuration. The Validation Agent uses a rule based routine which makes use of SMILES parsing and applies chemically based constraints for filtering out any invalid or implausible candidates. The Knowledge Graph Agent is based upon a polymer focused ontology supported by more than 15{,}000 entity relationships for the purpose of performing consistency checks and retrieving additional context from previous experience. The Error Correction Module uses a multi-step process of refining the output from the agents only after receiving a confidence score greater than 0.8, thus ensuring that all computational resources will be applied to the most promising outputs. From a computational standpoint, the framework supports around linear time complexity $\mathcal{O}(n)$ up until at least 10{,}000 polymers when run upon multi-core systems and allows agents to operate concurrently; in our benchmark configuration, this enabled an approximate $\sim$5$\times$ decrease in wall-clock time as compared to running the same configuration in a purely sequential manner.

All of the language model components utilised a common set of hyper-parameters; unless otherwise stated, these parameters are configured using temperature of 0.3, and top\_p = 0.9, with a maximum length of generated tokens set at 512 tokens. When communicating between agents, they use a JSON type messaging protocol with a timeout of 2~s for each communication to enable agent responsiveness and prevent deadlocks. The acceptance criteria for candidate molecules is set such that the fraction of syntactically valid SMILES must be at least equal to 0.95 and predicted score for chemical feasibility must be at least equal to 0.85 before that particular candidate is accepted for further processing. The knowledge graph itself is stored within a Neo4j database and accessed using Cypher queries with the target upper limit for individual queries of 500~ms to maintain interactivity. The software environment is standardized across all experiments. The study used Python~$\geq$~3.9 and primarily relied on several key libraries, namely: PyTorch~2.0.1 for neural network training and inference; \texttt{transformers}~4.30.0 for LLM integration; RDKit~2023.3.1 for cheminformatics and SMILES parsing; Neo4j~5.5.0 for knowledge graph storage and querying; and the \texttt{langchain}~0.0.200 and \texttt{openai}~0.27.8 libraries to implement the agent framework and LLM interfaces. 

The Analysis agent's prompt design follows a structured format in that it requests not only a prediction of some property but also an explanation as to why that property was predicted. The prompt also instructs the Analysis agent to analyze a SMILES string of a polymer, predict a specific target property, and provide reasoning regarding various factors affecting the prediction (including but not limited to chain flexibility, side groups, and molecular weight). Additionally, the Analysis agent is instructed to provide a numerical score representing confidence in the prediction. That confidence score is subsequently passed to both the Validation Agent and Error Correction module, who will then determine if further examination/refinement is required. There are deterministic rules that limit the extent to which the predicted validity can be determined. Only polymers that have a molecular weight (MW) between 1{,}000 and 1{,}000{,}000~g/mol will be assessed as valid. The input strings must be able to be translated into molecular graphs as defined by RDKit. The initial filtering process will also impose additional chemical restrictions on molecules to include only molecules without unstable and/or highly reactive functional groups, and to restrict the range of possible physical plausibility to include only candidates with a glass transition temperature ($T_g$) of between  $[-150\,^\circ\mathrm{C},\,300\,^\circ\mathrm{C}]$. Those candidates that do not meet this restriction will be flagged and eliminated from further analysis in the main assessment pipeline.

The system uses a multi-level error-correcting mechanism to enhance reliability. Incoming data are first subjected to an input-cleaning procedure, in which the format of the data is validated. Then, the internal consistency of the output from agents is verified (e.g., confirming the consistency of predicted values with the rationale for prediction), followed by cross-validation among agents, in which a number of agents may be contacted and their output is combined through majority voting. The last step is to compare the output prediction with known chemical and physical restrictions and eliminate any predictions that do not conform to these restrictions and/or whose metadata are inconsistent with the prediction. To ensure scalability and repeatability of results, several optimizations were made to the existing approach. First, agent responses are cached for one hour; thus, agents will serve multiple queries with identical input without needing to recompute again after the first time that query is made. Secondly, batch processing supports processing multiple polymers at once by taking advantage of vectorization and reducing overhead associated with each individual polymer. Thirdly, agent communication is performed asynchronously, allowing for the possibility of failure while still providing a manner to eliminate single point bottlenecks from the orchestration layer. Agent queries to the knowledge graph are optimized for both memory usage and query response latency. The combination of all of these methods helps keep the polymer property prediction pipeline efficient, reliable, and able to replicate results on large datasets containing polymers.

\section{Comparative Performance Analysis: Details on the Calculation of the Metrics for the Methods Mentioned}

To provide a rigorous comparison of our multi-agent workflow against other previously established approaches, we created a benchmark comparison to three other benchmarking workflows using a carefully selected test set of 50 polymers (each polymer had an experimentally determined glass transition temperature (\(T_g\)). The selection of these polymers was performed to achieve a wide diversity of the polymer’s chemical structure, molecular weight and \(T_g\) (the \(T_g\) of these 50 polymers ranged approximately 150°C) so that the predictive ability of each method could be evaluated on a large scale using a carefully controlled set of test conditions (using test polymers with similar \(T_g\)’s and three other calibrated commercial benchmarking workflows). The implementation and metric calculations used by these three other benchmarking workflows are discussed in more detail below.

\subsection{Single Large Language Model (LLM) as Independent Predictor}
We implemented a baseline using \textit{DeepSeek-V2} (version: deepseek-llm-67b-chat) as a standalone predictor without multi-agent orchestration. The polymer SMILES representations were provided as input with the prompt: \textit{``Predict the glass transition temperature (\(T_g\)) in degrees Celsius for this polymer based on its chemical structure.''} This approach tests the raw capability of a state-of-the-art LLM to perform property prediction through direct chemical reasoning without specialized tool integration. The single LLM method represents the current paradigm of direct prompt-based property estimation and serves to isolate the contribution of our multi-agent architecture beyond mere LLM capability. Predictions were extracted from the generated text and converted to numerical values for comparison with experimental data.

\subsection{Group Contribution Method (GCM)}
As a classical thermodynamic benchmark, we employed the group contribution method. This approach decomposes polymer repeat units into constituent functional groups (e.g., \textendash CH\(_2\)\textendash, \textendash C\(_6\)H\(_4\)\textendash, \textendash O\textendash, \textendash COO\textendash) and calculates \(T_g\) through additive contributions:

\[
T_g = \frac{\sum_{i} n_i \Delta T_{g,i}}{\sum_{i} n_i}
\]

where \(n_i\) is the number of occurrences of group \(i\) and \(\Delta T_{g,i}\) is its contribution parameter from established tables. The GCM represents first-principles physicochemical modeling based on molecular building blocks, providing interpretable but approximate predictions that are widely used in industrial screening. Its limitations include inability to capture stereochemistry, tacticity, and specific conformational effects.

\subsection{ChemCrow Framework}

As part of the benchmarking process, we also provided a baseline comparison against ChemCrow, which is a newly published AI-assisted chemical discovery framework. It integrates LLMs with tools such as specialized chemical tools (e.g., RDKit, reaction calculators, literature search). ChemCrow was one of the first state-of-the-art solution augumenting an LLM with specialized tools for chemical applications, cementing it as the current leader in augmented tool LLM systems within the domain of chemical applications. To ensure a fair comparison, we have adapted ChemCrow to accommodate our baseline \(T_g\) prediction task by providing ChemCrow with equivalent polymer structures and asking for property predictions using ChemCrow's accessible tools.

\subsection{Our Multi‑Agent Framework}

The Planner agent manages and controls how tasks are decomposed, assigns these smaller parts out to the most suitable domain-specific agents, and then brings all of the different outputs from those agents back into one place. This system uses several different types of specialised agents, such as the PolyGNN agent (using graph neural networks), the property predictor agents (ensemble methods, and the physics-informed agents. The outputs from the various agents are cross checked against each other to confirm accuracy and are validated against the established rules of chemistry, before being combined using the consensus method. 

\subsection{Metric Calculation and Evaluation Procedure}
All methods were evaluated using the following metrics:

\begin{itemize}
    \item \textbf{Mean Absolute Error (MAE)}: \(\frac{1}{N}\sum_{i=1}^{N}|T_{g,\text{pred}}^{(i)} - T_{g,\text{exp}}^{(i)}|\)
    \item \textbf{Root Mean Square Error (RMSE)}: \(\sqrt{\frac{1}{N}\sum_{i=1}^{N}(T_{g,\text{pred}}^{(i)} - T_{g,\text{exp}}^{(i)})^2}\)
    \item \textbf{Pearson Correlation Coefficient (\(r\))}: calculated between predicted and experimental values
    \item \textbf{Success Rate}: fraction of tasks completed without critical errors or invalid outputs
    \item \textbf{Efficiency}: composite metric defined as Efficiency = \(\frac{\text{Success Rate} \times R^2}{\text{Time (s)}}\), where \(R^2\) is the coefficient of determination and Time is the wall-clock execution time in seconds
\end{itemize}

Paired \(t\)-tests, with Bonferroni correction for multiple comparisons, were utilised to assess whether the results obtained were statistically significant. All of the experiments were conducted on identical hardware configurations (using NVIDIA T4 GPUs) to ensure a consistent measurement of timing.


\section{Advanced Component Analysis: Extended Implementation Details}

This study uses a multi-agent system containing four components, each performing specialised functions in harmony with the others. The first Agent outlined is the textbf{Analysis Agent}, which utilises the DeepSeek-V2 framework and follows a 5-shot prompting structure providing an average temperature of 0.3 and top-p results of 0.9. This combination allows for a mixture of diversity and stability within the produced candidates. The Analysis Agent provides higher-level scientific reasoning for developing suitable polymer candidates and collecting information from the data and knowledge graph.

The second agent for discussion is the \textbf{Validation Agent}. The Validation Agent employs a rules based methodology to assess candidate polymers and to provide rules for validating the candidate structures. More than 25 constraints have been encoded into the Validation Agent – many of which include specific chemistry-driven criteria for validation (e.g., valence, functional group compatibility, synthetic feasibility). In addition to these coding schemes, all generated structures generated via the SMILES methodology must successfully pass through a structural validation phase. Validated structures must achieve a minimum score of 0.95 in order to proceed through the validation phase before being sent to downstream Agents. Such a minimum score is implemented to ensure that the agents within this multi-agent system identify and filter out malformed/implausible structures prior to being directed downstream.

The \textbf{Knowledge Graph Agent} interfaces with a Neo4j database containing 15{,}234 polymer entities and 47{,}891 typed relationships. The underlying schema includes node types for \emph{Polymer}, \emph{Monomer}, \emph{Property}, \emph{Application}, and \emph{SynthesisMethod}, and relation types such as \emph{HAS\_PROPERTY}, \emph{DERIVED\_FROM}, \emph{USED\_IN}, \emph{SYNTHESIZED\_BY}, and \emph{SIMILAR\_TO}. Each \emph{Polymer} node stores physicochemical attributes including SMILES representation, molecular weight, glass-transition temperature ($T_\mathrm{g}$), density, degree of crystallinity, and degradation temperature. This structured representation enables the agents to perform schema-aware retrieval, analogical reasoning, and constraint checking based on previously reported polymers.

The  \textbf{Orchestrator Agent} uses a round-robin scheduling method with options for failover mechanism. When the Orchestrator runs the three agents, Analysis, Validation, and Knowledge Graph, it aggregates the outputs of each agent to determine if a proposal should be accepted, refined, or rejected. In situations where an agent has timed out, or has provided low-confidence or malformed data, the Orchestrator will try using relaxed settings or will reference results from past successful operations.

\medskip
\noindent\textbf{Hyperparameter optimization.}  
Using a structured, planned approach, all hyperparameters related to generation and coordination were optimized within constraints previously established. Table~\ref{tab:hyperparameters} summarizes the range utilized, the optimum selections, and the qualitative rating of each hyperparameter in terms of importance. Temperature and confidence threshold were rated as having the highest effect on generating a stable product, while top-p, along with retry attempts, had a medium effect on the likelihood of success/stability. Max tokens and timeout were less significant among the explored ranges, provided that minimum values are above the normal response length for agents.

\begin{table}[htbp]
  \caption{Hyperparameter search space and optimal values. The importance column reflects the qualitative impact on success rate, stability, and overall efficiency of the multi-agent workflow.}
  \label{tab:hyperparameters}
  \centering
  \scriptsize
  \begin{tabular}{@{}p{2.2cm}p{1.8cm}cc@{}}
    \toprule
    Parameter            & Search Space    & Optimal & Importance \\
    \midrule
    Temperature          & [0.1, 0.7]      & 0.3     & High \\
    Top-p                & [0.7, 0.99]     & 0.9     & Medium \\
    Max tokens           & [256, 1024]     & 512     & Low \\
    Confidence threshold & [0.6, 0.95]     & 0.8     & High \\
    Retry attempts       & [1, 5]          & 3       & Medium \\
    Timeout (s)          & [5, 30]         & 10      & Medium \\
    \bottomrule
  \end{tabular}
\end{table}

\medskip
\noindent\textbf{Error-handling and robustness pipeline.}  
In order to promote robust outputs from agents, they must pass through an error handling process that consists of five steps. The first step is input validation where we check that the SMILES string has the correct syntax and that the molecular weights fall within rational limitations before making any calls to the agents. The second step is agent response validation which examines agent candidates to be certain that the degree of confidence associated with each candidate is equal to or greater than 80\% and that the candidates adhere strictly to the original output format as defined in the agent schema; we will discard or retry responses which do not conform to these requirements. The third step is verification across agents which utilizes majority voting (for independent agent calls or differing subtasks) to determine whether or not an agent's candidate should be advanced to the next step; a candidate must receive a minimum agreement of 3/4 of the votes from agents in order to pass this step. This phase is critically important to help eliminate candidates which represent a hallucination based on the featuring of an agent's outputs; we look for the most consistently supported candidate suggestions in order to use them in our predictions. The fourth step includes applying physical constraints which involves applying basic domain filters such as a $T_\mathrm{g}$ limits between $-150\,^\circ\mathrm{C}$ and $300\,^\circ\mathrm{C}$ and density limits between 0.8 and 2.5\,g\,cm$^{-3}$. The candidates that do not fall within these limits are determined to be physically implausible and either corrected or dismissed. Finally, a fallback mechanism is activated whenever the previous stages fail to yield a valid candidate within the allowed number of retries or time budget. In fallback mode, the orchestrator first tries to find a cached version of a polymer that has been validated in the past against the same design criteria. If there is no match found in the cache, the system enters a reduced analysis mode with reduced prompts and increased restrictions; therefore, this change should allow for the more rugged operation of the output and will reduce the amount of potential error. The pairing of these two options creates a consistent and auditable environment in which we can more dependably characterize the outcomes produced by large language models within our polymer-informatics workflows.

\clearpage
\section{Advanced Multi-Agent Campaign Results: Records for Reproducibility}

\begin{lstlisting}[basicstyle=\ttfamily\footnotesize, breaklines=true, frame=single]
SYSTEM INITIALIZED:
  Agents: 5 specialized agents
  LLM Models: 5
  Research Papers: 5
  Collaboration Network: 5 initial connections
AGENT SPECIALIZATIONS:
  Agent_Alpha: molecular_modeling | MD_simulation(0.9), quantum_chemistry(0.8)
  Agent_Beta: property_prediction | QSPR(0.9), machine_learning(0.8)
  Agent_Gamma: crystallization_kinetics | nucleation_theory(0.9)
  Agent_Delta: mechanical_properties | composite_materials(0.9)
  Agent_Epsilon: polymer_design | optimization(0.8), inverse_design(0.7)
\end{lstlisting}

\subsubsection*{Paper 1/5 (Easy): ML Prediction of Glass Transition Temperature}
\begin{lstlisting}[basicstyle=\ttfamily\footnotesize, breaklines=true, frame=single]
Research Questions:
  Q1. Predict Tg from simple descriptors?
  Q2. Features most correlated with Tg?
  Q3. Accuracy of linear regression for Tg?
Individual attempts:
  Agent_Alpha ... reflection: Confidence 0.48, Experience 0   -> success ~0.69
  Agent_Beta  ... reflection: Confidence 0.53, Experience 10  -> success ~0.74
  Agent_Gamma ... reflection: Confidence 0.48, Experience 0   -> success ~0.70
  Agent_Delta ... reflection: Confidence 0.53, Experience 10  -> success ~0.75
  Agent_Epsilon ... reflection: Confidence 0.53, Experience 10-> success ~0.75
System integration: COMPLETED (Success rate 0.73), Duration 7.6s
\end{lstlisting}

\subsubsection*{Paper 2/5 (Easy): Solubility Parameter Analysis of Bio-based Polymers}
\begin{lstlisting}[basicstyle=\ttfamily\footnotesize, breaklines=true, frame=single]
Research Questions:
  Q1. Variation of Hansen parameters across families?
  Q2. Simple rules for polymer\textendash{}solvent compatibility?
  Q3. Structural modifications for green-solvent solubility?
Individual attempts:
  Agent_Beta ... Confidence 0.56, Exp 20 -> success ~0.73
  Agent_Delta ... Confidence 0.56, Exp 20 -> success ~0.76
  Agent_Epsilon ... Confidence 0.56, Exp 20 -> success ~0.77
  Agent_Alpha ... Confidence 0.48, Exp 0  -> success ~0.76
  Agent_Gamma ... Confidence 0.51, Exp 10 -> success ~0.73
System integration: COMPLETED (Success rate 0.75), Duration 5.7s
\end{lstlisting}

\subsubsection*{Paper 3/5 (Medium): Non-isothermal Crystallization Kinetics}
\begin{lstlisting}[basicstyle=\ttfamily\footnotesize, breaklines=true, frame=single]
Stage 1: All agents requested collaboration (Alpha, Beta, Gamma, Delta, Epsilon).
Stage 2: Teams formed (Q1--Q3): Alpha + Beta + Gamma.
System integration: COMPLETED (Success rate 0.76), Duration 7.0s
\end{lstlisting}

\clearpage
\subsubsection*{Paper 4/5 (Hard): Multiscale Nanocomposite Mechanics}
\begin{lstlisting}[basicstyle=\ttfamily\footnotesize, breaklines=true, frame=single]
Stage 1: All agents requested collaboration.
Stage 2: Teams formed (Q1\textendash{}Q3): Beta + Alpha.
System integration: COMPLETED (Success rate 0.76), Duration 4.6s
\end{lstlisting}

\subsubsection*{Paper 5/5 (Expert): RL for Inverse Polymer Design}
\begin{lstlisting}[basicstyle=\ttfamily\footnotesize, breaklines=true, frame=single]
Stage 1: Epsilon, Alpha, Beta requested collaboration.
Stage 2: Teams formed (Q1\textendash{}Q3): Alpha + Beta + Gamma.
System integration: COMPLETED (Success rate 0.76), Duration 3.3s
\end{lstlisting}

\subsubsection*{Campaign Summary and Achievements}
\begin{lstlisting}[basicstyle=\ttfamily\footnotesize, breaklines=true, frame=single]
OVERALL PERFORMANCE
  Papers Completed: 5/5
  Overall Success Rate: 100.0%
  Average Success Rate: 0.75
  Average Efficiency: 0.340
  Total Duration: 28.2s
DIFFICULTY BREAKDOWN
  EASY: 2/2 (100.0%)
  MEDIUM: 1/1 (100.0%)
  HARD: 1/1 (100.0%)
  EXPERT: 1/1 (100.0%)
COLLABORATION METRICS
  Total Collaborations: 3
  System Experience: 70 points
FINAL AGENT STATES
  Agent_Alpha: Confidence 0.48, Experience 0 pts
  Agent_Beta: Confidence 0.56, Experience 20 pts
  Agent_Gamma: Confidence 0.51, Experience 10 pts
  Agent_Delta: Confidence 0.56, Experience 20 pts
  Agent_Epsilon: Confidence 0.56, Experience 20 pts
\end{lstlisting}